\documentclass[11pt,oneside,a4paper]{article}
\usepackage{workingpaper}

\title{\textbf{Global Automation Atlas}}
\author{%
  Prashant Garg\textsuperscript{1}\thanks{Correspondence: \href{mailto:prashant.garg@unibocconi.it}{prashant.garg@unibocconi.it}.}
  \and Tommaso Crosta\textsuperscript{1}
  \and Jasmin Baier\textsuperscript{2}}
\date{%
  \small\textsuperscript{1}Department of Economics, Bocconi University, Milan, Italy\\
  \small\textsuperscript{2}Blavatnik School of Government, University of Oxford, United Kingdom\\[0.8em]
  October 2026}

\begin{document}
\addtocontents{toc}{\protect\setcounter{tocdepth}{-1}}
\maketitle
\begin{abstract}
\noindent

Automation can displace or complement labour, but this need not be constant across economies. Existing exposure measures typically assign fixed scores to tasks or occupations and capture cross-country variation through employment structure. We posit that exposure to automation depends jointly on task content and country-level conditions. We use a large language model to classify 18,797 work tasks in 124 economies by exposure, labour margin, technology channel and artificial-intelligence materiality. Country conditioning changes the classification of the same task and the occupation exposure rankings, especially in lower-income economies. Country-conditioned scores improve out-of-country predictions of firms’ production-technology use and AI adoption; physical-execution exposure tracks industrial-robot use and density, while Atlas scores align with observed work-related ChatGPT and Claude use across activities and countries. The share of tasks exposed to automation ranges from 3.3\% to 61.6\% across countries. It rises with income but varies considerably within income groups. Lower-income economies are more concentrated in rule-based and labour-substituting forms of automation, whereas physical execution, planning and inference channels, together with labour-augmenting uses of artificial intelligence, become more prominent with development. Combined with employment data, we find that women are disproportionately employed in occupations with substitution-facing exposure. Across countries, higher informality is associated with a smaller exposed-task share but a larger substitution-only share among exposed tasks. Machine-learning hypothesis generation methods point to both economic and institutional drivers: automation is more feasible where firms have machinery, digital records and connected systems, but less so where tasks depend on local knowledge and trusted relationships.
\end{abstract}
\clearpage
\section{Introduction}
The automation of tasks previously performed solely by humans is rapidly transforming the organisation of work and labour markets across the global economy. Which automation is feasible and in what form may vary within production systems, infrastructures, human capital and institutions \citep{CominHobijn2010,CominMestieri2018TechnologyIncomeDivergence,CaunedoKellerShin2021,CireraCominCruz2024}. Early work on industrial automation has shown that, historically, automation consists of a non-deterministic reallocation of tasks, discretion, and control within the production processes \citep{Noble1984ForcesProduction}. This insight runs through modern task-based theory, which models automation as a process of labour displacement or reallocation from existing tasks, and reinstatement through newly created tasks \citep{AutorLevyMurnane2003,AcemogluRestrepo2018,AcemogluRestrepo2019JEP,Restrepo2024AutomationOutlook}. The central question is therefore both whether a technology \emph{can} perform a task, and how firms choose to reorganise the division between workers and machines, given the constraints they face.

Such distinctions imply that automation exposure cannot be adequately represented by a single score attached to an occupation, and also that different technologies impact tasks differently. Existing measures have therefore focused on particular technological frontiers, including computerisation risk \citep{frey2017future}, robot exposure \citep{graetz2018robots,acemoglu2020robots}, software- and AI-related patent exposure \citep{Webb2020}, AI capability alignment \citep{BrynjolfssonMitchellRock2018,felten2021occupational}, and large-language-model (LLM) exposure \citep{EloundouEtAl2024}.

These measures recover distinct and important margins of automation, but they are also limited in their scope, both on the side of automation channels, as well as their geographic relevance. Many cross-country applications assign the same automation exposure to a given task or occupation in every country and introduce country variation only through employment weights, sectoral composition or readiness indices \citep{PizzinelliEtAl2023,GmyrekBergBescond2023,CazzanigaEtAl2024GenAI}. Country differences in exposure can arise because workers perform different tasks, or because the same task faces different conditions for automation. Survey-based measures capture differences in task content within occupations \citep{LewandowskiMadonPark2025DevelopmentStages}. But even with identical task content, differences in infrastructure, capital and production organisation can change which automation is economically feasible.

We therefore measure feasible automation exposure at the task-country level: whether currently available technology can reallocate a non trivial share of labour input for a given task in a specific country setting. Using the O*NET task dictionary \citep{onet_database_29_1}, we classify $18{,}797$ standardised tasks across $124$ country contexts. These economies account for about $99\%$ of world population and GDP \citep{worldbank_wdi}. For each task-country pair, we then jointly classify four dimensions: (i) exposure records the extent of economically credible labour reallocation, (ii) the technology channel identifies the main way the task could be automated, (iii) the labour margin distinguishes between substituting for workers, augmenting their work, and cases where both are feasible and neither clearly predominates in typical deployment in that country, and finally, (iv) AI materiality identifies whether learned models are central to the automation route, rather than treating AI as synonymous with automation \citep{BrynjolfssonMitchellRock2018,EloundouEtAl2024,AmershiEtAl2019HumanAIInteraction,SvanbergLiFlemingGoehringThompson2024}. This design allows us to aggregate the same labels to occupations, industries, and countries while preserving the underlying composition of exposure. %
Assessing the validity of these scores requires evidence beyond agreement across models and prompts, which can reflect shared errors \citep{DellRambachan2026MeasurementRevolution}. We therefore compare them with established exposure measures and observed technology use.%

The same task can be classified as exposed in one country and unexposed in another. These differences also change which occupations rank as most exposed, particularly in lower-income economies. To assess whether these differences capture economically meaningful variation, we test whether they improve predictions of observed technology use. Country-conditioned scores predict firms’ technology sophistication across business functions more accurately than fixed industry scores in 11 of 12 countries \citep{CireraCominCruz2024,WorldBankFAT2024}. Adding country-specific AI-material exposure to fixed scores and readiness measures also improves predictions of firms’ reported AI use in 19 of 23 countries \citep{Eurostat2024AIAdoption}. Each test country is excluded from estimation, so these gains reflect predictive performance beyond the countries used to fit the models. In Anthropic’s data, the same work activity receives relatively more Claude use in countries where the Atlas assigns it greater exposure and AI materiality \citep{MassenkoffEtAl2026AnthropicEconomicIndex}. Even after accounting for GDP per capita, internet use and region, Atlas scores are more strongly associated with the work-related share of ChatGPT and Claude use than with overall platform use \citep{OpenAI2026SignalsData,MassenkoffEtAl2026AnthropicEconomicIndex}. Manufacturing physical-execution exposure also correlates positively with industrial-robot use and robot density across countries \citep{Eurostat2022RobotUse,IFR2026RobotDensity}. 
Figure~\ref{fig:country_conditioning_frontier_labs} presents the central comparisons. Supplementary Appendix, section~\ref{sec:validation_construct}, reports the full analysis, and Supplementary Appendix, Table~\ref{tab:validation_external_summary}, summarises the external comparisons.

Our country-level measurement of automation exposure shows development changes not only the amount of work exposed, but also its form. The exposed-task share ranges from $3.3\%$ in South Sudan to $61.6\%$ in China. Crucially, substitution-only exposure exceeds augmentation-only exposure in every country, with lower-income economies especially skewed towards substitution-facing automation, such as rule-based work, while AI-material and augmentation-facing routes become more prominent at higher income levels. A measure of automation exposure exclusively focusing on AI would therefore miss much of the economically feasible automation frontier in poorer countries and would obscure how the labour margin of AI-related exposure varies with development.

Linking our measures to employment data shows that women are disproportionately employed in occupations with substitution-facing exposure. Moreover, across countries, higher informality is associated with lower overall exposure, but substitution-only tasks constitute a larger share of the tasks that are exposed. Comparing at similar income levels by linearly adjusting for log GDP per capita, higher informality is associated with a lower substitution-only share among exposed tasks. By analysing the rationales of the classifications, we also retrieve some hypotheses on why the same task may be automatable in one country but not another: suitable machinery, digital records and connected systems can make work easier to standardise, while reliance on local institutional knowledge and trusted relationships can make it harder to replace its human content.

\section{Results}
\label{sec:results}
\subsection{Country context changes which work is exposed}
\label{sec:results_country_conditioning}

A common country-level adjustment can raise or lower overall exposure without changing occupation rankings within a country. Our task-country measure allows local conditions to affect the exposure of each task differently, and therefore to change which occupations appear most exposed.

\begin{figure}[!t]
\centering
\caption{Country context changes which tasks appear exposed and improves prediction of firms' technology use.}
\includegraphics[width=0.98\textwidth]{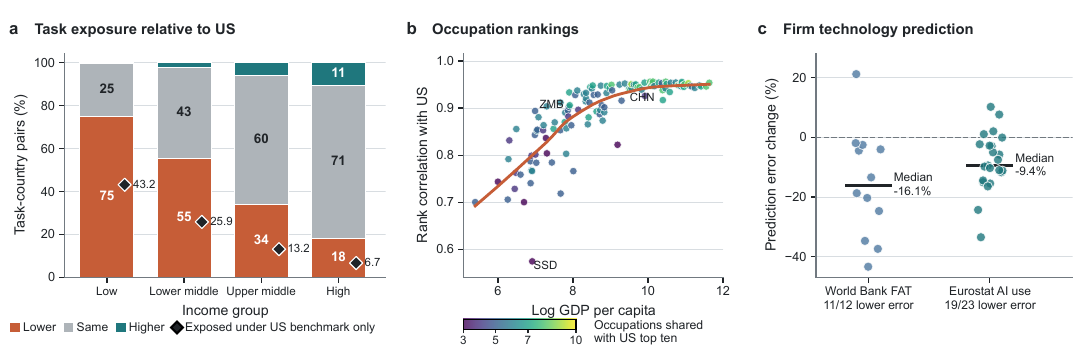}
\caption*{\scriptsize Notes: Panel~(a) compares exposure levels for identical tasks under local and US conditions. Diamonds mark the share of all task-country pairs exposed under the US benchmark but unexposed locally. Panel~(b) reports Spearman occupation-ranking correlations with the US. Colours count occupations shared with the US top ten. The curve is a locally weighted linear fit (LOWESS, span $0.5$). Panel~(c) reports percentage changes in root-mean-square prediction error for excluded test countries. Country-specific scores replace fixed scores in FAT and supplement the fixed-score-and-readiness baseline in Eurostat. Points are countries and black lines mark medians. Negative values indicate lower error.}
\label{fig:country_conditioning_frontier_labs}
\end{figure}

Figure~\ref{fig:country_conditioning_frontier_labs}a uses the US classification as a fixed benchmark and compares it with the country-conditioned label for the same task across every other country. The restriction matters most in low-income economies: $43.2\%$ of all task-country pairs are classified as exposed under the US benchmark but not under the country-specific classification, compared with $6.7\%$ in high-income economies. Applying a fixed score estimated on high income economies would therefore count many tasks as exposed even where local production conditions do not support economically viable automation.

Heterogeneity in exposure at the task level can then reflect in differential exposure of occupations across countries. In panel b, the median Spearman correlation between each country’s occupation ranking and the U.S. ranking goes from $0.95$ in high-income countries to $0.76$ in low-income countries. A variance decomposition attributes $7.3\%$ of exposure variation to country-occupation interactions, i.e. differences specific to an occupation in a given country, beyond the average country and occupation exposure levels (Supplementary Appendix, Fig.~\ref{fig:validation_country_contribution}).

These country-specific differences in exposure raise a further question: do country-conditioned scores predict firms’ technology use beyond what fixed scores can capture? Panel~(c) tests this in two firm datasets. In the World Bank Firm Adoption of Technology survey, country-conditioned labour-margin scores predict firms’ technology sophistication across business functions more accurately than fixed industry scores: prediction error is lower in 11 of 12 held-out-country comparisons, and pooled RMSE is $12.6\%$ lower \citep{CireraCominCruz2024,WorldBankFAT2024}. In Eurostat data, adding the country-conditioned AI-material score improves predictions of firms’ reported AI use in 19 of 23 held-out countries \citep{Eurostat2024AIAdoption}. Because each test country is excluded from estimation, these results show that the country-conditioned scores predict firm technology use in countries not used to fit the models.

A complementary test uses Anthropic’s Claude-use data across 95 countries, where the same O*NET activity accounts for a larger share of Claude use relative to its global share in countries where its country-specific Atlas score is higher, after controlling for country and activity-by-month fixed effects. The standardised association is small at the individual-task level (about $0.02$) but larger for general work activities: $0.19$ for overall exposure and $0.21$ for AI materiality (Supplementary Appendix, Fig.~\ref{fig:validation_anthropic_activity_hierarchy}) \citep{MassenkoffEtAl2026AnthropicEconomicIndex}.

\subsection{Automation exposure rises with income but varies considerably within income groups}
\label{sec:results_global_trends}

\begin{figure}[!t]
\centering
\caption{Cross-country patterns in automation exposure.}

\label{fig:country_opening}

\begin{subfigure}[t]{0.525\textwidth}
    \centering
    \caption{Economically exposed task share.}
    \includegraphics[width=\textwidth]{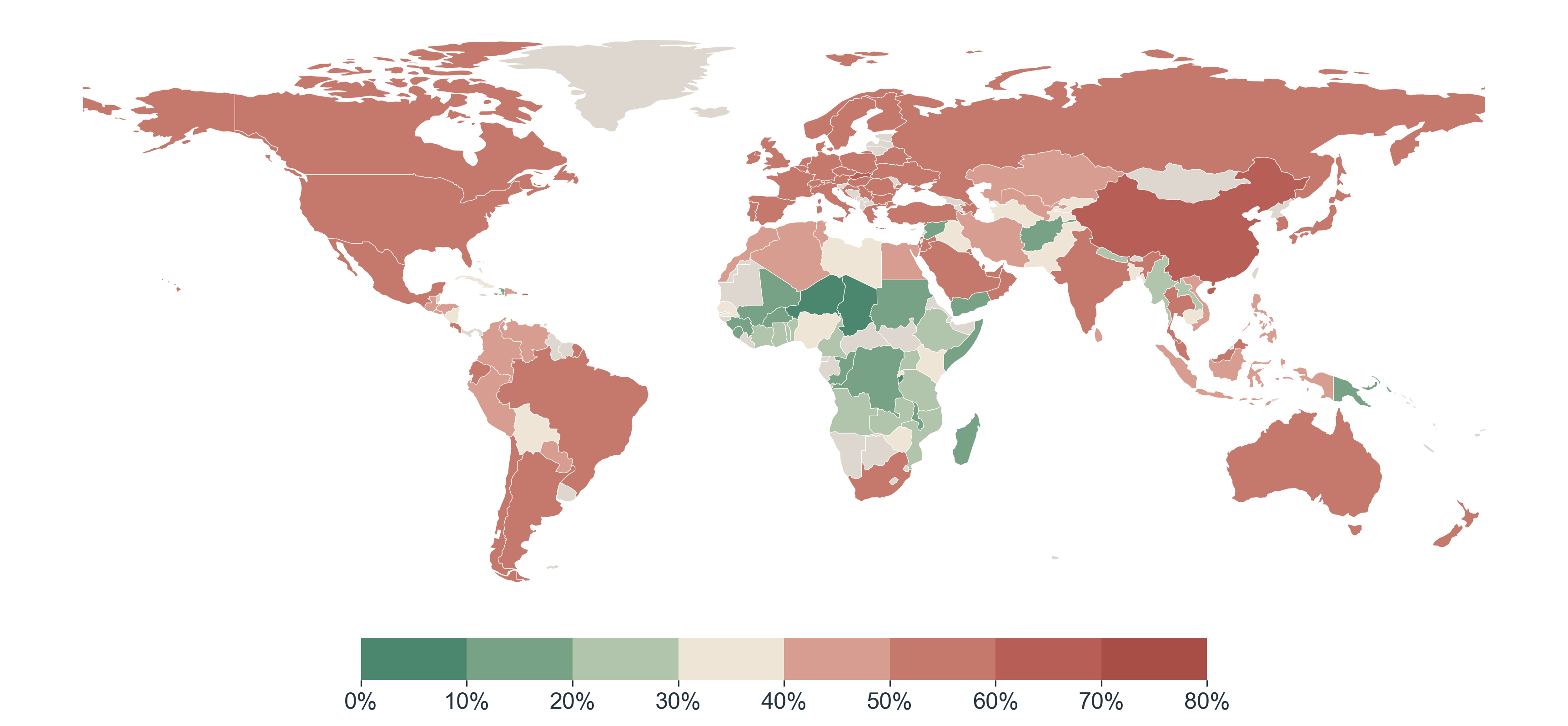}
\end{subfigure}\hfill
\begin{subfigure}[t]{0.42\textwidth}
    \centering
    \caption{Exposed task share and GDP per capita.}
    \includegraphics[width=\textwidth]{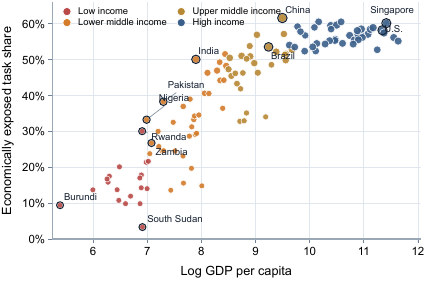}
\end{subfigure}
\caption*{\scriptsize Notes: The map covers 124 countries. The scatter includes 122 with an assigned World Bank income group. Exposure is the share of tasks at levels 2 or 3. Point size in panel~(b) denotes the level-3 share.}
\end{figure}

Figure~\ref{fig:country_opening} shows that the exposed-task share rises strongly with GDP per capita, ranging from $3.3\%$ in South Sudan to $61.6\%$ in China. Yet income does not fully determine exposure. Substantial variation remains within income tiers, especially among middle-income countries (Supplementary Appendix, Fig.~\ref{fig:appendix_within_income_variability}).

\subsection{Lower-income countries are associated with higher substitution shares of automation exposure}
\label{sec:results_margin_heterogeneity}

\begin{figure}[!t]
\centering
\caption{Substitution-only exposure exceeds augmentation-only exposure across countries.}
\includegraphics[width=0.98\textwidth]{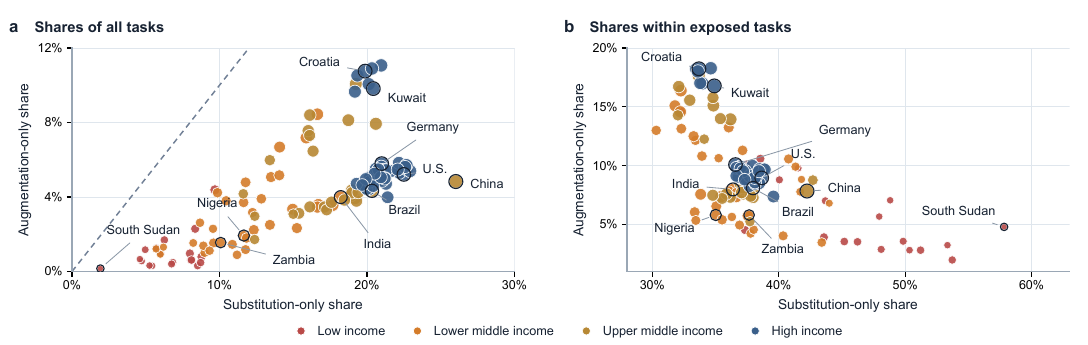}
\caption*{\scriptsize Notes: Points represent 122 countries with assigned income groups. Shares use all tasks in panel~(a) and exposed tasks in panel~(b), where the remainder is balanced-both. Point size denotes the exposed-task share. The dashed line marks equal substitution-only and augmentation-only shares.}
\label{fig:task_country_pathways}
\end{figure}

Figure~\ref{fig:task_country_pathways}a indicate that substitution-only exposure exceeds augmentation-only exposure in every country. The transition matrices in Supplementary Appendix, Fig.~\ref{fig:appendix_income_group_pathway_heatmaps}, show that tasks classified as non-exposed in lower-income groups are mainly classified as balanced-both or substitution-only when exposed in higher-income groups. Augmentation-only exposure remains comparatively small. 

Within exposed tasks, however, the margin composition varies markedly across countries (Figure~\ref{fig:task_country_pathways}b). Most low-income countries are substitution-heavy, with few purely augmentation-oriented paths, even though substitution-only exposure accounts for a smaller share of all tasks than in high-income countries. High-income countries lie closer to the middle of the distribution, with several smaller high-income economies near the upper tail of augmentation exposure. Middle-income countries show the widest dispersion. Supplementary Appendix, Fig.~\ref{edfig:appendix_polarisation_p}, shows that the balanced-both margin nevertheless remains large, covering roughly $45\%$ to $55\%$ of exposed tasks. %

\subsection{Non-AI channels remain the most prevalent source of automation for low-income countries}
\label{sec:results_channels_ai}

\begin{figure}[!t]
\centering
\caption{Rule-based workflows dominate exposure in poorer economies, while AI materiality differs sharply across channels.}
\includegraphics[width=0.98\textwidth]{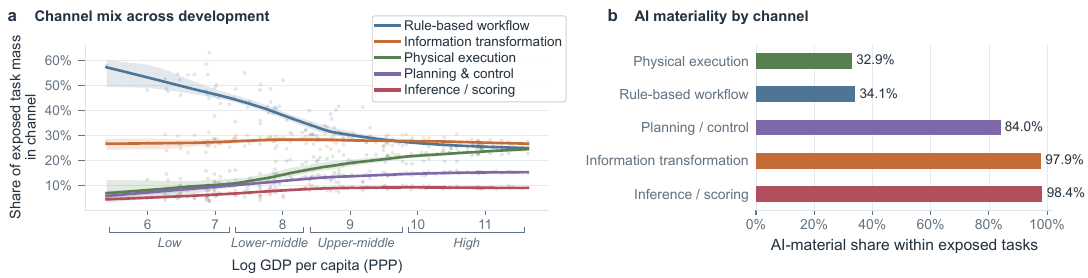}
\caption*{\scriptsize Notes: Panel~(a) plots channel shares among exposed tasks against log GDP per capita; curves are LOESS fits with $95\%$ bootstrap intervals and brackets mark World Bank income tiers. Panel~(b) reports the AI-material share within each channel. Channels are mutually exclusive, while AI materiality can occur in any channel.}
\label{fig:channel_ai_mechanism}
\end{figure}

The technologies carrying exposed work differ across income groups (Figure~\ref{fig:channel_ai_mechanism}a). Rule-based workflows account for more than $50\%$ of exposed tasks in low-income countries, about $40\%$ in lower-middle-income countries, and about $25\%$ in high-income countries. Physical execution shows the reverse pattern, rising from roughly $5\%$ to $25\%$ between low- and high-income countries, consistent with evidence linking industrial robots to more capital-intensive production \citep{graetz2018robots,acemoglu2020robots}. AI is less often central to rule-based workflows and physical execution, and more often central to information transformation, inference/scoring, and planning/control (Figure~\ref{fig:channel_ai_mechanism}b). Thus, AI accounts for only part of exposed work; many tasks, especially in lower-income countries, remain non-AI or only weakly AI-mediated.

Across all exposed tasks, AI is materially involved in roughly $35\%$ at the low-income end of the income distribution and more than $70\%$ at the high-income end (Supplementary Appendix, Fig.~\ref{fig:ai_materiality_margins}). This pattern is consistent with evidence linking AI adoption and effective use to digital, organisational, and human-capital complements \citep{CazzanigaEtAl2024GenAI,McElheranEtAl2023AIAdoptionAmerica}. But greater AI involvement does not necessarily mean greater worker augmentation. In low- and middle-income countries, AI-material exposure remains predominantly labour-substituting, while augmentation is concentrated in a smaller group of mostly high-income economies.

\subsection{Women are concentrated in substitution-facing occupations; more informal economies have less exposed work but a more substitution-heavy exposure mix.}
\label{sec:results_occupation_industry}

We aggregate task-level exposure to ISCO-08 occupations and ISIC Rev.~4 industries, the level present in many administrative datasets and structural-transformation analyses \citep{mcmillan2011_globalization,Rodrik2016PrematureDeindustrializationJEG,CaunedoKellerShin2021,LewandowskiMadonPark2025DevelopmentStages}. Replacing context-free labels with country-conditioned labels changes exposure levels substantially at both the ISCO-2 and ISIC-2 levels, with larger ranking changes for industries (Supplementary Appendix, section~\ref{sec:validation_country_contribution}).

Clerical support occupations are exposed even in low-income settings, consistent with routine-task accounts of codifiable transactional work \citep{AutorLevyMurnane2003,goos2014explaining}. Business-administration, ICT and professional occupations become more exposed with income, matching evidence on the exposure of information-processing work \citep{felten2021occupational,EloundouEtAl2024,GmyrekBergBescond2023}. Plant-operator and manufacturing exposure rise at higher GDP levels, when capital intensity makes physical automation more deployable \citep{graetz2018robots,acemoglu2020robots}. Transactional and clerical pockets are mainly substitution-facing, while professional, care, analytical and service-upgrading pockets are more augmentation-facing (Supplementary Appendix, Figs.~\ref{fig:isco_income_rotation} and~\ref{fig:industry_isic_income_main} and Tables~\ref{tab:isco_income_pockets_joint}--\ref{tab:isic_income_pockets_joint}).

\begin{figure}[!t]
\centering
\caption{Labour-margin exposure linked to gender composition and informality.}
\includegraphics[width=0.96\textwidth]{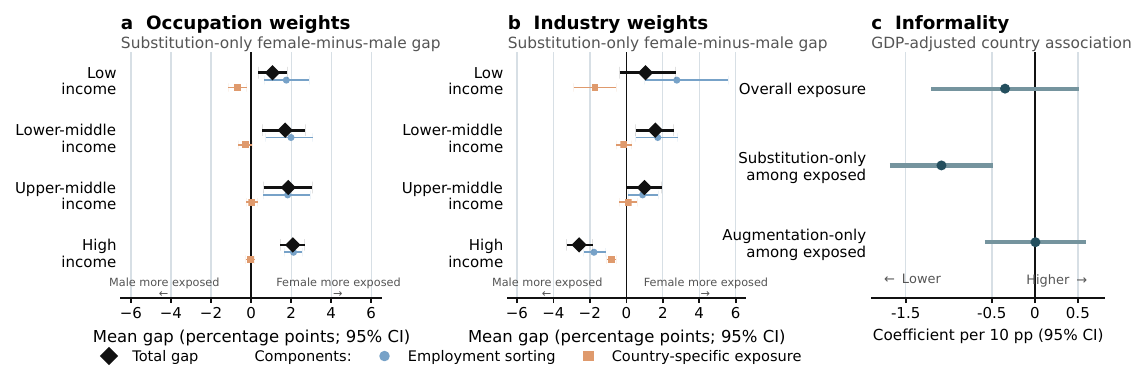}
\caption*{\scriptsize Notes: Panels~(a)--(b) decompose female-minus-male substitution-only exposure into employment sorting and country-specific exposure. Points are income-group means with country-bootstrap $95\%$ confidence intervals. Employment shares are normalised by sex within country. Both sexes receive the same country--cell exposure score. Samples cover 88 countries for two-digit ISCO-08 occupations and 72 for two-digit ISIC Rev.~4 industries. Panel~(c) reports OLS coefficients per 10 percentage point increase in informality, controlling for log GDP per capita, with HC1 $95\%$ intervals (92 countries). Labour-margin shares use exposed tasks as the denominator.}
\label{fig:gender_informality_margin_applications}
\end{figure}

Because women and men are distributed differently across occupations and industries, their employment-weighted exposure may also differ. We ask whether female employment is more concentrated in substitution-facing or augmentation-facing work, building on research on gender differences in automation and AI exposure \citep{BorrowmanKlasen2020,GmyrekBergBescond2023,GmyrekEtAl2025,CazzanigaPantonLiPizzinelliTavares2025GenderLens,AlbanesiDiasDaSilvaJimenoLamoWabitsch2025WomenAI,Lane2024WorkersAffectedAI}.
Figure~\ref{fig:gender_informality_margin_applications} shows that, in aggregate, female employment is more exposed to labour-substituting automation in both occupation and industry comparisons. The exception is industry exposure in high-income economies, where the gap reverses. The occupation difference largely reflects where women work: they are more concentrated in occupations with higher substitution-facing exposure. Country-specific exposure differences slightly offset this pattern.\footnote{Supplementary Appendix, Fig.~\ref{fig:appendix_ilostat_country_weighting} compares linkage-weighted and ILOSTAT employment-weighted occupation exposure across countries.}

Across countries, higher informality is associated with lower overall exposure but a higher substitution-only share among exposed tasks. The latter association reverses after controlling for GDP per capita: higher informality is then associated with a lower substitution-only share (Figure~\ref{fig:gender_informality_margin_applications}c).

\subsection{Emerging hypotheses propose that automation is more likely where firms have machinery, digital records and connected systems, but less so where tasks depend on local knowledge and trusted relationships}
\label{sec:results_country_predictors}

Income is strongly associated with exposure, but that relationship may partly reflect other country characteristics that vary alongside income and may also shape automation feasibility. We link country-level exposure measures to GDP per capita, internet use, investment, capital stock per worker, human capital, schooling, government effectiveness, regulatory quality and trade openness \citep{worldbank_wdi,FeenstraInklaarTimmer2015PWT,BarroLee2013Education,worldbank_wgi,GaulierZignago2010BACI}. We compare these covariates’ predictive contributions to the exposed-task share and to the substitution–augmentation mix among exposed tasks. We also analyse the rationales accompanying task-country labels to identify more specific conditions associated with differences in exposure for the same task across countries. Candidate concepts are identified in a discovery sample and evaluated in a separate held-out sample using text-based hypothesis-generation methods \citep{GentzkowKellyTaddy2019,LudwigMullainathan2024HypothesisGeneration,MovvaEtAl2025HypotheSAEs}. These analyses describe predictive associations and generate hypotheses, without establishing causal effects.

\begin{figure}[!t]
\centering
\caption{Country covariates and rationale concepts behind country-conditioned exposure and labour margins.}
\includegraphics[width=0.98\textwidth]{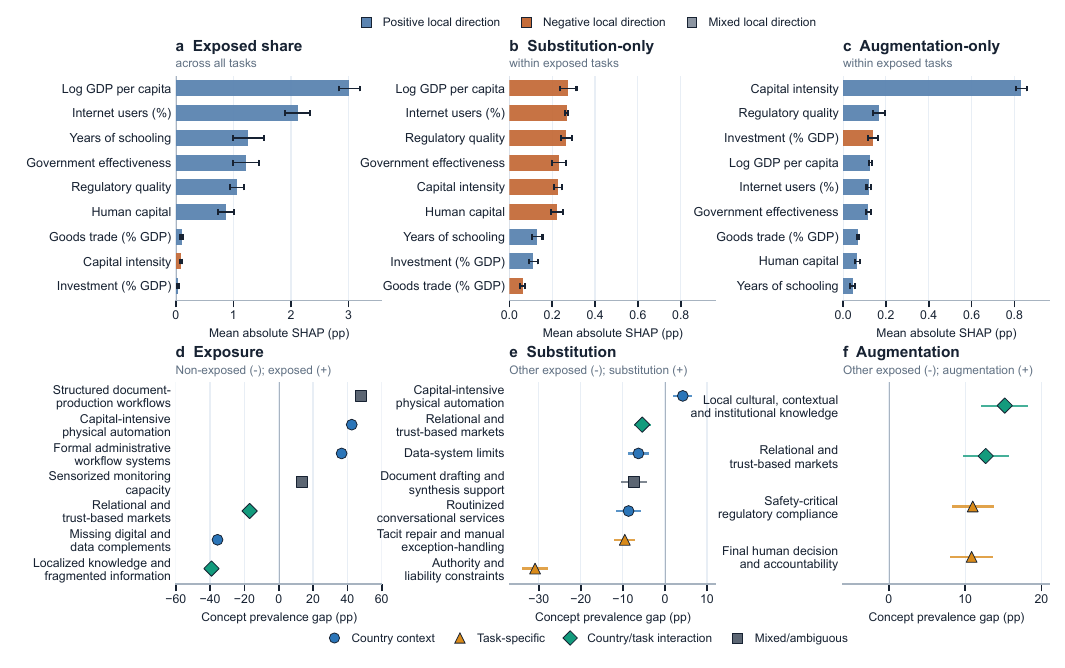}
\caption*{\scriptsize Notes: Panels~(a)--(c) report mean absolute TreeSHAP values across five forests in the 68-country complete-case sample; values are percentage points and colours show the accumulated-local-effect direction. Whiskers show $\pm1.96$ standard deviations across seeds, not sampling confidence intervals. Outcomes are exposed share among all tasks and substitution-only and augmentation-only shares among exposed tasks. Panels~(d)--(f) report first-group-minus-second-group differences in pooled rationale-concept families, with $95\%$ confidence intervals from paired standard errors. Concepts are selected in discovery samples and evaluated in separate heldout samples. Neither analysis estimates causal effects.}
\label{fig:country_covariate_feature_importance}
\end{figure}

In Figure~\ref{fig:country_covariate_feature_importance}a, log GDP per capita and internet use make the largest predictive contributions to the exposed-task share, and both are positively associated with predicted exposure, followed by level of schooling, government effectiveness, regulatory quality, and human capital. %

Within exposed tasks, substitution-only automation is negatively associated with per capita income, regulatory quality, government effectiveness and capital intensity. Augmentation-only automation is instead most strongly associated with capital intensity, with a smaller role for regulatory quality. Alternative samples, margin decompositions and importance measures are reported in Supplementary Appendix, section~\ref{sec:appendix_residual_country_explanations}.

Panels~(d)--(f) of Figure~\ref{fig:country_covariate_feature_importance} distinguish hypotheses operating at three levels. Country-level concepts describe broad conditions such as capital equipment, digital infrastructure and formal administrative capacity. Task-level concepts describe whether the work is routine, relational, judgement-intensive or physically dexterous. Interaction concepts capture cases in which the feasibility of automating a particular task depends on how it is organised in a country. For exposure, Panel~(d) shows that the main recurring rationales on the exposed side concern capital equipment, structured records and digital systems, consistent with evidence on the role of infrastructure, management practices and firm capabilities in technology adoption \citep{AndrewsNicolettiTimiliotis2020DigitalDiffusionEER,BloomVanReenen2007,CireraCominCruz2024}. Vendor-payment processing and customs documentation are hypothesized to be more exposed when invoices, approvals, tariff codes and shipment records are codified. On the non-exposed side, rationales refer more often to dispersed information, local institutional knowledge and trust-based exchange \citep{North1990Institutions,BakerGibbonsMurphy2002RelationalContracts}. To provide an illustrative example, in land-title checks and credit intermediation automation is judged more feasible where registries and verification rules are formalised, and less feasible where the work depends on local office knowledge, broker relationships or interpersonal trust.

Within exposed work, substitution and augmentation are associated with different concepts. Panel~(e) shows that substitution-only labels are less common when rationales refer to relational and trust-based markets or to authority and liability constraints, consistent with replacement being harder when performance depends on non-contractible information or responsibility for mistakes \citep{AutorLevyMurnane2003,AcemogluRestrepo2018,BakerGibbonsMurphy2002RelationalContracts}. In staff scheduling, for example, software can replace much of routine roster construction once availability, demand and constraints are codified. In engineering calculations, computation and drafting can be automated, but authority, liability and final approval limit complete human replacement. Panel~(f) shows that augmentation-only labels are more common when rationales refer to local institutional knowledge, relational markets, safety and regulatory compliance or human accountability. Consider the cases of food-safety monitoring, clinical referrals and community-resource referrals. Here, technology can flag risks or screen cases, but workers remain central to judgement, trust and accountability. Supplementary Appendix, Figs.~\ref{edfig:appendix_hypothesaes_exposure_20_concepts}--\ref{edfig:appendix_hypothesaes_augmentation_20_concepts} report the corresponding child-concept estimates.

\section{Discussion}

\label{sec:discussion}

A new wave of automation technologies is rapidly changing labour markets around the world. In this paper, we show how this technological process can look very different across countries, and how solely focusing on one particular type of technology or not accounting for meaningful differences across countries can miss important trends. We therefore construct a country-specific, task-based measurement of automation using LLMs. We show that the extent and type of automation exposure change with economic development, as well as their implication for the role of human labour. Whereas lower-income settings are more concentrated in rule-based type of automation, that are heavily labour-substituting, higher-income settings instead have a more nuanced mix of physical-execution, planning and inference types of automation, which imply more augmentation of labour. An exclusive focus on AI would overlook much of the feasible automation in lower-income economies.

While low- and middle-income countries are less exposed to automation overall, their exposure is more concentrated in labour-substituting technologies. This raises concerns about how workers would adjust if substitution led to job loss, particularly where income support, job-search assistance and protection against displacement are limited \citep{BanerjeeHannaOlkenLisker2024SocialProtection, AbebeCariaEtAl2020JobSearch, Freeman2010LaborRegulations}. Women, who already face significant barriers in low- and middle-income labour markets \citep{BorrowmanKlasen2020}, are also disproportionately employed in occupations with substitution-facing exposure. On the other hand, labour-augmenting exposure can be successfully leveraged by countries with training, certification and matching policies that help workers use new systems and move towards complementary tasks \citep{oecd2018_automation_training,LassebieQuintini2022}. Exploiting such opportunities requires a realistic picture of which tasks are currently most exposed in one's own context. The Atlas, released as a public resource, can provide first guidance on this question. 

Room for government policies to affect the direction of automation seems possible, as across countries, roughly 45–55\% of exposed tasks could plausibly be organised either to reduce labour input or to support workers who remain involved. The contrasting associations for government effectiveness (positive with overall exposure but negative with substitution-only exposure) further suggest that conditions associated with the reach of automation need not predict whether exposed work is organized to replace labour. This deployment-contingent exposure means that labour market consequences of these technologies crucially depend on how firms organise work around the use of these technologies. For example, during the computerization era, IT use was more strongly associated with demand for skilled workers in firms that decentralized decision-making and reorganized work \citep{BresnahanBrynjolfssonHitt2002}. Workplace institutions can matter as well: unions and works councils may encourage training and job redesign that strengthen the complementarity between workers and new technologies \citep{BellocEtAl2022AutomationRiskWorkplaceGovernance,BellocBurdinLandini2023AdvancedTechnologiesWorkerVoice}. Lastly, regulations may shape how tasks involving compliance, accountability and judgement are divided between workers and automated systems, seeing that regulatory quality appears negatively associated with substitution-only exposure. Our informality results raise related questions about how formalisation might affect automation opportunities. Greater informality is associated with a higher substitution-only share among exposed tasks before income adjustment, but a lower share after linear adjustment for log GDP per capita. While this cannot establish a causal relationship, one hypothesis is that formalisation strengthens incentives for labour substitution when effective labour costs rise faster than productivity. Evidence from Brazil, for example, shows that stricter enforcement against informal hiring raises effective labour costs, particularly for less productive formal firms \citep{Ulyssea2018FirmsInformalityDevelopment}. Another is that improvements in financing, record keeping and organisational capabilities enable automation, with management choices potentially shaping the extent of substitution or augmentation, as discussed above. This motivates testing whether bundling formalisation policies with training and productivity-enhancing support changes technology adoption, labour reallocation and workers’ continued participation in formal employment.

Taken together, the patterns in this paper suggest that income captures only part of the country differences in automation exposure. Digital infrastructure and workforce capacity appearing as important factors is consistent with accounts that emphasize connectivity and workers’ ability to learn and adapt as complements to technology diffusion \citep{AndrewsNicolettiTimiliotis2020DigitalDiffusionEER,NelsonPhelps1966HumanCapitalDiffusion}. The strong association between capital intensity and augmentation, alongside the smaller contribution of regulatory quality, aligns with evidence in which capital and complementary investment support technology use that raises worker productivity \citep{BresnahanBrynjolfssonHitt2002,RaischKrakowski2021AutomationAugmentation}. These relationships may help explain why AI-material exposure is more substitution-facing in lower-income settings and more augmentation-facing in higher-income settings \citep{BrynjolfssonLiRaymond2023GenerativeAIatWork,NoyZhang2023ProductivityEffectsGenAI,DellAcquaEtAl2026JaggedFrontier}.

Recent survey evidence from Europe shows that country context shapes the tasks workers perform \citep{JimenezMartinezEtAl2026DistributionTasksEurope}. We complement this evidence by asking whether the same standardised task is automatable in different production settings, and whether automation replaces or supports workers. By locating digital records, formal procedures, local knowledge, trust and accountability in specific task-country comparisons, our rationale analysis connects research on technology diffusion and firm capabilities \citep{CominHobijn2010,AcemogluAghionZilibotti2006DistanceToFrontier,CireraCominCruz2024} with institutional and relational-contract accounts of how information and trust shape exchanges \citep{North1990Institutions,BakerGibbonsMurphy2002RelationalContracts}. Future work could exploit staggered administrative-digitisation reforms to test whether automation rises most in record-intensive tasks, while tasks involving dispersed information, local institutional knowledge and trust remain less exposed and, when exposed, are more often augmentation-only. The greater prevalence of physical-execution exposure in higher-income settings, alongside capital intensity’s strong association with augmentation, further suggests an opportunity to test whether changes in machinery tariffs alter capital investment and physical automation, and with what labour-margin consequences.

Finally, our measure has two principal limitations. First, it captures feasible exposure under current country contexts, and not realised adoption, real employment, wage or productivity changes. Second, a common task dictionary imposes a boundary: O*NET makes comparison possible but was developed for the United States and cannot capture every country-, sector-, firm- or workplace-specific variation in task content, especially in informal or locally organised work. Extending task dictionaries and validating individual task-country labels with local evidence are therefore important next steps. The Atlas provides a common starting point for that work and can be linked to labour-force, firm, education and policy data to study when feasible exposure becomes adoption, through which labour margin and with what labour market consequences for workers.

\section{Materials and Methods}
\label{sec:data_measurement}

We measure automation exposure at the task-country level. The task universe consists of $18{,}797$ standardised work tasks from O*NET version 29.1 and their occupation links \citep{onet_database_29_1}. 

We apply a fixed classification protocol using \texttt{gemini-3.1-flash-lite} to every task in 124 country contexts. These economies account for $99.0\%$ of world population and $99.1\%$ of world GDP. The resulting dataset contains $2{,}330{,}776$ task-country observations. The complete classifier prompt, model configuration and data inventory are reported in Supplementary Appendix, section~\ref{sec:appendix_measurement_inventory} and Table~\ref{tab:data_inventory}. Supplementary Appendix, Fig.~\ref{edfig:task_country_exposure_measures} summarises the measurement and aggregation pipeline.

\paragraph{Exposure levels} Exposure is defined as whether currently available technology can perform, transform, or materially reorganise a nontrivial share of the task core at sufficient quality, reliability, and effective cost. This definition follows the task-based view that automation reallocates task content between labour and technology \citep{AutorLevyMurnane2003,AcemogluRestrepo2018}. We use four exposure levels:
\methoddefinition{Level 0. no credible economic automation margin}{current technology cannot create a credible labour-reallocation margin for the task core.}
\methoddefinition{Level 1. assistive-only contact}{technology may help, but does not materially reduce human labour input on the task core.}
\methoddefinition{Level 2. meaningful partial economic exposure}{a nontrivial share of the task core can be reassigned away from labour, while humans remain central.}
\methoddefinition{Level 3. extensive economic exposure}{a large share of the task core can be reassigned away from labour, or the task is mostly automatable in typical settings.}

Throughout the paper, the economically exposed share comprises levels 2 and 3, while high exposure refers to level 3 alone. 

\paragraph{Dominant technology channel} Each labelled task is assigned one dominant channel from a fixed taxonomy of five technology archetypes, with \textit{none} reserved for cases where no automation mechanism is identified:
\methoddefinition{Physical execution}{robotics, mechanical actuation, and embodied systems that physically perform the task, such as warehouse picking or assembly automation \citep{graetz2018robots,acemoglu2020robots}.}
\methoddefinition{Rule-based workflow}{scripted software, robotic process automation, and deterministic business rules that execute structured workflows without learning, such as claims routing or payroll rules \citep{AutorLevyMurnane2003,SyedEtAl2020RPA}.}
\methoddefinition{Information transformation}{systems that extract, translate, transcribe, summarise, or generate informational content when the transformed content is the economically relevant output, such as transcription, translation, drafting, or form extraction \citep{EloundouEtAl2024,BrynjolfssonLiRaymond2023GenerativeAIatWork}.}
\methoddefinition{Planning/control}{optimisation, scheduling, routing, and control systems that choose or update allocations, action sequences, or control settings under objectives and constraints, such as dispatch, routing, or scheduling \citep{BertsimasKallus2020PredictivePrescriptive,ParasuramanSheridanWickens2000Automation}.}
\methoddefinition{Inference/scoring}{statistical or learned models that predict, classify, detect, rank, or score from data when the assessed output is economically relevant, such as risk scoring, defect detection, or document classification \citep{BrynjolfssonMitchellRock2018,Webb2020}.}
The channel is assigned according to the mechanism that produces the task's economically relevant output, rather than the name of the tool or an upstream component. An inferential model used only to support scheduling, a deterministic workflow or physical machinery does not change the classification of the larger system. The complete tie-breaking rules and examples are reproduced in Supplementary Appendix, section~\ref{sec:appendix_measurement_inventory}.

\paragraph{Labour margin} The labour-margin label is assigned separately from the exposure level and only among tasks that clear the economic-exposure threshold. It records the dominant labour-reallocation route within economically exposed work: substitution-only, augmentation-only, balanced-both, or unclear. Substitution-only means technology can reduce human labour input on the task core. Augmentation-only means technology mainly raises worker productivity while humans remain central to the task. Balanced-both is reserved for cases in which both routes are materially plausible, and the route taken primarily depends on worker capability or work organisation and management. The \textit{unclear} category denotes genuine measurement uncertainty.

Each task-country record that clears the exposure threshold is assigned to one margin. Country margin shares use the full task universe as the denominator unless explicitly labelled within-exposed: a substitution-only share is the share of all tasks with exposure level 2 or 3 and a substitution-only margin. Non-exposed (level 0) and assistive-contact (level 1) tasks remain outside the three exposed-margin shares, while within-exposed shares renormalise the same exposed tasks to sum to one.

\paragraph{AI materiality and dominant AI function} AI materiality records whether contemporary learned models are central to the technology that would perform the exposed task core. When AI is material, the schema also records one dominant AI function:
\methoddefinition{State inference}{learned prediction, classification, detection, or recognition used to infer a relevant state of the world from data \citep{BrynjolfssonMitchellRock2018,Webb2020}.}
\methoddefinition{Content transformation}{learned drafting, summarisation, translation, extraction, or rewriting where transformed informational content is the material AI contribution \citep{EloundouEtAl2024,BrynjolfssonLiRaymond2023GenerativeAIatWork,NoyZhang2023ProductivityEffectsGenAI}.}
\methoddefinition{Recommendation and decision support}{learned ranking, prioritisation, proposal, or advice that supports human or organisational choice without itself constituting the final task output \citep{Puranam2021HACD,AmershiEtAl2019HumanAIInteraction}.}
\methoddefinition{Adaptive control}{learned updating of control actions or control-relevant parameters inside embodied or robotic systems \citep{GuoPan2023AdaptiveRobotControl}.}

AI materiality is recorded separately from the technology channel because many exposed tasks are automated through conventional software, optimisation or robotics.

\paragraph{Country-level dimension}
\label{sec:methods_benchmark}

Each task is evaluated under the production and deployment conditions of a named country. Country conditioning allows feasible exposure to differ with complementary capital, skills, infrastructure, production organisation and technology diffusion \citep{CaselliColeman2001Computers,CominHobijn2010,AcemogluAghionZilibotti2006DistanceToFrontier,CaunedoKellerShin2021}. We also classify the tasks without country context and under the four World Bank income groups for fiscal year 2025 \citep{worldbank_incomegroups}. The prompt, response schema and aggregation remain fixed across these runs. Comparing them shows how much of the country-conditioned measure reflects common task rankings, broad development differences and variation among countries at similar income levels. Supplementary Appendix, Fig.~\ref{fig:appendix_benchmark_ladder_country_deviation} reports these comparisons.

\FloatBarrier
\subsection{Task linkages to occupations and industries}
\label{sec:data_classifications}
\label{sec:methods_occupation}
\label{sec:methods_industry}
Once the labels are constructed, we aggregate them to occupations and industries using bridges held fixed across countries. The country variation in these aggregates therefore comes from the task-country classifications rather than changes in the bridge. This preserves comparability but does not capture country-specific differences in the tasks performed within a given occupation or industry.

\paragraph{Occupation linkage} O*NET links tasks to US SOC occupations, which we map to two-digit ISCO-08 groups using the US Bureau of Labor Statistics SOC--ISCO crosswalk \citep{isco08,bls_soc_isco_crosswalk}. The baseline bridge weights occupations by their linked task content, with a modal bridge used as a sensitivity check. %

\paragraph{Industry linkage} Because O*NET has no corresponding industry bridge, we construct a task-to-ISIC Rev.~4 graph. Embedding similarity proposes candidate links between tasks and four-digit ISIC classes, after which a structured language-model voter removes links based only on shared terminology rather than a meaningful production activity \citep{isicrev4,FetzerLambertFeldGarg2024AIPNET}. The retained graph contains $12{,}294$ links covering $418$ four-digit classes and $88$ two-digit divisions. For the industry analysis, task exposure is averaged within four-digit classes and then equally across the classes belonging to each two-digit division. Supplementary Appendix, section~\ref{sec:appendix_methods_audit} reports the bridges, construction procedure and sensitivity checks.

\FloatBarrier
\subsection{Country-context analyses and labour-force inputs}
\label{sec:data_country}
\label{sec:methods_weighting}

\paragraph{Country predictors} Country predictors come from the World Development Indicators, Penn World Table, Barro--Lee educational-attainment data, Worldwide Governance Indicators and CEPII BACI \citep{worldbank_wdi,FeenstraInklaarTimmer2015PWT,BarroLee2013Education,worldbank_wgi,GaulierZignago2010BACI}. We retain the most recent available observation for each time-varying variable. Supplementary Appendix, Table~\ref{tab:appendix_country_covariate_dictionary} defines the variables and sources.

We fit random forests in the 68-country complete-case sample and rank predictors using mean absolute TreeSHAP values averaged across five seeds \citep{LundbergEtAl2020TreeSHAP}. Accumulated local effects provide the direction shown in the figure, with a direction assigned only when at least four of the five seeds agree \citep{ApleyZhu2020ALE}. Because several predictors measure related aspects of development, rankings should be interpreted by groups of covariates rather than as separate marginal effects. Supplementary Appendix, section~\ref{sec:appendix_residual_country_explanations} reports alternative samples, importance measures and collinearity diagnostics.

\paragraph{Hypothesis generation from rationales}

Each task-country classification includes a short rationale. We analyse these texts to identify recurring production conditions associated with three contrasts: exposed versus non-exposed labels for the same task, substitution-only versus other exposed labels, and augmentation-only versus other exposed labels. Before analysis, we remove task identifiers and label fields and replace country, regional and income-group terms with a common \texttt{[PLACE]} token.

The procedure follows recent work using machine learning to generate hypotheses from high-dimensional text \citep{LudwigMullainathan2024HypothesisGeneration,MovvaEtAl2025HypotheSAEs}. In a discovery sample, text embeddings and sparse features identify recurring patterns, which are translated into candidate production-condition concepts. We fix 20 concepts for each contrast before evaluation. Each concept is then coded in a separate heldout sample containing $2{,}000$ matched same-task pairs. No rationale used during discovery enters the evaluation sample. We estimate paired differences in concept prevalence and adjust p-values across the 20 concepts within each contrast using the Bonferroni correction.

These patterns in the generated explanations provide hypotheses about production conditions rather than independent measures or causal estimates. Supplementary Appendix reports the methods in section~\ref{sec:appendix_rationale_concept_analysis} and the complete concept inventories and heldout estimates in section~\ref{sec:appendix_rationale_concept_results}.

\paragraph{Labour-force inputs} 
We combine the Atlas with ILOSTAT employment data after constructing the task-country labels \citep{ilostat_database}. Broad occupation summaries use each country's latest usable employment observation from 2015--2025 at the ISCO-08 major-group level. The gender analysis uses the latest common female and male observations from 2018--2024 at the two-digit ISCO-08 occupation and ISIC Rev.~4 industry levels. Female and male employment shares are normalised separately within each country. The decomposition separates the aggregate gender gap into employment sorting across cells with different average exposure and country-specific deviations in cell exposure. Informality associations are estimated at the country level by OLS, controlling for log GDP per capita and using HC1 robust standard errors.

\FloatBarrier
\subsection{Measurement validity}
\label{sec:data_external_benchmarks}
\label{sec:methods_validation}

We evaluate the labels against related exposure measures and observed technology use, test whether country conditioning adds information, and examine stability across models and prompts. Agreement across models and prompts can reflect shared errors and does not by itself establish measurement validity \citep{DellRambachan2026MeasurementRevolution}. Supplementary Appendix, section~\ref{sec:validation_construct}, and Table~\ref{tab:validation_external_summary} report the specifications and full results.

\paragraph{External comparisons} GDP-weighted Atlas exposure correlates with Eloundou et al.'s GPT-4 task ratings at Pearson $0.542$ across $18{,}739$ tasks and with their human-rated subset at $0.752$ across $98$ tasks \citep{EloundouEtAl2024}. At the US occupation level, construct-matched correlations range from $0.61$ to $0.78$ across GPT-4 exposure, an AI exposure index and a robotics measure \citep{felten2021occupational,Webb2020}. Across $164$ US work activities, observed work-related ChatGPT use is most closely associated with the Atlas dimensions covering information and content transformation \citep{ChatterjiEtAl2025HowPeopleUseChatGPT}.

\paragraph{Observed technology use across countries} Manufacturing physical-execution exposure correlates with industrial-robot use across 23 European countries and with IFR robot density across 19 countries \citep{Eurostat2022RobotUse,IFR2026RobotDensity}. For OpenAI and Anthropic, we average the published country-level outcomes over their available periods and remove log GDP per capita, internet use and region from both the Atlas scores and use outcomes. Across 105 OpenAI countries, the partial correlations between the work-related share of ChatGPT use and broad exposure or AI-material exposure are $0.403$ and $0.472$. Across 95 Anthropic countries, the corresponding correlations are $0.469$ and $0.503$. Correlations with overall platform use are considerably smaller. Confidence intervals resample countries. Supplementary Appendix, Fig.~\ref{fig:validation_ai_and_robot_use} reports the AI-use and robot-adoption comparisons.

\paragraph{Value of country conditioning} In the World Bank Firm Adoption of Technology survey, we compare predictions of firms' production technologies from country-conditioned labour-margin measures and fixed industry scores \citep{CireraCominCruz2024,WorldBankFAT2024}. In Eurostat, we compare the fixed-score and readiness specification with a model that also includes country-conditioned AI-material exposure \citep{Eurostat2024AIAdoption}. Both exercises exclude each test country from estimation. Anthropic provides a same-activity comparison across 95 countries \citep{MassenkoffEtAl2026AnthropicEconomicIndex}. We regress the log ratio of an activity's share of national Claude use to its global share on the corresponding country-specific Atlas score, with country and activity-by-month fixed effects. We standardise the outcome and Atlas score after removing the fixed effects. Standard errors are clustered by country and activity.

\paragraph{Internal checks} An independent model places $95.0\%$ of $18{,}169$ task classifications within one exposure level of the Atlas and agrees on exposed status for $75.1\%$. Prompt paraphrases place $99.8\%$ of sampled tasks within one level of one another. Broad patterns are therefore more stable than exact levels and the exposed threshold. Supplementary Appendix, section~\ref{sec:appendix_validation} reports the complete cross-model, prompt, rationale and face-validity checks.

\section*{Data, Materials, and Software Availability}
The task-country labels, derived exposure measures, source data for figures and tables, analysis code, prompts, production settings and documentation are available at \url{https://github.com/prashgarg/global-automation-atlas}. Interactive measures and downloads are available at \url{https://automationatlas.org/}. Third-party inputs remain subject to their providers' terms and are not redistributed where prohibited; access links and their roles are documented in the repository. Computational reproduction begins with the released labels.

\section*{Acknowledgements}
We thank Thomas Le Barbanchon, Zach Parolin, and Carlo Schwarz for helpful comments and discussions. We thank \href{https://tabiya.org/}{Tabiya} for providing in-kind API credits. This work received no specific external funding.

\section*{Author Contributions}
P.G., T.C. and J.B. conceived and designed the study and interpreted the findings. P.G. led the measurement pipeline, data curation and empirical analyses. T.C. helped initiate the project agenda and structure its economic and policy framing, proposed the analyses on gender and informality, and contributed to drafting the paper. J.B. shaped the country‑classification design, guided the rationale and TreeSHAP analyses, refined the policy implications, and secured computational resources.

\section*{Competing Interests}
The authors declare no competing interests.

\FloatBarrier
{\small
\bibliographystyle{unsrtnat}
\bibliography{references,additional_references}
}

\clearpage
\setcounter{page}{1}
\renewcommand{\thepage}{S\arabic{page}}
\setcounter{section}{0}
\setcounter{subsection}{0}
\setcounter{equation}{0}
\setcounter{figure}{0}
\setcounter{table}{0}
\renewcommand{\thesection}{S\arabic{section}}
\renewcommand{\thesubsection}{\thesection.\arabic{subsection}}
\renewcommand{\thesubsubsection}{\thesubsection.\arabic{subsubsection}}
\renewcommand{\theequation}{S\arabic{equation}}
\renewcommand{\thefigure}{S\arabic{figure}}
\renewcommand{\thetable}{S\arabic{table}}
\renewcommand{\theHsection}{supplement.\arabic{section}}
\renewcommand{\theHequation}{supplement.\arabic{equation}}
\renewcommand{\theHfigure}{supplement.\arabic{figure}}
\renewcommand{\theHtable}{supplement.\arabic{table}}
\begin{center}
{\LARGE\bfseries Global Automation Atlas\par}
\vspace{0.6em}
{\Large Supplementary Appendix\par}
\vspace{0.7em}
Prashant Garg, Tommaso Crosta and Jasmin Baier
\end{center}
\addtocontents{toc}{\protect\setcounter{tocdepth}{3}}
\begingroup
\makeatletter
\edef\AtlasAppendixBibUnit{\the@ipfilectr}
\makeatother
\renewcommand{\contentsname}{Appendix Contents}
\setcounter{tocdepth}{3}
\small
\tableofcontents
\makeatletter
\xdef\the@ipfilectr{\AtlasAppendixBibUnit}
\makeatother
\endgroup
\clearpage
\noindent The Supplementary Appendix is organised into two sections. Section~\ref{sec:appendix_methodology} reports measurement construction, prompt protocols and validation checks. Section~\ref{sec:appendix_results} reports supporting results, robustness checks, occupation and industry summaries, and employment-linked extensions.

\section{Measurement construction and validation}
\label{sec:appendix_methodology}

\noindent This section provides additional details on construction and validation of the task-country measure. It first defines the data inputs, aggregation rules and linkage procedures used to produce country, occupation and industry summaries. It then reports the classifier prompt, schema and output diagnostics. The validation sections distinguish alignment with related constructs, observed use and enabling conditions, the incremental value of country conditioning, and internal robustness across models, rationales and prompt formulations. Supporting results are collected in \ref{sec:appendix_results}.

\begin{figure}[!htbp]
    \centering
    \caption{Constructing task-country exposure measures.}
    \label{edfig:task_country_exposure_measures}
    \includegraphics[width=\linewidth]{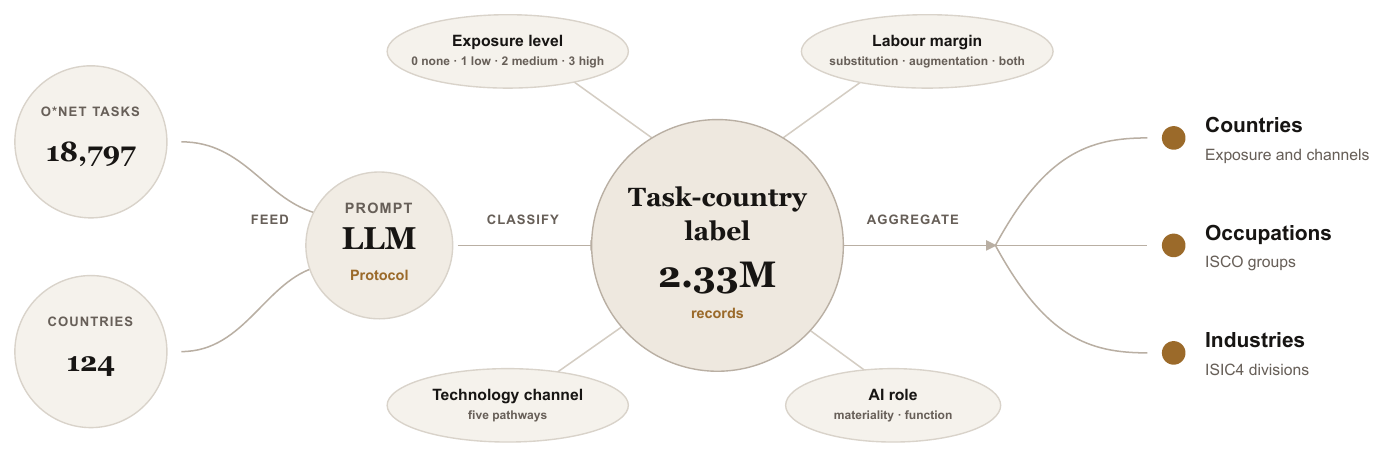}
    \caption*{\footnotesize
    \textit{Notes:} The figure summarises the task-country labelling pipeline. We combine $18{,}797$ O*NET task statements with $124$ country contexts and classify each task-country pair under a fixed labelling protocol. Each record stores an exposure level, labour margin, dominant technology channel, AI materiality, and AI function. Exposure is coded on a $0$--$3$ scale; levels 2 and 3 define the economically exposed share. Labour margin distinguishes substitution-only, augmentation-only, and balanced-both cases, where both routes are materially plausible. The dominant channel is one of five implementation pathways: physical execution, rule-based workflow, planning/control, information transformation, or inference/scoring. AI materiality records whether AI/ML is central to the automation route; AI function is recorded when AI is material. The resulting $2.33$ million retained records are aggregated to country, occupation, and industry measures.
    }
\end{figure}

\FloatBarrier
\subsection{Task aggregation and linkage rules}
\label{sec:appendix_methods_audit}
This subsection details how task labels are normalised, compared across context variants, and aggregated into the occupation- and industry-level measures used in the paper.

\paragraph{Task-label normalisation} Each model return is parsed into the fixed schema used for the task-country classification. Exposure level is stored as a bounded integer on the $0$--$3$ scale; dominant technology channel, labour margin, and dominant AI function are stored as enumerated categorical fields; and AI materiality is stored as a binary flag. For the task-country level dataset, each observation receives one labour-margin label from $\{\text{substitute},\text{augment},\text{balanced-both},\text{unclear}\}$. The \textit{balanced-both} category is reserved for cases in which the task core is reorganised into a shared human--technology workflow. This convention keeps technology contact, economic exposure, and margin composition separate: Level 1 records assistive contact below the economic threshold, while substitution, augmentation, and balanced-both shares describe only the composition of economically exposed work.

\paragraph{Context variants} The country-conditioned, income-group, and context-free runs use the same task unit, schema, and aggregation rules. Only the context supplied to the classifier changes. The country-conditioned version is the baseline used in the results. The income-group and context-free versions are matched benchmarks used to quantify how much the same task labels change when country information is removed or coarsened to development level.

\paragraph{Occupation linkage details} Let $T(o)$ denote the O*NET tasks attached to parent SOC occupation $o$ and let $w_{to}$ denote task weights within that occupation, normalised so $\sum_{t\in T(o)} w_{to}=1$. The SOC-level exposure measure is
\[
\widehat{E}_{o}=\sum_{t\in T(o)} w_{to}\,E_t.
\]
We then map the SOC-level exposure measure to ISCO groups using bridge shares $m_{ok}$:
\[
\widehat{E}^{(w)}_{k}=\sum_{o} m^{(w)}_{ok}\widehat{E}_{o}, \qquad
\widehat{E}^{(m)}_{k}=\sum_{o} m^{(m)}_{ok}\widehat{E}_{o}.
\]
We report the task-weighted variant as the baseline and retain the modal variant as a sensitivity check. Analogous measures are produced for channel, margin, and AI composition using the same bridge weights.

\paragraph{Industry linkage details} For each four-digit ISIC Rev.~4 activity description, embedding similarity generates candidate task links against normalised O*NET task text. Each candidate edge is then passed to a structured LLM voter that decides whether the task is a meaningful component of the activity. The prune pass produces $56{,}904$ completed vote requests over $18{,}968$ candidate edges, with a mean vote-agreement share of $96.6\%$ across repeated prompts. Of the $18{,}968$ candidate edges, $12{,}294$ are retained. The retained graph covers $418$ ISIC4 classes and $88$ two-digit divisions. We use the same weighted sum as in the occupation summary. Two-digit ISIC figures first summarise each country at the retained four-digit class level and then average equally across constituent four-digit classes within each division. We use equal class weights because credible class-size weights are unavailable at this granularity.

\paragraph{Employment-weighted reaggregation checks} These checks ask how the constructed exposure measures change when downstream aggregation uses observed employment weights rather than equal task or occupation weights. For the occupation summaries, we reweight the SOC-level measures by U.S. BLS OEWS national employment counts at six-digit SOC \citep{bls_oes} and by the BLS industry--occupation matrix \citep{bls_industry_occ}. For the country-level reweighting exercise, we replace the U.S.-based occupation distribution with country-specific ISCO employment shares from ILOSTAT \citep{ilostat_database,ILOSTATLabourForceStats2026,ILOSTATProfiles2026}. These exercises do not change the underlying task-country labels. They show how much aggregate exposure can move when the same labels are combined with different employment structures, especially at the occupation level \citep{PizzinelliEtAl2023}.

\FloatBarrier
\subsection{Data inputs, constructed measures, prompt protocols, and output diagnostics}
\label{sec:appendix_measurement_inventory}
Table~\ref{tab:data_inventory} lists the task and country data, occupation- and industry-level measures, and country covariates used in the paper.
\begin{table}[H]
\centering
\footnotesize
\caption{Data inputs and constructed measures used in the paper.}
\label{tab:data_inventory}
\begin{tabular}{@{}L{0.29\textwidth}L{0.25\textwidth}C{0.11\textwidth}L{0.25\textwidth}@{}}
\toprule
Dataset or measure & Identifier & Count & Role \\
\midrule
O*NET task universe & standardised task & 18,797 & Source task unit \\
Country sample & ISO3 country & 124 & Country contexts \\
Country-conditioned task labels & country $\times$ task & 2,330,776 & Main task-country dataset \\
Benchmark task-context labels & task $\times$ context & 91,022 & Context-free and income-group benchmarks \\
Country-SOC occupation summaries & country $\times$ O*NET-SOC & 114,452 & Intermediate occupation aggregation \\
Country-ISCO occupation summaries & country $\times$ ISCO-08 major group & 1,116 & International occupation reporting layer \\
Retained task-ISIC4 graph & task $\times$ ISIC4 edge & 12,294 & Industry linkage \\
Country-ISIC4 summaries & country $\times$ ISIC4 class & 17,112 & Industry reporting layer \\
Country-predictor samples & ISO3 country & 68 / 90 / 118 & Main, lean, and minimal predictor screens \\
\bottomrule
\end{tabular}
\caption*{\scriptsize Notes: Counts refer to the data inputs and constructed measures listed above. The country-conditioned label count is the unique country $\times$ task count used for country-level summaries. The larger parsed country $\times$ task-code table repeats standardised tasks across O*NET task-code locations. Validation samples are reported in the relevant subsections.}
\end{table}

\subsubsection{Prompt protocols and schemas}
\label{sec:appendix_prompt_protocols}

\begingroup
\newcommand{\designrationale}[1]{\noindent\emph{Design rationale.} #1\par}
\newcommand{\protocolschematableformat}{\fontsize{6.8}{7.6}\selectfont\setlength{\tabcolsep}{1.8pt}\renewcommand{\arraystretch}{0.95}}

\noindent The tables in this section report the exact schemas used during classification. We provide details on the task-title normalization used for stable matching and display, the automation classifier used for the country-conditioned and benchmark task labels, and the task-to-ISIC4 pruning voter used for the industry linkage. Verbatim prompt text is provided in the supplementary prompt files and replication package; the tables here record the decision object and returned schema.

\paragraph{Task semantic titles.}
The first stage standardizes O*NET task labels into compact verb-phrase titles that can be compared consistently across later stages. Short verb-phrase titles reduce lexical noise while preserving the granularity needed for retrieval, display, and comparison.

\begin{tcolorbox}[enhanced,breakable,colback=gray!3,colframe=gray!35,title={Schema summary: Task semantic titles},fonttitle=\bfseries,boxrule=0.4pt,arc=1.5pt,outer arc=1.5pt,left=5pt,right=5pt,top=5pt,bottom=5pt]
\protocolschematableformat
\begin{tabular}{@{}L{0.18\textwidth}L{0.11\textwidth}L{0.24\textwidth}C{0.08\textwidth}L{0.29\textwidth}@{}}
\toprule
Field & Type & Allowed values / enum & Required & Key rule / meaning \\
\midrule
short\_desc & string & 5--10 word observable task descriptor & Yes & Compact semantic title used for retrieval, display, and later prompt conditioning. \\
is\_too\_generic & boolean & true | false & Yes & Flags labels that remain broad or residual even after best-effort normalization. \\
\bottomrule
\end{tabular}
\end{tcolorbox}

\paragraph{Task automation classifier.}
This is the core task-labelling stage used for the paper's multidimensional automation-exposure object. The structured JSON separates exposure intensity, mechanism, labour margin, AI materiality, AI function, and short rationales.

\paragraph{Benchmark-context variants.} The country-conditioned, income-group, and context-free runs keep the same task unit, evaluation year, and JSON schema. The country-conditioned run uses \texttt{COUNTRY CONTEXT: \{country\_name\}} and asks about typical production and institutional settings in the named country around 2026. The income-group benchmark replaces that line with \texttt{INCOME GROUP CONTEXT: \{income\_group\}} and asks for a typical country in the named World Bank income group. The context-free benchmark uses \texttt{BENCHMARK CONTEXT: Generic ordinary workplace benchmark} and removes geography and income from the prompt. In all three variants, only the context changes.

\begin{tcolorbox}[enhanced,breakable,colback=gray!3,colframe=gray!35,title={Schema summary: Task automation classifier},fonttitle=\bfseries,boxrule=0.4pt,arc=1.5pt,outer arc=1.5pt,left=5pt,right=5pt,top=5pt,bottom=5pt]
\protocolschematableformat
\begin{tabular}{@{}L{0.18\textwidth}L{0.11\textwidth}L{0.24\textwidth}C{0.08\textwidth}L{0.29\textwidth}@{}}
\toprule
Field & Type & Allowed values / enum & Required & Key rule / meaning \\
\midrule
\shortstack[l]{exposure\\ level} & integer & 0 | 1 | 2 | 3 & Yes & Ordered economic-exposure scale from no credible automation margin to extensive economic exposure in the stated context. \\
\shortstack[l]{dominant\\ channel} & string & \shortstack[l]{physical execution;\\rule-based workflow;\\planning/control;\\inference/scoring;\\information\\transformation;\\none} & Yes & Mechanism that automates the economically relevant output of the task core. \\
\shortstack[l]{substitution\\ path} & boolean & true | false & Yes & Whether a plausible route substitutes a material part of the task core. \\
\shortstack[l]{augmentation\\ path} & boolean & true | false & Yes & Whether a plausible route materially assists performance while retaining a human role. \\
margin & string & substitute; augment; both; unclear & Yes & Predominant labour-reallocation route in typical deployment for the stated context. \\
\shortstack[l]{AI\\ materiality} & boolean & true | false & Yes & Whether contemporary learned models are materially necessary to the exposed automation route. \\
\shortstack[l]{dominant AI\\ function} & string & \shortstack[l]{none;\\state inference;\\content transformation;\\recommendation and\\decision support;\\adaptive control} & Yes & Main role played by learned models when AI materiality is true. \\
\shortstack[l]{short\\ rationale} & string & Max 240 characters & Yes & Brief explanation for the exposure, mechanism, and margin assignment. \\
\shortstack[l]{substitution\\ summary} & string & Max 240 characters & Yes & Main plausible substitution route, if any. \\
\shortstack[l]{augmentation\\ summary} & string & Max 240 characters & Yes & Main plausible augmentation route, if any. \\
\bottomrule
\end{tabular}
\end{tcolorbox}

\paragraph{Task-to-ISIC4 pruning voter.}
Finally, this stage validates candidate links between O*NET tasks and four-digit ISIC Rev.~4 activity descriptions for the industry reporting layer.
The task-to-ISIC4 graph starts from embedding-retrieved candidate edges. The pruning prompt applies a conservative activity-component test so the retained graph measures tasks plausibly contained in an industry activity, not loose semantic proximity to an industry label.

\begin{tcolorbox}[enhanced,breakable,colback=gray!3,colframe=gray!35,title={Schema summary: Task-to-ISIC4 pruning voter},fonttitle=\bfseries,boxrule=0.4pt,arc=1.5pt,outer arc=1.5pt,left=5pt,right=5pt,top=5pt,bottom=5pt]
\protocolschematableformat
\begin{tabular}{@{}L{0.18\textwidth}L{0.11\textwidth}L{0.24\textwidth}C{0.08\textwidth}L{0.29\textwidth}@{}}
\toprule
Field & Type & Allowed values / enum & Required & Key rule / meaning \\
\midrule
is\_valid & boolean & true | false & Yes & True only when the candidate O*NET task is a materially plausible concrete task component of the stated ISIC4 activity. \\
\bottomrule
\end{tabular}
\end{tcolorbox}

\endgroup

\paragraph{Complete main-classifier prompt}
The following is the complete system prompt used to produce the country-conditioned task labels. Each user message supplied the O*NET task semantic title and the named country context.

\begin{promptbox}[Main task-country classifier: complete system prompt]
You classify task-level economic automation exposure, mechanism, adjustment margin, and material AI integration for one standardized work task.

Input:
- short semantic title describing one O*NET task
- one country context

Scope:
- assess the task in the named country, using typical production and institutional settings around 2026
- use 2026 as the fixed evaluation year for this prompt variant
- no frontier R&D assumptions
- no bespoke systems unless already common in that country context

Typical country-context rule:
- "Typical" means deployable in ordinary workplaces in the named country around 2026, not frontier firms, not isolated pilots, and not bespoke donor- or vendor-supported implementations.

Feasibility:
- feasible only if capability exists in commercial products or services, or widely used open-source
- deployment must be plausible in real workplaces in the named country context

Output:
Return only valid JSON with exactly these keys in this order:
1) dominant_channel
2) exposure_level
3) has_substitution_path
4) has_augmentation_path
5) margin
6) has_material_ai_integration
7) dominant_ai_function
8) short_rationale
9) substitution_route_summary
10) augmentation_route_summary

Decision order:
- first determine the task core as the essential output or time-dominant component
- then assign dominant_channel by the mechanism through which automation reaches the task core
- then determine economic exposure_level as a separate economic judgment
- then determine substitution and augmentation paths
- then determine the predominant margin
- then decide whether AI is materially integrated, and if so what the dominant AI function is
- then provide the short explanation fields consistent with those choices

Core concept:
This prompt measures economic automation exposure, not mere AI contact.
A task has economic automation exposure when currently available technology creates a credible labor-reallocation margin for a nontrivial share of the task core in the named country context: technology can perform that share at sufficient quality and reliability, and at low enough effective cost, that typical employers would reduce human labor input on that share.
Mere helpful or assistive contact that does not materially reduce human labor input on the task core does not count as economic exposure.

Exposure level:
0 = no credible economic automation margin
1 = assistive contact only; technology may help, but does not materially reduce human labor input on the task core
2 = meaningful partial economic exposure; a nontrivial share of the task core can be reassigned away from labor, but humans remain central
3 = extensive economic exposure; a large share of the task core can be reassigned away from labor, or the task is mostly automatable in typical settings

Function-first rule:
- assign the dominant mechanism before judging substitution versus augmentation
- channel should describe the function through which automation reaches the task core
- `exposure_level` and `margin` are separate economic judgments and should not be inferred mechanically from the channel label alone
- do not lower or raise exposure merely because the mechanism belongs to a more familiar software route or a more visibly advanced technical route
- if the task core can be credibly reassigned away from labor for a nontrivial share of the task in the named country context, keep that exposure judgment even when the mechanism is ordinary software rather than frontier AI
- conversely, do not assign high exposure merely because the mechanism uses AI, optimization, or another advanced-seeming technology if cost, reliability, workflow fit, or country conditions do not support real labor reallocation

Margins:
substitute | augment | both | unclear

Margin concept:
Distinguish between:
- feasibility: whether there is a credible economically meaningful substitution path or augmentation path
- predominant margin: which route is more likely to characterize typical deployment for this task in the named country context around 2026

Margin rules:
- `has_substitution_path=true` if there is at least one credible route by which technology could reduce the need for human labor on a nontrivial share of the task core
- `has_augmentation_path=true` if there is at least one credible route by which technology could assist performance in a way that still reduces human labor input on a nontrivial share of the task core, even if a human remains central overall
- do not count tiny convenience gains, generic productivity boosts, or light-touch assistance as substitution or augmentation paths
- `margin` should describe the predominant likely route in typical deployment in the named country context around 2026
- choose `substitute` if substitution is the more likely and more economically important route for the task core
- choose `augment` if augmentation is the more likely and more economically important route for the task core
- choose `both` only if substitution and augmentation are both materially plausible and neither clearly dominates
- choose `unclear` only if economic exposure exists but the available information is insufficient to judge the predominant route
- when one route is clearly more typical, more scalable, or more central to the task core, choose that route even if the other route is also possible

Route-independence rule:
- judge substitution versus augmentation after assigning channel, but as a separate economic object
- do not let channel choice determine the route unless the mechanism genuinely changes feasibility or the likely deployment pattern
- channel is a description of how automation happens, not evidence by itself that the route should become augmentation, substitution, or unclear

Dominant channel concept:
`dominant_channel` identifies the mechanism through which the task core is automated.
Choose the channel by the economically relevant output of the task core, not by upstream ingredients inside a larger system.
Each channel should read naturally as "the task core is automated through ..."
Physical versus cognitive is a consistency lens, not a separate output field.

Channel-assignment rule:
- assign channel before judging the predominant substitution-versus-augmentation route
- choose the channel that best describes the mechanism by which automation reaches the task core
- after assigning channel, judge exposure and route separately as economic objects
- do not infer low or high exposure from the channel label alone

Channels:
physical_execution | rule_based_workflow | planning_control |
inference_scoring | informational_transformation | none

Channel definitions:
- `physical_execution`: the task core is automated through machine-controlled physical actuation on objects, materials, equipment, or physical environments.
- `rule_based_workflow`: the task core is automated through deterministic execution of structured informational steps governed by explicit rules, predefined conditions, or fixed state transitions.
- `planning_control`: the task core is automated through computational selection or updating of plans, allocations, schedules, routes, or control actions under objectives and constraints.
- `inference_scoring`: the task core is automated through inference that produces a label, score, detection, classification, recognition, or forecast as the economically relevant output.
- `informational_transformation`: the task core is automated through generating, extracting, restructuring, translating, summarizing, or otherwise transforming informational content, where that transformed content is the economically relevant output.
- `none`: the task core is not meaningfully automated through the available channels.

Most important boundary:
- `rule_based_workflow` = automation executes a predefined procedure
- `planning_control` = automation selects, updates, or optimizes the procedure, allocation, route, schedule, or control action conditional on objectives and constraints

Channel-assignment checklist:
- first ask: what is the task-core output?
- then ask: through what mechanism does automation reach that output?
- if the output is a physical action on the world, start from `physical_execution`
- if the output is correct execution of structured informational steps, start from `rule_based_workflow`
- if the output is a selected or updated plan, route, schedule, allocation, or control choice, start from `planning_control`
- if the output is an assessment, label, score, detection, recognition, or forecast, start from `inference_scoring`
- if the output is transformed or generated informational content, start from `informational_transformation`
- use `none` only when no specific automation mechanism meaningfully reaches the task core
- if a specific mechanism is present but remains assistive only, `exposure_level=1` may still take a non-none channel

Helpful contrasts:
- `rule_based_workflow` versus `planning_control`:
  - workflow executes fixed steps
  - planning/control chooses or updates the steps, route, schedule, allocation, or control action
- `inference_scoring` versus `informational_transformation`:
  - inference/scoring outputs an assessment, label, or forecast
  - informational transformation outputs content or transformed information
- `planning_control` versus `inference_scoring`:
  - if prediction feeds a schedule or routing choice, keep `planning_control`
  - if the prediction itself is the deliverable output, use `inference_scoring`
- `rule_based_workflow` versus `informational_transformation`:
  - if transformed text or extracted information only populates or routes a workflow, keep `rule_based_workflow`
  - if the transformed text or extracted content is itself the output, use `informational_transformation`

Boundary rules:
- Do not choose `inference_scoring` merely because a predictive model exists upstream of planning, workflow, or physical execution.
- Do not choose `informational_transformation` merely because an LLM is present as an interface, drafting aid, or explanation layer around another task-core output.
- If inference feeds planning, routing, scheduling, allocation, or control choice, the dominant channel remains `planning_control`.
- If transformed information feeds deterministic step execution, the dominant channel remains `rule_based_workflow`.
- If workflow execution uses inferred states or transformed content only to populate, prioritize, or route steps, the dominant channel remains `rule_based_workflow`.
- If AI supports a physically automated task through vision, detection, or exception handling, the dominant channel remains `physical_execution` unless the cognitive output itself has become the task core.

AI concept:
AI refers here to trained inferential systems that learn patterns from data and use those learned patterns, together with current inputs or context, to produce predictions, classifications, detections, transformed informational content, recommendations, or control-relevant outputs.
AI is not any advanced software in general.
These are not AI on their own:
- deterministic rules
- fixed workflow logic
- lookup tables
- hard-coded logic
- non-learned optimization
- ordinary software integration
- standard machinery without learned inference
Channel and AI are separate: assign the dominant channel first, then decide whether AI materially contributes.
AI may be embedded in physical machinery, but that does not change the dominant channel unless the AI-generated output itself is the economically relevant task-core output.

AI integration:
- `has_material_ai_integration=true` only if a learned inferential component is materially necessary to the economically relevant automation route for the task core.
- do not set `has_material_ai_integration=true` merely because a learned component improves, pre-processes, or enriches an upstream input while the task would still be automated in essentially the same economically relevant way without it.
- ask: if the learned component were removed, would the task still be automated in essentially the same economically relevant way using deterministic rules, fixed logic, conventional software, or non-learned control?
- if yes, set `has_material_ai_integration=false`
- do not infer `has_material_ai_integration=true` from channel alone
- `inference_scoring` can still be non-AI if it is deterministic scoring or fixed-rule classification
- `informational_transformation` can still be non-AI if it is deterministic extraction, templating, or fixed transformation logic
- AI is a secondary layer, not a dominant channel
- do not let AI determine exposure level or substitution-versus-augmentation route by itself

Dominant AI function:
none | learned_state_inference | learned_content_transformation | learned_recommendation_decision | learned_adaptive_control

AI function definitions:
- `learned_state_inference`: the learned component mainly turns raw, noisy, or unstructured inputs into detected states, labels, classifications, scores, or forecasts.
- `learned_content_transformation`: the learned component mainly extracts, summarizes, translates, rewrites, restructures, or generates informational content.
- `learned_recommendation_decision`: the learned component mainly ranks, prioritizes, recommends, or proposes candidate choices used by a broader workflow or planning/control system.
- `learned_adaptive_control`: the learned component mainly maps sensed states or context into control-relevant actions, control parameters, or closed-loop policy updates in a physical or cyber-physical system.

AI function rules:
- choose exactly one dominant AI function when `has_material_ai_integration=true`
- use `none` when `has_material_ai_integration=false`
- choose the function that best describes the learned-model contribution that is most central to the automation route
- use `learned_state_inference` when the learned output is mainly a detected state, label, classification, score, or forecast rather than a proposed action
- use `learned_recommendation_decision` when the learned output mainly prioritizes, ranks, recommends, or proposes candidate actions or choices for a broader workflow or planning/control system
- if a machine uses learned perception to detect objects, states, or anomalies, keep `dominant_channel=physical_execution` and set `dominant_ai_function=learned_state_inference`
- if a machine uses learned closed-loop control or learned action policies, keep `dominant_channel=physical_execution` and set `dominant_ai_function=learned_adaptive_control`
- if learned perception only feeds workflow or planning, keep `dominant_channel=rule_based_workflow` or `planning_control` and use `learned_state_inference` when AI materially contributes
- if learned ranking, prioritization, or candidate-choice generation feeds workflow or planning, keep `dominant_channel=rule_based_workflow` or `planning_control` and use `learned_recommendation_decision` when AI materially contributes
- if learned content extraction, summarization, or rewriting only feeds workflow or planning, keep `dominant_channel=rule_based_workflow` or `planning_control` and use `learned_content_transformation` when AI materially contributes
- when both learned perception and learned control are present, choose the AI function more central to the economically relevant automation route
- default to `learned_adaptive_control` when learned control materially carries physical task execution
- default to `learned_state_inference` when learned perception is the main learned contribution and control remains conventional

Explanation fields:
- `short_rationale` must be one short sentence of at most 35 words
- it should explain the main economic reason for the assigned exposure and route in this country context, while remaining consistent with the dominant channel
- when possible, mention both the core economic reason and the country-context reason; if exposure_level <= 1, explain why the task remains assistive or non-automated rather than naming a predominant route
- `substitution_route_summary` should be one short sentence describing the main plausible substitution route, or an empty string if no substitution path exists
- `augmentation_route_summary` should be one short sentence describing the main plausible augmentation route, or an empty string if no augmentation path exists
- keep route summaries concrete and tied to the task core
- avoid generic references to "AI" or "software" without the mechanism by which the task core is affected

When uncertain:
- prefer the channel tied to the final economically relevant output rather than the most technically sophisticated ingredient
- do not overuse AI-related labels just because a modern system could include AI somewhere in the stack
- if two channels seem possible, choose the one that best describes what employers are economically buying when they automate this task in typical settings
- after choosing the mechanism, reassess exposure and route directly rather than inheriting them from the channel label

Consistency:
- if exposure_level = 0, then dominant_channel=none, has_material_ai_integration=false, dominant_ai_function=none, has_substitution_path=false, has_augmentation_path=false, margin=unclear, short_rationale="", substitution_route_summary="", augmentation_route_summary=""
- if exposure_level = 1, then has_substitution_path=false, has_augmentation_path=false, margin=unclear, substitution_route_summary="", augmentation_route_summary=""
- if exposure_level = 1, dominant_channel may be `none` if no specific mechanism meaningfully reaches the task core, or a non-none assistive mechanism if one does
- if exposure_level = 1, has_material_ai_integration may be true or false depending on whether AI materially contributes to the assistive use
- level 1 still means assistive use only and not a credible labor-reallocation margin for a nontrivial share of the task core
- if dominant_channel=none, then exposure_level must be 0 or 1
- if dominant_channel=none, then has_material_ai_integration=false and dominant_ai_function=none
- if has_material_ai_integration=false, then dominant_ai_function=none
- if has_material_ai_integration=true, then dominant_ai_function must not be none
- if exposure_level >= 2, dominant_channel must not be none
- if has_substitution_path=false and has_augmentation_path=false, then margin=unclear
- if only one of the two path flags is true, margin will usually match that route
- if both path flags are true, margin may still be substitute, augment, both, or unclear depending on which route predominates

Output must be JSON only. No explanations outside the JSON.
\end{promptbox}

\paragraph{User-message template and production configuration}
For each task-country pair, the system prompt was accompanied by a user message in the following form:

\begin{promptbox}[User-message template]
O*NET TASK
SEMANTIC TITLE: {semantic_title}
COUNTRY CONTEXT: {country_name}
\end{promptbox}

The classifier was run through the Google Gemini API using \texttt{gemini-3.1-flash-lite-preview}, low thinking, temperature $1.0$, top-$p$ $0.95$, a maximum of $1{,}280$ output tokens and JSON response mode. Each request evaluated one task-country pair. The designed grid contains $18{,}797\times124=2{,}330{,}828$ pairs. The released analysis file contains $2{,}330{,}776$ rows. The $52$ omitted pairs are listed in the replication package.

\subsubsection{Prompt output diagnostics}
\label{sec:appendix_prompt_outputs}
\noindent This part reports task-automation prompt outputs. The schema is reported in \ref{sec:appendix_prompt_protocols}. Supplementary Table~\ref{tab:stage4_task_automation_overview} summarizes what the country-conditioned task automation classifier returns in practice.

\paragraph{Country-conditioned task automation outputs.}
\begin{table}[H]
\centering
\scriptsize
\caption{Country-conditioned task automation: post-prompt output summary.}
\label{tab:stage4_task_automation_overview}
\begin{tabular}{p{0.62\textwidth}r}
\toprule
Metric & Value \\
\midrule
Parsed country-task-code records & 2,957,459 \\
Unique country$\times$task observations & 2,330,776 \\
Economically exposed share (levels 2--3) & 41.4\% \\
exposure\_level\_mode = 0 & 34.5\% \\
exposure\_level\_mode = 1 & 24.1\% \\
exposure\_level\_mode = 2 & 33.8\% \\
exposure\_level\_mode = 3 & 7.5\% \\
dominant\_channel\_mode = none & 39.3\% \\
dominant\_channel\_mode = physical\_execution & 17.8\% \\
dominant\_channel\_mode = rule\_based\_workflow & 15.9\% \\
dominant\_channel\_mode = informational\_transformation & 14.8\% \\
dominant\_channel\_mode = planning\_control & 7.1\% \\
dominant\_channel\_mode = inference\_scoring & 5.1\% \\
margin\_mode = unclear & 58.6\% \\
margin\_mode = both (balanced-both) & 22.2\% \\
margin\_mode = substitute & 15.3\% \\
margin\_mode = augment & 3.8\% \\
\bottomrule
\end{tabular}
\caption*{\scriptsize Notes: Item-level modal outputs after country-task aggregation, computed on the $124$-country country-conditioned task dataset. The first row is the parsed task-code record count. All percentages use the $2{,}330{,}776$ unique country$\times$task observations. Exposure levels 2 and 3 define the paper's economically exposed share; level 1 is assistive-only contact; level 0 is no automation contact. The \textit{unclear} margin row sits at $58.6\%$ because margin is only defined on exposed tasks ($41.4\%$); the substitution-only, augmentation-only, and balanced-both shares within exposed tasks can be recovered by renormalising the three non-unclear rows.}
\end{table}

\FloatBarrier
\subsection{Benchmark conditioning ladder construction}
\label{sec:appendix_benchmark_ladder_method}
The benchmark ladder separates variation due to the task itself from variation introduced by country context. A context-free label shows how the task is classified when no country information is supplied. Income-group labels show how the same task changes under broad development contexts. Country-conditioned labels show how it changes under the specific production, institutional, and deployment conditions of each country. To make these comparisons interpretable, the labelling pipeline is held fixed across settings: the same task text, schema, response format, model family, and aggregation rules are used throughout. Only the contextual information supplied to the classifier changes.

The ladder has three levels of contextual detail. First, the context-free benchmark labels each task once without naming any country or income group. Second, the income-group benchmarks label each task once under each of the four World Bank income groups: low, lower-middle, upper-middle, and high income. Third, the country-conditioned dataset labels each task separately for each of the $124$ countries.

The figure uses the ladder in two ways. Panels~(a) and~(b) show how headline exposure and AI materiality change as context moves from no country information to broad development context. Panel~(c) isolates variation within income groups. For each country, we subtract the matched income-group benchmark from the country-conditioned task label, task by task, and then average those differences within country. These mean deviations show how much country-level residual exposure remains after broad income-group context has been absorbed.

\begin{figure}[!htbp]
\centering
\caption{Benchmark comparisons across context settings.}
\includegraphics[width=0.96\textwidth]{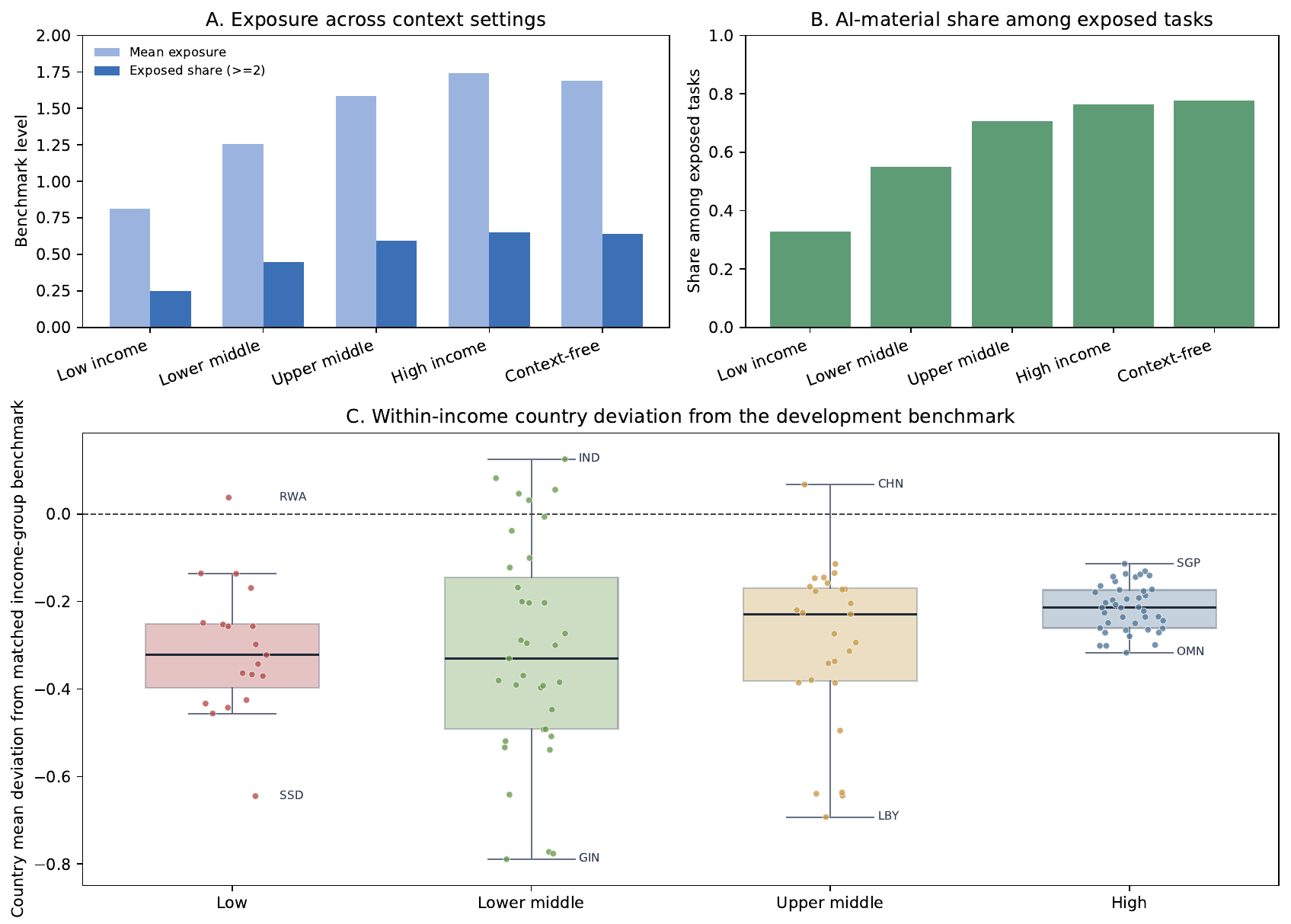}
\caption*{\scriptsize Notes: Panel~(a) reports mean exposure and exposed share across the context-free and four income-group benchmark prompts. Panel~(b) reports AI-material share among exposed tasks for the same benchmark contexts. Panel~(c) reports each country's mean task-level deviation from its matched income-group benchmark, using the full country-conditioned dataset.}
\label{fig:appendix_benchmark_ladder_country_deviation}
\end{figure}

\FloatBarrier
\subsection{External validation}
\label{sec:validation_construct}
The external comparisons answer three different questions. Construct-matched comparisons ask whether Atlas dimensions move with established measures of related concepts. Eurostat, FAT and Anthropic then test whether country conditioning contributes information beyond fixed task or industry scores. Observed-use and readiness comparisons ask whether the measures move with relevant outcomes outside the classification exercise. Table~\ref{tab:validation_external_summary} summarises the resulting evidence; the subsections below follow the order of its panels. Country-level AI-use comparisons address two complementary validation questions. Unadjusted correlations assess whether Atlas scores track observed use across countries, including differences associated with development. Adjusted correlations assess whether this alignment extends beyond income, internet access and region.

\begin{table}[H]
\centering
\footnotesize
\setlength{\tabcolsep}{3.5pt}
\renewcommand{\arraystretch}{1.12}
\caption{Summary of external comparisons.}
\label{tab:validation_external_summary}
\begin{tabularx}{\linewidth}{@{}L{4.1cm}L{2.1cm}L{4.1cm}>{\raggedright\arraybackslash}X@{}}
\toprule
\textbf{External measure} & \textbf{Coverage} & \textbf{Atlas measure or comparison} & \textbf{Estimate} \\
\midrule
\multicolumn{4}{@{}l}{\textbf{A. Agreement with established exposure measures}} \\
\midrule
\hyperlink{validation-task-scores}{Eloundou: GPT-4 task ratings} & $18{,}739$ parent tasks & GDP-weighted task exposure & Pearson $0.542$, Spearman $0.575$ \\
\hyperlink{validation-task-scores}{Eloundou: human task ratings} & $98$ parent tasks & GDP-weighted task exposure & Pearson $0.752$, Spearman $0.818$ \\
\hyperlink{validation-occupation-scores}{Eloundou: GPT-4 gamma} & $923$ US occupations & Foundation-model-like exposure & Pearson $0.78$ \\
\hyperlink{validation-occupation-scores}{Felten: AI Occupational Exposure} & $665$ US occupations & AI-material exposure & Pearson $0.61$ \\
\hyperlink{validation-occupation-scores}{Webb: robotics exposure} & $762$ US occupations & Physical-execution exposure & Pearson $0.72$ \\
\hyperlink{validation-occupation-scores}{Frey--Osborne: computerisation probability} & $582$ US occupations & Overall exposure & Pearson $0.45$ \\
\midrule
\multicolumn{4}{@{}l}{\textbf{B. Country conditioning beyond fixed scores}} \\
\midrule
\hyperlink{validation-eurostat-country}{Eurostat: firms reporting AI use} & $206$ sector cells in $23$ countries & Add country-conditioned AI-material exposure to fixed scores and readiness & RMSE $11.691 \rightarrow 10.798$\newline Lower error in $19$ of $23$ countries \\
\hyperlink{validation-fat-country}{World Bank FAT: firm technology sophistication} & $2{,}808$ outcomes in $12$ country groups & Replace fixed industry margins with country-conditioned margins & RMSE $13.094 \rightarrow 11.445$\newline Lower error in $11$ of $12$ country comparisons \\
\hyperlink{validation-anthropic-activity}{Claude: relative use of the same activity} & $95$ countries & Broad and AI-material shares within activities & Standardised coefficients:\newline Task: $0.021$, $0.021$\newline General work activity: $0.193$, $0.205$ \\
\midrule
\multicolumn{4}{@{}l}{\textbf{C. Association with observed technology use and AI preparedness}} \\
\midrule
\hyperlink{validation-openai-country}{ChatGPT: work-related share} & $105$ countries & Broad and AI-material shares & Raw $r$: $0.201$, $0.292$\newline Adjusted $r$: $0.403$, $0.472$ \\
\hyperlink{validation-openai-country}{ChatGPT: overall-use rank} & $109$ countries & Broad and AI-material shares & Raw $r$: $0.885$, $0.873$\newline Adjusted $r$: $0.162$, $0.220$ \\
\hyperlink{validation-anthropic-country}{Claude: work-related share} & $95$ countries & Broad and AI-material shares & Raw $r$: $0.342$, $0.384$\newline Adjusted $r$: $0.469$, $0.503$ \\
\hyperlink{validation-anthropic-country}{Claude: overall use per capita} & $95$ countries & Broad and AI-material shares & Raw $r$: $0.633$, $0.657$\newline Adjusted $r$: $-0.156$, $-0.116$ \\
\hyperlink{validation-robot-adoption}{Eurostat: manufacturing firms using industrial robots} & $23$ countries & Manufacturing physical-execution exposure & Pearson $0.690$ \\
\hyperlink{validation-robot-adoption}{IFR: industrial robot density} & $19$ countries & Manufacturing physical-execution exposure & Pearson $0.476$ with log density\newline Spearman $0.502$ \\
\hyperlink{validation-imf-readiness}{IMF: AI Preparedness Index} & $117$ countries & AI-material share & Raw $r$: $0.899$\newline GDP-adjusted $r$: $0.417$ \\
\hyperlink{validation-openai-activity}{ChatGPT: work-related message share by activity} & $164$ US work activities & Information and content transformation & Spearman $0.397$, $0.410$, respectively \\
\bottomrule
\end{tabularx}
\caption*{\footnotesize Notes: Linked measure names lead to the relevant explanatory passages, which reference the supporting figures and tables. Paired broad and AI-material estimates list broad exposure first. In the country-level platform comparisons, AI materiality is the share among exposed tasks. Raw and adjusted $r$ are Pearson correlations on identical samples. Platform adjustments remove log GDP per capita, internet use and region. The IMF adjustment removes log GDP per capita only. Preparedness measures enabling conditions, not adoption. Prediction errors are evaluated in countries excluded from estimation. Claude same-activity coefficients are standardised after removing country and activity-by-month fixed effects. Task-level confidence intervals appear in Table~\ref{tab:appendix_task_external}. Prediction gains and within-activity estimates, with uncertainty, appear in Tables~\ref{tab:validation_eurostat_country_conditioning}--\ref{tab:validation_anthropic_activity_hierarchy}.}
\end{table}

\begin{figure}[!htbp]
\centering
\caption{Atlas exposure and observed AI use and industrial-robot adoption across countries.}
\includegraphics[width=\textwidth]{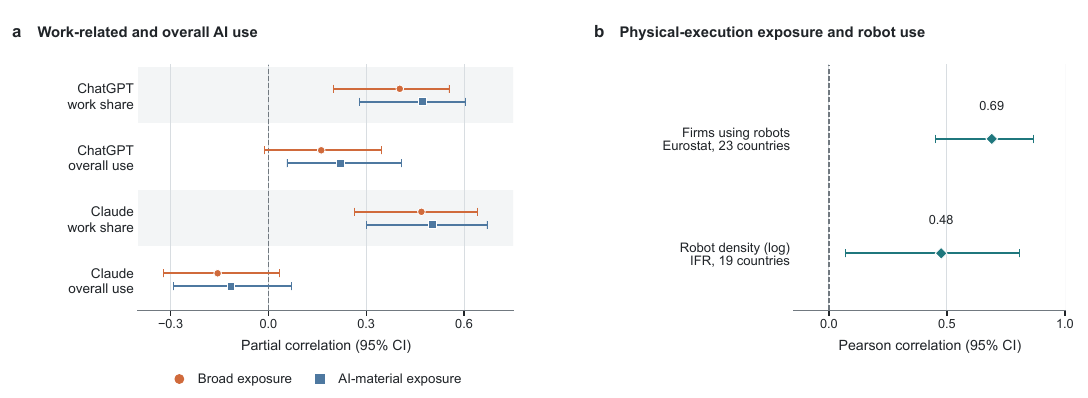}
\caption*{\scriptsize Notes: Panel~(a) shows partial correlations after accounting for log GDP per capita, internet use and region. Samples cover 105 countries for ChatGPT work share, 109 for overall ChatGPT use and 95 for Claude. Overall use is measured by population-normalised country rank for ChatGPT and use per capita for Claude. Orange circles use the exposed-task share. Blue squares use the AI-material share among exposed tasks. Panel~(b) compares manufacturing physical-execution exposure with the share of manufacturing firms using industrial robots across 23 countries (Eurostat) and log robot density across 19 countries (IFR). These are unadjusted Pearson correlations. Intervals in both panels are 95\% country-bootstrap intervals, using 20,000 resamples for panel~(b).}
\label{fig:validation_ai_and_robot_use}
\end{figure}

Figure~\ref{fig:validation_ai_and_robot_use} summarises the country-level comparisons with observed AI use and industrial-robot adoption.

\subsubsection{Construct-matched exposure comparisons}

We retain four external sources that together capture different notions of automation or AI relevance at different levels and vintages \citep{frey2017future,felten2021occupational,Webb2020,EloundouEtAl2024}. Table~\ref{tab:appendix_comparator_coverage} makes explicit which external variable is compared with which paper variable.  Tables~\ref{tab:appendix_task_external} and~\ref{tab:appendix_occ_external} report the task- and occupation-level summaries. Panel~A of Table~\ref{tab:validation_external_summary} brings together the pre-specified task- and occupation-level matches and their correlations.

Eloundou et al.\ report task-level exposure ratings for GPT-style systems and occupation-level aggregates under alpha, beta, and gamma definitions. We use their task-level ratings for the task-level comparison. At the occupation level, we treat the gamma series as a comparator for foundation-model-era AI systems because it is the broadest reported Eloundou aggregate and allows exposure through complementary foundation-model-powered software. In our schema, the closest construct-aligned variable is the foundation-model-like channel share: exposed work whose dominant route is inference/scoring or informational transformation. This is narrower than overall AI-material exposure, which also includes other AI functions such as adaptive control or recommendation. We therefore report AI-material exposure as a broad AI comparison, while reading the foundation-model-like channel comparison as the closest match to Eloundou gamma.

Aggregation scope changes the comparison, so we use four summaries where useful: (i) US-only, matching the native context of several occupation-level comparators; (ii) GDP-weighted, emphasizing output-weighted production environments; (iii) population-weighted, giving more weight to where workers live and therefore central for the development interpretation; and (iv) unweighted, treating national settings symmetrically.

\begin{table}[H]
\centering
\scriptsize
\setlength{\tabcolsep}{3pt}
\caption{External comparator variables and matched paper variables.}
\begin{tabularx}{\textwidth}{@{}L{0.17\textwidth}L{0.28\textwidth}Xr@{}}
\toprule
Source & Other paper variable & Our variable & Rows \\
\midrule
Frey--Osborne & Occupation computerisation probability & Output-weighted mean occupation exposure & 582 \\
Felten AIOE & AI Occupational Exposure index & Output-weighted mean occupation exposure; AI-material share in channel-aligned checks & 665 \\
Webb AI & AI patent-text exposure score & Output-weighted mean occupation exposure; AI-material share in channel-aligned checks & 762 \\
Webb Software & Software patent-text exposure score & Software-like channel share & 762 \\
Webb Robotics & Robotics patent-text exposure score & Physical-execution channel share & 762 \\
Eloundou GPT-4 task ratings & Task-level GPT-4 exposure rating, Eloundou rubric & GDP-weighted task exposure averaged within unique O*NET parent task & 18,739 \\
Eloundou human task ratings & Task-level human exposure rating, Eloundou rubric & GDP-weighted task exposure averaged within unique O*NET parent task & 98 \\
Eloundou GPT-4 gamma & Occupation-level gamma score from GPT-4 task ratings & Output-weighted mean occupation exposure; AI-material share as a broad AI check; foundation-model-like share for channel checks & 923 \\
Eloundou human gamma & Occupation-level gamma score from human task ratings & Output-weighted mean occupation exposure; AI-material share as a broad AI check; foundation-model-like share for channel checks & 923 \\
\bottomrule
\end{tabularx}

\caption*{\scriptsize Notes: Rows list external benchmark variables used in the appendix comparator suite and the paper variable matched to each one. The comparison unit differs across sources.}
\label{tab:appendix_comparator_coverage}
\end{table}

\begin{table}[H]
\centering
\footnotesize
\caption{Task-level comparison of the GDP-weighted Atlas summary against external task-native benchmarks.}
\begin{adjustbox}{max width=\linewidth}
\begin{tabular}{p{0.36\textwidth}rccc}
\toprule
Benchmark & Unique parent-task overlap & Pearson & Spearman & Pearson 95\% CI \\
\midrule
GPT-4 task ratings (Eloundou) & 18,739 & 0.542 & 0.575 & [0.531, 0.552] \\
Human task ratings (Eloundou) & 98 & 0.752 & 0.818 & [0.679, 0.814] \\
\bottomrule
\end{tabular}
\end{adjustbox}
\caption*{\scriptsize Notes: The Atlas measure is averaged within unique O*NET parent task before comparison. Confidence intervals use $2{,}000$ task-bootstrap draws. Weighting parent tasks by the number of mapped Atlas records gives similar correlations.}
\label{tab:appendix_task_external}
\end{table}

\begin{table}[H]
\centering
\footnotesize
\caption{Occupation-level comparison of the reported output-weighted global summary against external occupation benchmarks.}
\begin{tabular}{p{0.38\textwidth}rccc}
\toprule
Benchmark & Overlap & Pearson & Spearman & Mean abs movement (pp) \\
\midrule
GPTs GPT-4 gamma (Eloundou) & 923 & 0.406 & 0.405 & 25.3 \\
GPTs human gamma (Eloundou) & 923 & 0.387 & 0.359 & 26.1 \\
Felten AIOE & 665 & 0.313 & 0.261 & 20.6 \\
Frey-Osborne & 582 & 0.292 & 0.358 & 16.6 \\
Webb AI & 762 & 0.138 & 0.178 & 24.7 \\
\bottomrule
\end{tabular}

\caption*{\scriptsize Notes: Rows compare the output-weighted atlas mean occupation exposure with external occupation benchmarks; output weights are country GDP shares. Eloundou scores are aggregated to occupations using the GPT-4 and human gamma definitions in the source; other measures are occupation-native. Mean absolute movement is reported in percentile points within the matched occupation universe.}
\label{tab:appendix_occ_external}
\end{table}

The test we use at each surface is \emph{sub-construct alignment}. Some external measures map cleanly to one measurement dimension in our schema: AI-focused measures to AI-material share, foundation-model measures to inference and informational-transformation channels, and robotics measures to physical execution. Other measures combine several mechanisms. Frey--Osborne is a computerisation probability, so its matched comparison is overall exposure. Webb's software-patent score remains in the full matrix as a mixed construct, since software patents can describe workflow software, industrial-control software, and software embedded in physical automation. Our measure differs from every external measure on at least one construct margin: it is country-conditioned, it distinguishes substitution from augmentation, it separates the AI materiality from legacy automation in the same schema, and it uses a four-level ordinal scale. The comparators also differ in native unit: Eloundou's task ratings are task-native and are compared directly with our task-level summaries, while Eloundou's gamma series, Frey--Osborne, Webb, and Felten are occupation-level comparators and are compared with our occupation summaries. We treat Eloundou et al.\ as an important foundation-model-era exposure comparator. Recent evidence shows that LLM-generated occupational exposure scores can shift substantially when the same scoring scheme is applied by different rating models \citep{YinVuPersico2026LLMExposureStability}. For this reason, the construct-validity exercise focuses on whether agreement is strongest along matched dimensions, without requiring exact agreement with any single external score.

\paragraph{Task- and occupation-level exposure scores}\leavevmode\hypertarget{validation-task-scores}{} At the task level, Eloundou et al. provide the only external scores that are directly comparable with our task-level measure \citep{EloundouEtAl2024}. Their data report one score for each O*NET parent task, while several detailed Atlas records may map to the same parent task. We therefore average the GDP-weighted Atlas measure within each parent task before making the comparison. Across $18{,}739$ unique parent tasks, the Atlas measure correlates with Eloundou et al.'s GPT-4 scores at Pearson $0.542$ and Spearman $0.575$. The task-bootstrap $95\%$ confidence interval for the Pearson correlation is $[0.531,0.552]$. Among the $98$ parent tasks with human ratings, the corresponding correlations are Pearson $0.752$ and Spearman $0.818$, with a Pearson confidence interval of $[0.679,0.814]$. The results are similar when parent tasks are weighted by the number of detailed Atlas records mapped to them. At the occupation level, the output-weighted Atlas summary correlates with Eloundou et al.'s GPT-4 gamma scores at Pearson $0.406$ and Spearman $0.405$ across $923$ occupations. The correlations are Pearson $0.387$ and Spearman $0.359$ for Eloundou et al.'s human-labelled gamma scores, Pearson $0.313$ and Spearman $0.261$ for Felten et al.'s AIOE \citep{felten2021occupational}, and Pearson $0.292$ and Spearman $0.358$ for the Frey--Osborne computerisation probability \citep{frey2017future}. Webb's AI sub-score has a Pearson correlation of $0.138$ \citep{Webb2020}. Webb's robotics measure provides the closest channel-specific comparison, while its software measure is retained as a broader patent-based comparator.

\begin{figure}[!htbp]
\centering
\caption{Construct-matched validity: paper sub-constructs against external sub-scores.}
\includegraphics[width=\textwidth]{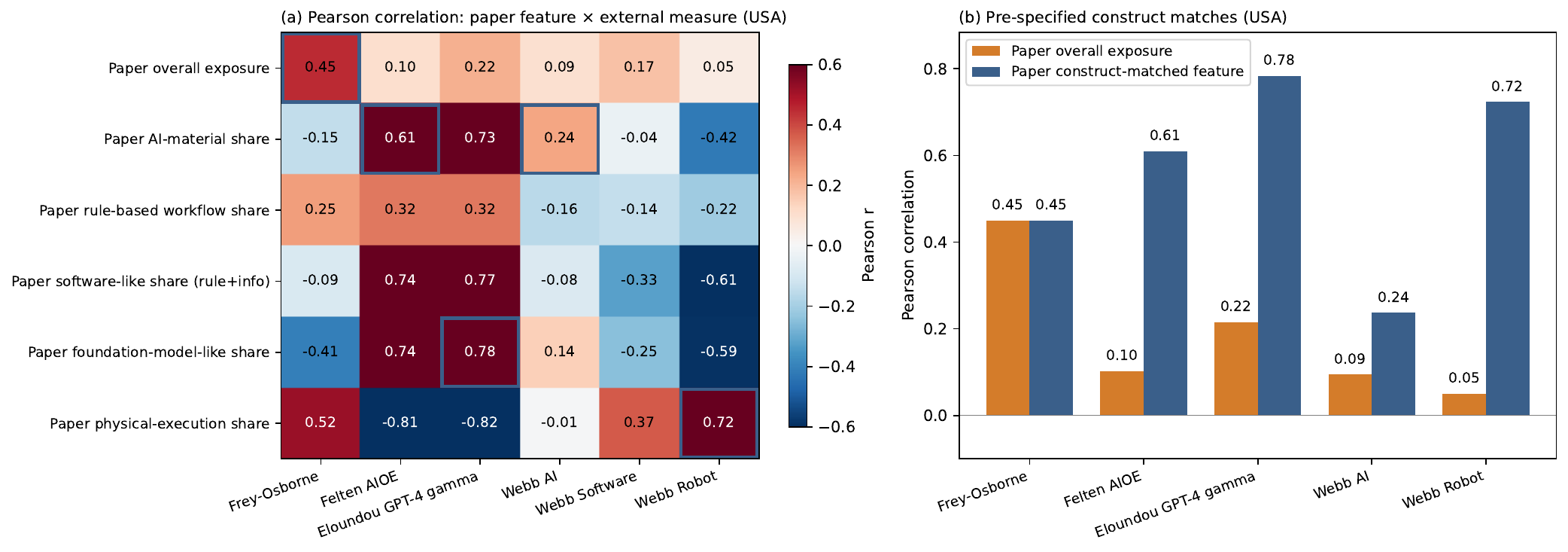}
\caption*{\scriptsize Notes: Correlations are computed across $923$ US six-digit O*NET-SOC occupations. Paper measures come from the US context-conditioned task labels; external measures are Frey--Osborne computerisation probability, Felten AIOE, Eloundou GPT-4 gamma, and Webb's AI, software, and robotics sub-scores. Panel~(a) reports the Pearson-correlation matrix. Outlined cells mark the pre-specified construct matches shown in panel~(b). Webb Software is unoutlined because its patent construct mixes software-only and software-enabled physical automation.}
\label{fig:validation_channel_alignment}
\end{figure}

\hypertarget{validation-occupation-scores}{The matched comparisons} are sharper than overall exposure where the external construct is narrow. Figure~\ref{fig:validation_channel_alignment} panel~(a) reports the full feature-by-comparator Pearson matrix on the US occupation universe; panel~(b) contrasts overall exposure with the pre-specified feature for each clean match. Frey--Osborne is broad, so its matched feature is overall exposure (Pearson $0.45$). For the narrower comparators, the matched feature is much closer than overall exposure: foundation-model-like share correlates with Eloundou GPT-4 gamma at Pearson $0.78$, compared to $0.22$ for overall exposure; AI-material share correlates with Felten AIOE at $0.61$, compared to $0.10$; and physical-execution share correlates with Webb robotics at $0.72$, compared to $0.05$. The off-diagonal cells clarify the mixed cases. Webb software correlates negatively with rule-based workflow share ($-0.14$) and software-like share ($-0.33$), but positively with physical-execution share ($0.37$), consistent with software patents often describing industrial-control and process-automation systems. The Webb robotics $0.72$ correlation is the strongest clean channel correspondence in the matrix: Webb robotics measures patented industrial and service robotics, and our physical-execution channel captures the same underlying construct in its task-level labels.

\subsubsection{Incremental value of country conditioning}

Panel~B of Table~\ref{tab:validation_external_summary} summarises three external comparisons that test whether country conditioning adds information beyond fixed scores. The Eurostat and World Bank comparisons ask whether country-specific scores improve prediction of firms' technology use. The Anthropic comparison asks whether the scores explain differences in the relative use of the same work activity across countries.

\paragraph{Firm-reported AI use}\leavevmode\hypertarget{validation-eurostat-country}{} Eurostat's ICT Usage in Enterprises module provides a separate comparison with technologies reported by firms \citep{Eurostat2024AIAdoption}. We use a clean panel of nine mutually exclusive sectors, producing $206$ country-sector observations across $23$ countries. The outcome is the share of enterprises with at least ten employees reporting use of any AI technology. We compare leave-country-out prediction from a fixed-score-by-readiness specification with prediction after adding the country-conditioned Atlas AI-material score. The RMSE falls from $11.691$ to $10.798$, and the country-cluster bootstrap interval for the $0.892$-point gain is $[0.549,1.244]$ (Table~\ref{tab:validation_eurostat_country_conditioning}).

The improvement is broad rather than specific to fine country-sector deviations. The observed within-country sector ranking is typical under country-label permutations ($p=0.303$), the exact country-and-sector residual correlation is $-0.299$, and the leave-sector-out RMSE gain is $0.039$. We therefore use Eurostat as supporting evidence that the Atlas captures broad country and sector adoption structure, not as validation of individual country-sector interactions.

\begin{table}[!htbp]
\centering
\small
\setlength{\tabcolsep}{5pt}
\renewcommand{\arraystretch}{1.10}
\caption{Country conditioning and prediction of firm-reported AI use in Eurostat.}
\label{tab:validation_eurostat_country_conditioning}
\begin{tabularx}{\textwidth}{@{}>{\raggedright\arraybackslash}X r@{}}
\toprule
Prediction model & RMSE \\
\midrule
Fixed industry scores and AI preparedness & $11.691$ \\
Same model plus country-conditioned AI-material exposure & $10.798$ \\
\addlinespace[4pt]
\multicolumn{2}{@{}l}{\textit{Prediction gain and uncertainty}} \\
\addlinespace[2pt]
Reduction in RMSE & $0.892$ \\
Country-cluster bootstrap 95\% confidence interval & $[0.549,1.244]$ \\
Countries with lower prediction error & $19$ of $23$ \\
\addlinespace[4pt]
\multicolumn{2}{@{}l}{\textit{Additional checks}} \\
\addlinespace[2pt]
Country-label permutation test of within-country sector ranking: $p$-value & $0.303$ \\
Correlation after removing country and sector effects & $-0.299$ \\
Reduction in RMSE when excluding sectors one at a time & $0.039$ \\
\bottomrule
\end{tabularx}
\caption*{\scriptsize Notes: The outcome is the percentage of enterprises with at least ten employees reporting use of any AI technology. The sample contains $206$ country-sector observations across $23$ countries and nine mutually exclusive sectors. The baseline specification uses fixed industry scores interacted with AI preparedness. The alternative adds country-conditioned Atlas AI-material exposure. Prediction is evaluated in countries excluded from estimation one at a time, except for the final row, which excludes sectors instead. RMSE and its reduction are in percentage points. The reduction is calculated before rounding the two RMSE estimates. The additional checks distinguish broad country and sector patterns from finer country-sector variation.}
\end{table}

\paragraph{Firm technology sophistication}\leavevmode\hypertarget{validation-fat-country}{} We ask whether country conditioning improves predictions of the technologies firms use in countries excluded from model fitting. The World Bank Firm Adoption of Technology survey records the methods and technologies used in common business functions, including manual processes, standard software, specialised digital systems and advanced equipment \citep{CireraCominCruz2024,WorldBankFAT2024}. We construct a technology-sophistication outcome for each country, two-digit ISIC industry and business function. The analysis contains $2{,}808$ such observations, built from $364$ distinct country-industry Atlas predictor cells in $12$ country groups: Bangladesh, Brazil, Burkina Faso, Croatia, Ethiopia, Ghana, India, Kenya, Korea, Poland, Senegal and Viet Nam. Cambodia, Chile and Georgia do not enter because their released FAT files lack usable two-digit industry codes, and the two Indian survey waves are assigned to the same heldout-country fold.

Both specifications use the same three labour-margin measures: substitution-only, balanced-both and augmentation-only. The first assigns every country the same industry scores. The second uses industry scores measured separately for each country. We evaluate the models by leaving out one country at a time, assigning the two India survey waves to the same fold.

\begin{figure}[!htbp]
\centering
\caption{Country conditioning lowers heldout-country prediction error for firm technology sophistication.}
\includegraphics[width=0.78\textwidth]{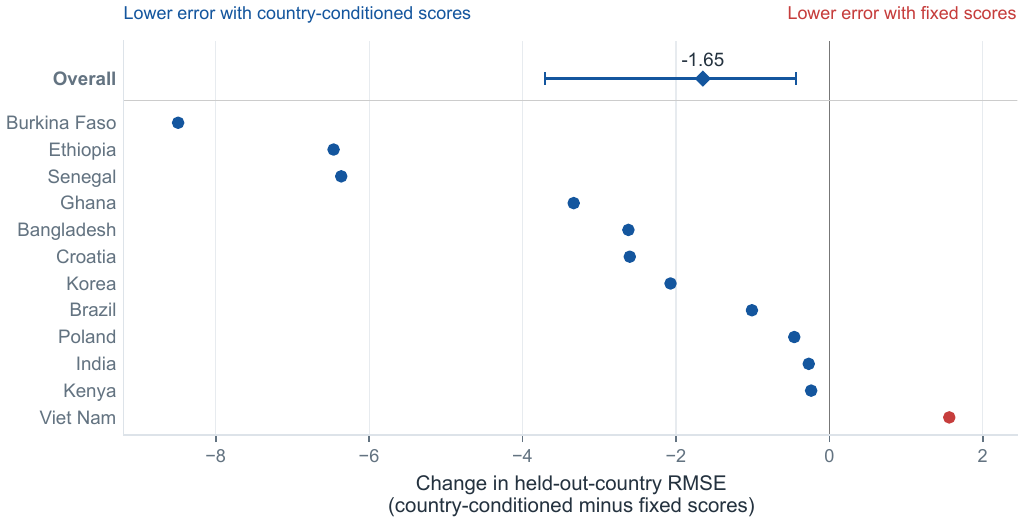}
\caption*{\scriptsize Notes: Data are from the World Bank Firm Adoption of Technology (FAT) survey. The matched sample contains $12$ country groups. Cambodia, Chile and Georgia are excluded because their released files lack usable two-digit industry codes. The two India waves form one country group. Country points report the difference in leave-country-out weighted RMSE between models using country-conditioned and fixed industry labour-margin scores. Negative values indicate lower prediction error with country conditioning. Both models include the same business-function controls and labour-margin measures; only the industry scores differ. The overall diamond reports the pooled difference across $2{,}808$ country--industry--business-function outcomes, and the horizontal line gives its country-cluster bootstrap 95\% confidence interval. The pooled RMSE difference is $-1.649$ points (95\% CI: $[-3.705,-0.433]$). Country conditioning lowers RMSE in $11$ of the $12$ country folds (exact one-sided sign test $p=0.003$; exhaustive sign-randomisation $p=0.002$).}
\label{fig:validation_fat_country_conditioning}
\end{figure}

The model based on fixed industry scores has a weighted RMSE of $13.094$ and an out-of-sample $R^2$ of $0.346$. Using the country-conditioned scores lowers the RMSE to $11.445$ and raises the out-of-sample $R^2$ to $0.500$. Prediction loss falls in $11$ of the $12$ country folds (Table~\ref{tab:validation_fat_country_conditioning} and Fig.~\ref{fig:validation_fat_country_conditioning}). The exact one-sided sign test gives $p=0.003$, and exhaustive sign randomisation gives $p=0.002$. Permuting the country labels gives $p=0.001$. The result is also present when we impose minimum firm counts, remove the second India wave, estimate the relationship at the firm level and use alternative technology outcomes.

The predictive gain comes from broad differences across countries. Once country, industry and business-function means are removed, the remaining interaction explains almost none of the outcome variation ($R^2=0.0008$, $p=0.807$). The FAT comparison therefore supports the country calibration of the Atlas measures. It provides little evidence about individual country-industry-function deviations.

\begin{table}[!htbp]
\centering
\begin{minipage}{0.87\linewidth}
\small
\setlength{\tabcolsep}{4pt}
\renewcommand{\arraystretch}{1.10}
\caption{Country conditioning and prediction of firm technology sophistication in the World Bank FAT survey.}
\label{tab:validation_fat_country_conditioning}
\begin{tabularx}{\linewidth}{@{}>{\raggedright\arraybackslash}X C{2.55cm} C{2.90cm}@{}}
\toprule
Prediction model & Weighted RMSE & Out-of-sample $R^2$ \\
\midrule
Fixed industry labour-margin scores & $13.094$ & $0.346$ \\
Country-conditioned industry labour-margin scores & $11.445$ & $0.500$ \\
\midrule
\multicolumn{3}{@{}l}{\textit{Prediction difference and uncertainty}} \\
\addlinespace[2pt]
RMSE difference: country-conditioned minus fixed & \multicolumn{2}{c@{}}{$-1.649$} \\
Country-cluster bootstrap 95\% confidence interval & \multicolumn{2}{c@{}}{$[-3.705,-0.433]$} \\
Country comparisons with lower prediction error & \multicolumn{2}{c@{}}{$11$ of $12$} \\
Exact one-sided sign test: $p$-value & \multicolumn{2}{c@{}}{$0.003$} \\
Exhaustive sign-randomisation test: $p$-value & \multicolumn{2}{c@{}}{$0.002$} \\
Country-label permutation test: $p$-value & \multicolumn{2}{c@{}}{$0.001$} \\
\midrule
\multicolumn{3}{@{}l}{\textit{After removing country, industry and business-function means}} \\
\addlinespace[2pt]
$R^2$ for the remaining interaction component & \multicolumn{2}{c@{}}{$0.0008$} \\
$p$-value & \multicolumn{2}{c@{}}{$0.807$} \\
\bottomrule
\end{tabularx}
\caption*{\scriptsize Notes: FAT denotes the World Bank Firm Adoption of Technology survey. The outcome measures the sophistication of technologies used in business functions. The sample contains $2{,}808$ country-industry-business-function observations with $364$ distinct country-industry Atlas predictor cells in $12$ country groups. Both models include the same business-function controls and three labour-margin measures: substitution-only, balanced-both and augmentation-only. Only the industry scores differ. Prediction is evaluated in countries excluded from estimation one at a time, with the two India waves in the same group. The final two rows assess finer variation after removing country, industry and business-function means.}
\end{minipage}
\end{table}

\paragraph{Use of the same work activity across countries}\leavevmode\hypertarget{validation-anthropic-activity}{} The Anthropic Economic Index permits a more direct test because it reports the composition of Claude use for the same O*NET work activities in different countries \citep{MassenkoffEtAl2026AnthropicEconomicIndex}. The public data cover April and May 2026. After matching to the Atlas, the sample contains $32{,}877$ published country-activity-month cells across $95$ countries and $1{,}730$ O*NET task identifiers. The main model uses the $31{,}289$ cells and $1{,}137$ tasks for which the log location quotient can be estimated.

The location quotient measures the share of a country's published Claude use devoted to an activity relative to that activity's global share. Country fixed effects absorb differences in overall use across countries, while activity-by-month fixed effects absorb the global popularity of each activity in each month. The remaining variation asks whether the same activity occupies a larger share of Claude use where its country-specific Atlas score is higher.

\begin{figure}[!htbp]
\centering
\caption{Country-specific exposure and relative Claude use across work-activity levels.}
\includegraphics[width=0.78\textwidth]{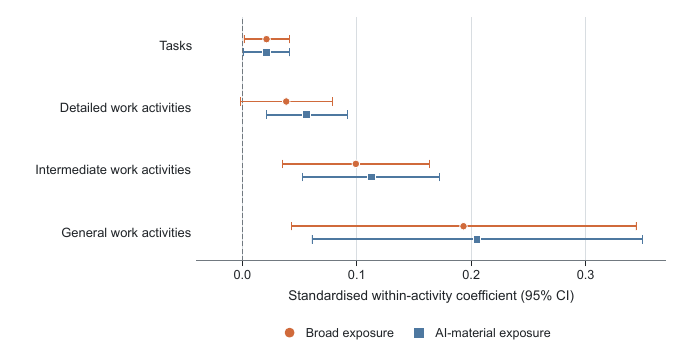}
\caption*{\scriptsize Notes: Points report standardised associations between country-specific Atlas scores and relative Claude use within the same O*NET activity across 95 countries in April and May 2026. The outcome is the log ratio of an activity's share of a country's published Claude use to its global share. Models include country and activity-by-month fixed effects. Both variables are standardised after removing these fixed effects. Orange circles denote broad exposure and blue squares AI-material exposure. Intervals are 95\% confidence intervals based on standard errors clustered by country and activity. Samples contain 31,289 task-level cells, 19,762 detailed-work-activity cells, 11,605 intermediate-work-activity cells and 3,727 general-work-activity cells.}
\label{fig:validation_anthropic_activity_hierarchy}
\end{figure}

Anthropic publishes the O*NET hierarchy at the task, detailed-work-activity, intermediate-work-activity and general-work-activity levels. We estimate the same specification at each level and standardise the outcome and Atlas score after removing the fixed effects. For broad exposure, the coefficients are $0.021$, $0.038$, $0.099$ and $0.193$ from the task to the general-work-activity level. For AI-material exposure, they are $0.021$, $0.056$, $0.112$ and $0.205$. The association is therefore weak for narrowly stated tasks and clearer for broader, pre-defined work categories. Figure~\ref{fig:validation_anthropic_activity_hierarchy} compares the coefficients across activity levels. Table~\ref{tab:validation_anthropic_activity_hierarchy} reports the samples and uncertainty at each level.

\begin{table}[!htbp]
\centering
\small
\setlength{\tabcolsep}{4pt}
\renewcommand{\arraystretch}{1.12}
\caption{Country-specific exposure and relative Claude use within the same work activity.}
\label{tab:validation_anthropic_activity_hierarchy}
\begin{tabularx}{\textwidth}{@{}>{\raggedright\arraybackslash}X r r r r r r@{}}
\toprule
& & & \multicolumn{2}{c}{Broad exposure} & \multicolumn{2}{c@{}}{AI-material exposure} \\
\cmidrule(lr){4-5}\cmidrule(l){6-7}
Activity level & Observations & Activities & Coefficient (SE) & $p$-value & Coefficient (SE) & $p$-value \\
\midrule
Task & $31{,}289$ & $1{,}137$ & $0.021\;(0.010)$ & $0.036$ & $0.021\;(0.010)$ & $0.042$ \\
Detailed work activity & $19{,}762$ & $432$ & $0.038\;(0.021)$ & $0.062$ & $0.056\;(0.018)$ & $0.002$ \\
Intermediate work activity & $11{,}605$ & $156$ & $0.099\;(0.033)$ & $0.002$ & $0.112\;(0.031)$ & $<0.001$ \\
General work activity & $3{,}727$ & $34$ & $0.193\;(0.077)$ & $0.012$ & $0.205\;(0.074)$ & $0.005$ \\
\bottomrule
\end{tabularx}
\caption*{\scriptsize Notes: The Anthropic Economic Index sample covers $95$ countries in April and May 2026 at four levels of the O*NET hierarchy. Each observation is a published country-activity-month cell. The outcome is the log ratio of an activity's share of a country's published Claude use to its global share. Each coefficient comes from a separate regression with country and activity-by-month fixed effects. The outcome and Atlas predictor are standardised after removing the fixed effects, and cells receive equal weight. Standard errors, in parentheses, are clustered by country and activity. Reported $p$-values are nominal, without adjustment for multiple comparisons.}
\end{table}

\begin{figure}[!htbp]
\centering
\caption{Country-conditioned Atlas exposure is associated with relative Claude use within the same work activity.}
\includegraphics[width=0.95\textwidth]{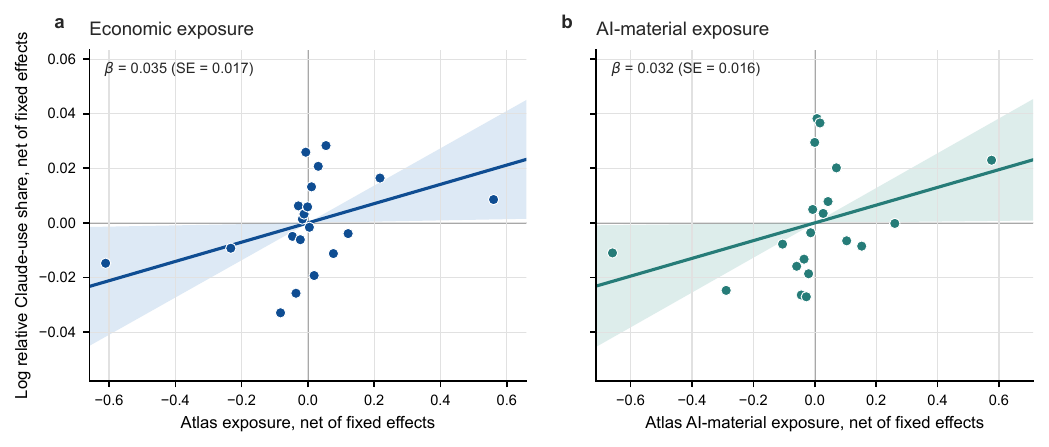}
\caption*{\scriptsize Notes: The horizontal and vertical axes show values after country and task-by-month fixed effects have been removed. Each dot averages one twentieth of the sample, ordered by the horizontal-axis value. The two end dots each contain about $1{,}565$ observations, not single outliers. The outcome is the log share of each country's published Claude use devoted to an activity relative to that activity's global share. Panel~(a) uses Atlas exposed share and panel~(b) uses Atlas AI-material share. Lines are fitted using all $31{,}289$ cells, spanning $95$ countries and $1{,}137$ O*NET activities in April and May 2026. Shaded regions are 95\% confidence intervals based on standard errors two-way clustered by country and activity. Maximum-statistic country-label permutation tests across two outcomes and five Atlas predictors give familywise $p=0.035$ in panel~(a) and $p=0.042$ in panel~(b). This figure reports the task-level specification in the original residualised units. Figure~\ref{fig:validation_anthropic_activity_hierarchy} reports the standardised coefficients across activity levels.}
\label{fig:validation_anthropic_same_activity}
\end{figure}

The within-$R^2$ is small, and weighting the published cells by their share of Claude use produces imprecise estimates. We therefore interpret the result as evidence of association in the relative composition of published Claude-use cells. It does not provide a high-accuracy prediction of Claude use or evidence about the use of automation technologies more generally.

Anthropic also classifies Claude interactions by whether they automate work or augment workers. At the broader work-activity level, Atlas substitution-only exposure is positively associated with Anthropic's automation share when cells receive equal weight. The coefficient is $21.4$ percentage points, with a standard error of $10.7$ and $p=0.044$. The estimate becomes imprecise when cells are weighted by their Claude-use shares. We treat this as secondary evidence for the distinction between labour substitution and augmentation.

\subsubsection{Observed use and enabling conditions}

Panel~C of Table~\ref{tab:validation_external_summary} summarises the comparisons with observed technology use, firm adoption and enabling conditions. The paragraphs below describe each source and specification.

\paragraph{Global country-level ChatGPT use}\leavevmode\hypertarget{validation-openai-country}{}
OpenAI Signals reports both a population-normalised measure of overall consumer ChatGPT use and the share of consumer messages classified as work-related by country \citep{OpenAI2026SignalsData}. The overall-use panel contains $663$ country-quarter observations for $112$ Atlas countries over six quarters, of which $109$ appear in every quarter. The work-related-share panel covers $105$ matched countries over 24 months. OpenAI does not release country-by-work-activity cells outside the United States, so the global comparison concerns country-level use rather than the relative use of the same activity across countries.

The raw six-quarter average use rank correlates with Atlas exposed share at Spearman $0.909$ and with AI-material share at $0.902$. These associations partly reflect the common relationship of platform use, internet access and Atlas exposure with development. We therefore remove log GDP per capita, internet use and region from both the external outcome and the Atlas measure. The partial correlations with the work-related share are $0.403$ for exposed share and $0.472$ for AI-material share. The corresponding correlations with overall use are $0.162$ and $0.220$. The stronger relationship with work-related use is the construct-specific prediction for a measure built from work tasks.

\paragraph{Global country-level Claude use}\leavevmode\hypertarget{validation-anthropic-country}{}
Anthropic also reports each country's use per capita and the share of use classified as work-related. After removing log GDP per capita, internet use and region, the partial correlations between work-related share and broad exposure or AI-material exposure are $0.469$ and $0.503$. The corresponding correlations with overall use per capita are $-0.156$ and $-0.116$. The work-related association therefore does not appear to be a general platform-penetration relationship.

\paragraph{Industrial robot adoption}\leavevmode\hypertarget{validation-robot-adoption}{} The physical-execution component measures exposure through technologies that act directly on physical tasks. Industrial robot use provides an observed counterpart to this part of the Atlas. In Eurostat, the share of manufacturing firms using industrial robots is higher in countries with greater physical-execution exposure (Pearson $r=0.690$ across $23$ countries) \citep{Eurostat2022RobotUse}. IFR measures operational industrial robots relative to manufacturing employment. Across the $19$ countries covered by both sources, physical-execution exposure correlates with log robot density at Pearson $r=0.476$ and with the country ranking at Spearman $\rho=0.502$ \citep{IFR2026RobotDensity}. Both comparisons therefore support this specific component of the measure.

\begin{figure}[!htbp]
\centering
\caption{Physical-execution exposure is higher where manufacturers use more industrial robots.}
\includegraphics[width=0.95\textwidth]{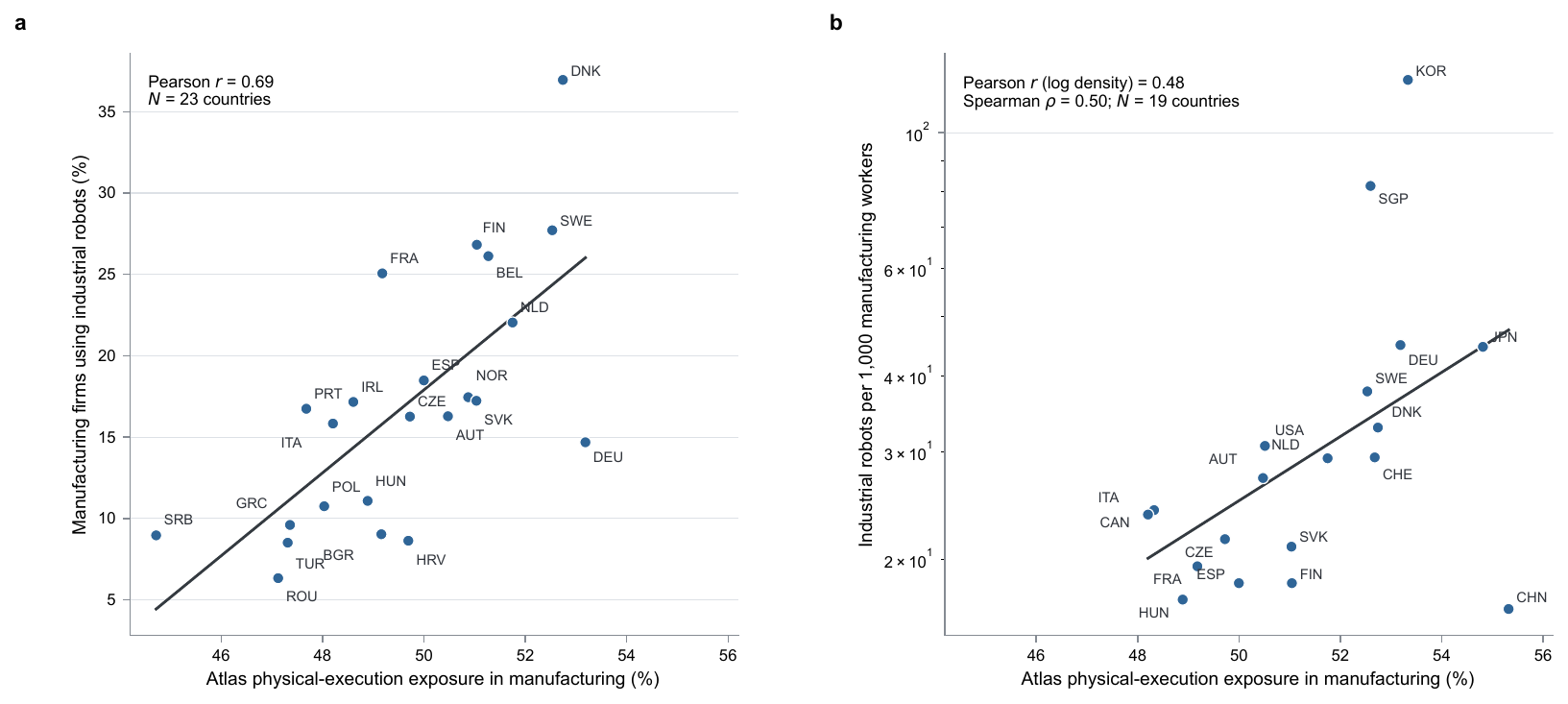}
\caption*{\scriptsize Notes: Points represent countries. The Atlas measure is the share of manufacturing task content classified as exposed through physical execution. Panel~(a) compares this measure with the Eurostat share of manufacturing enterprises with at least ten workers reporting Industrial robot use in 2022 across $23$ countries. Panel~(b) compares it with IFR operational industrial robots per $1{,}000$ manufacturing workers in 2024 across $19$ countries. The vertical axis in panel~(b) uses a logarithmic scale. Lines show fitted relationships in levels in panel~(a) and with log robot density in panel~(b).}
\label{fig:validation_robot_adoption}
\end{figure}

\begin{figure}[!htbp]
\centering
\caption{Country-level AI-material share against the IMF AI Preparedness Index (AIPI).}
\includegraphics[width=0.75\textwidth]{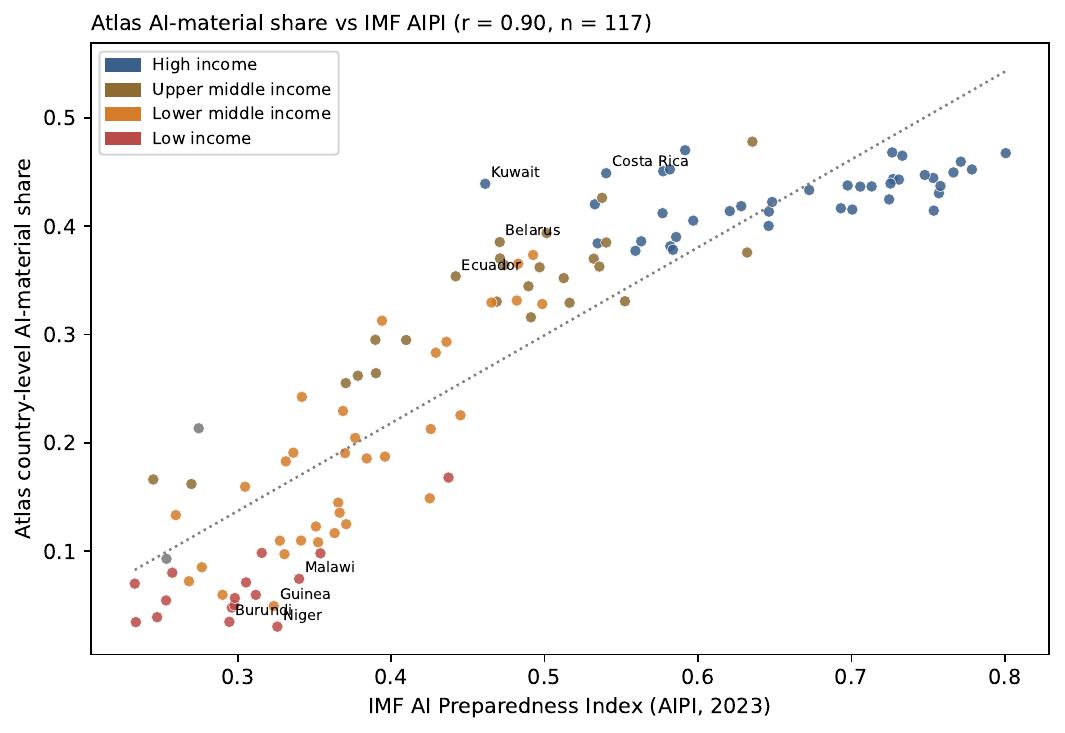}
\caption*{\scriptsize Notes: Points are the 117 countries with coverage in both the atlas and the IMF AIPI composite. The x-axis is the IMF AIPI composite for 2023; the y-axis is country-level AI-material share. The dotted line is a linear fit. Pearson correlation is $0.90$ and Spearman correlation is $0.93$. Colours mark World Bank income groups; Table~\ref{tab:appendix_aipi_stress_tests} reports robustness checks.}
\label{fig:validation_imf_aipi}
\end{figure}

\paragraph{Country-level AI readiness}\leavevmode\hypertarget{validation-imf-readiness}{} A further external anchor asks whether our country-level AI-material pattern agrees with an independently constructed country-level readiness measure. The IMF AI Preparedness Index (AIPI) \citep{CazzanigaEtAl2024GenAI} is the closest comparator: it scores economies on digital infrastructure, human capital and labour-market policies, innovation and economic integration, and regulation and ethics. We compare the AIPI composite and its sub-components with country-level AI-material share, foundation-model-like share, and overall mean exposure across the countries with both paper and IMF coverage. Figure~\ref{fig:validation_imf_aipi} reports the main comparison; Table~\ref{tab:appendix_aipi_stress_tests} reports the raw, partial, within-income, rank, and leave-one-out checks.

The raw relationship is strong. AI-material share correlates with the AIPI composite at Pearson $0.90$ (Spearman $0.93$); overall mean exposure correlates at Pearson $0.88$ (Spearman $0.95$). The comparison is deliberately demanding because both variables are closely related to development: AI-material share correlates with log GDP per capita at Pearson $0.92$, and AIPI correlates with log GDP per capita at Pearson $0.91$. We therefore report the log-GDP-residual association as a stricter check on whether the two measures align beyond their common development gradient. After residualising both variables on log GDP per capita, the AI-material--AIPI partial Pearson remains positive at $0.42$.

Within-income comparisons give the same reading. AI-material share still correlates with AIPI at Pearson $0.72$ in low-income countries, $0.84$ in lower-middle-income countries, and $0.90$ in upper-middle-income countries; the high-income correlation is lower ($0.50$), where the AIPI approaches its ceiling. The residual disagreements are interpretable: our task-level country conditioning gives relatively more AI-material exposure to economies with specific sectoral technology deployment, and less to economies where composite readiness scores run ahead of task-level deployment conditions. The AIPI check therefore supports the country-level ordering while making clear that the raw $0.90$ correlation partly reflects shared development-stage information.

Table~\ref{tab:appendix_aipi_stress_tests} compiles the raw and partial Pearson correlations, the within-income-group Pearson correlations, the rank-robustness check, and a leave-one-out stability check into one reference panel. Leave-one-out robustness is tight: dropping any single country of the $117$ changes the Pearson within $[0.897, 0.905]$ (SD $0.001$), and dropping the US and China jointly moves the coefficient from $0.899$ to $0.898$. No single country shifts the correlation materially.

\begin{table}[H]
\centering
\scriptsize
\setlength{\tabcolsep}{4pt}
\renewcommand{\arraystretch}{1.08}
\caption{Atlas--AIPI cross-country correlation checks.}
\begin{tabularx}{\textwidth}{@{}L{0.24\textwidth} C{0.08\textwidth} C{0.08\textwidth} C{0.10\textwidth} C{0.06\textwidth} >{\raggedright\arraybackslash}X@{}}
\toprule
\multicolumn{6}{l}{\textbf{Panel A.} Raw and log-GDP-residual Pearson correlations, atlas AI-material share $\times$ IMF AIPI series.} \\
\midrule
IMF series & Raw $r$ & Residual $r$ & Raw-to-residual change & $n$ & Interpretation \\
\midrule
AIPI composite                     & 0.899 & 0.417 & 54\% & 117 & Positive residual association beyond the common GDP gradient. \\
Digital Infrastructure (DI)        & 0.914 & 0.498 & 46\% & 122 & Largest residual association. \\
Human Capital \& LM Policies (HCLMP) & 0.858 & 0.378 & 56\% & 121 & Similar profile to composite. \\
Innovation \& Economic Integration (IEI) & 0.750 & 0.070 & 91\% & 118 & Closely tied to the GDP gradient in this sample. \\
Regulation \& Ethics (RE)          & 0.835 & 0.319 & 62\% & 122 & Positive residual association after GDP adjustment. \\
\addlinespace
\multicolumn{6}{l}{\textbf{Panel B.} Within-income-group Pearson correlations, atlas features $\times$ AIPI composite.} \\
\midrule
Income group & $n$ & AI-material vs AIPI & FM-like vs AIPI & & Interpretation \\
\midrule
Low income                 & 19 & 0.718 & 0.710 & & Strong residual correspondence despite compressed range. \\
Lower middle income        & 35 & 0.839 & 0.826 & & Strongest within-group agreement. \\
Upper middle income        & 25 & 0.897 & 0.814 & & Strongest within-group agreement. \\
High income                & 41 & 0.504 & 0.038 & & AIPI approaches ceiling; atlas still varies. \\
\bottomrule
\end{tabularx}
\caption*{\scriptsize Notes: Rows report alternative correlations between country-level AI-material share and the IMF AIPI composite. FM-like denotes the foundation-model-like channel bundle. Residual correlations are computed by residualising both series on the named controls and then correlating the residuals. Within-income rows remove World Bank income-group means. Leave-one-out rows report the range after dropping one country at a time.}
\label{tab:appendix_aipi_stress_tests}
\end{table}

\paragraph{Observed use at the work-activity level}\leavevmode\hypertarget{validation-openai-activity}{}
We also compare the atlas labels for the United States with observed ChatGPT use in the same work-activity taxonomy. The unit of the comparison is an O*NET intermediate work activity (IWA). For each IWA, the OpenAI measure is its share of U.S. work-related consumer ChatGPT messages, not an adoption rate, task frequency, or measure of employment exposure. Public OpenAI observed-use data report monthly shares of U.S. consumer ChatGPT messages by IWA, separately for all messages and for messages classified as work-related \citep{ChatterjiEtAl2025HowPeopleUseChatGPT}. The IWA files are U.S.-only and include an `Other IWA' residual category for rare activities. We use the work-related series as the primary outcome, average April--June 2026, drop `Other IWA', and match the 164 named IWA categories to the O*NET task hierarchy. These categories account for 97.7\% of work-related U.S. messages in 2026 Q2. We then aggregate atlas task labels for the United States to the IWA level by averaging over task--IWA edges. For each IWA, broad exposure is the share of linked U.S.-conditioned task edges at exposure levels 2 or 3. Channel measures are the corresponding exposed shares whose dominant channel is the named channel. AI-function measures are defined analogously for exposed task edges whose dominant AI function is the named function, and AI-material share is the exposed share for which AI is material to the automation channel.

\begin{figure}[!htbp]
\centering
\caption{Observed ChatGPT work use and Atlas exposure across U.S. work activities.}
\includegraphics[width=\textwidth]{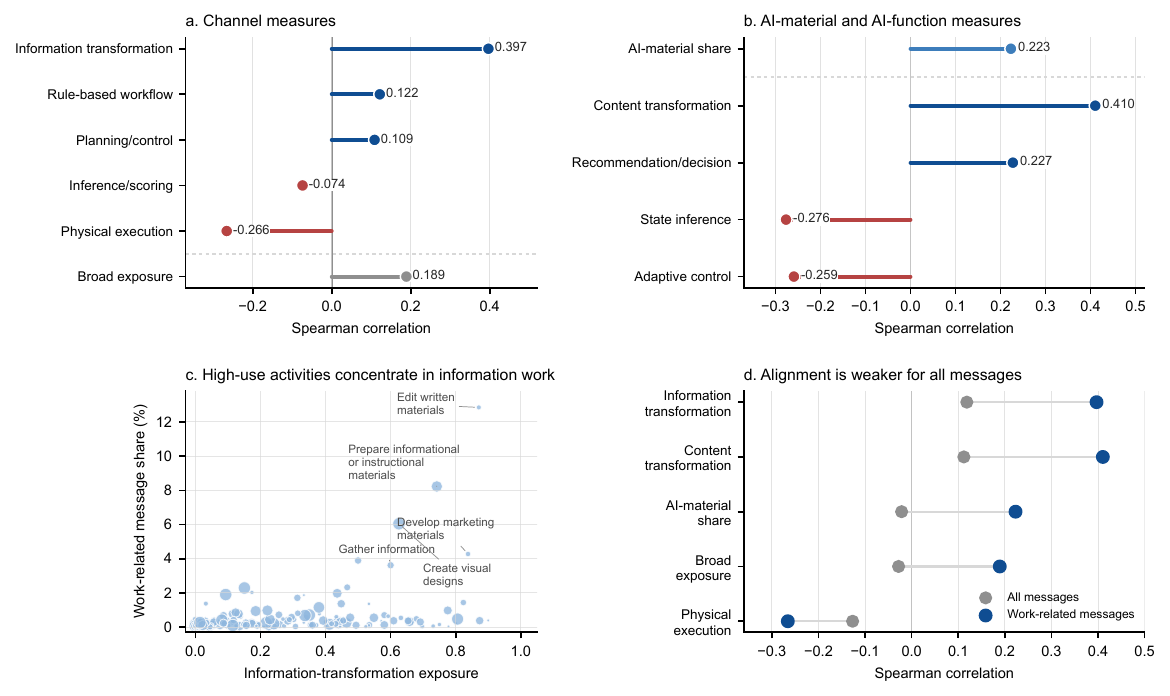}
\caption*{\scriptsize Notes: The unit of observation is an O*NET intermediate work activity (IWA). OpenAI observed-use data are monthly shares, released with differential privacy protections, of U.S. consumer ChatGPT messages by IWA. The main outcome is the average share among messages classified as work-related in April--June 2026. The figure drops `Other IWA', leaving 164 named IWA categories. Atlas measures use U.S.-conditioned task labels aggregated to IWA categories by averaging over task--IWA edges. Panel~(a) reports Spearman rank correlations between work-related message share and Atlas channel measures; broad exposure is separated as an aggregate measure. Panel~(b) reports correlations for AI-material share and AI-function measures. Panel~(c) plots work-related message share against information-transformation exposure. Panel~(d) compares selected correlations for all messages and work-related messages.}
\label{fig:validation_openai_observed_use_iwa_alignment}
\end{figure}

Figure~\ref{fig:validation_openai_observed_use_iwa_alignment} reports the observed-use correlations, which follow the expected ordering. In the channel panel, work-related message shares correlate most strongly with information transformation (Spearman $\rho=0.397$). Rule-based workflow ($0.122$) and planning/control ($0.109$) are positive but much smaller, inference/scoring is close to zero ($-0.074$), and physical execution is negative ($-0.266$). Broad economic exposure is positive but smaller ($0.189$). In the AI panel, content transformation has the strongest correlation ($0.410$), while AI-material exposed share is positive but broader ($0.223$). Recommendation/decision is also positive ($0.227$), whereas state inference ($-0.276$) and adaptive control ($-0.259$) are negative. The exercise therefore supports a narrow interpretation: public OpenAI observed-use data line up with the information and content-transformation parts of the taxonomy, rather than with AI exposure in general.

The information and content-transformation pattern is stable across work-related alternatives. It appears when we use June 2026 alone, average over the full July 2024--June 2026 public period, divide message shares by the number of atlas task edges in the IWA, or residualise log message share on log task-edge count (Table~\ref{tab:appendix_openai_observed_use_iwa_robustness}). The correlations are much smaller for the all-message IWA series, which suggests that the check is picking up work-related use rather than general consumer activity. The scope of the check is limited. OpenAI releases IWA shares only for the United States, and the public sample covers consumer ChatGPT rather than enterprise messages or Codex. We therefore read the exercise as a U.S. observed-use construct-validity check, not as evidence on adoption outside the United States or on labour-market effects.

\begin{table}[!htbp]
\centering
\footnotesize
\caption{Robustness of observed-use correlations across U.S. work activities.}
\label{tab:appendix_openai_observed_use_iwa_robustness}
\begin{tabular}{lrrrrrr}
\toprule
Atlas measure & Work Q2 & Work June & Work full & Per edge & Residual & All Q2 \\
\midrule
Information transformation & 0.397 & 0.373 & 0.457 & 0.376 & 0.397 & 0.118 \\
Rule-based workflow & 0.122 & 0.129 & 0.118 & 0.022 & 0.080 & -0.077 \\
Planning/control & 0.109 & 0.121 & 0.081 & 0.046 & 0.088 & -0.064 \\
Inference/scoring & -0.074 & -0.096 & -0.050 & -0.145 & -0.116 & -0.030 \\
Physical execution & -0.266 & -0.265 & -0.280 & -0.406 & -0.346 & -0.127 \\
Broad economic exposure & 0.189 & 0.177 & 0.200 & 0.157 & 0.175 & -0.028 \\
AI-material exposed share & 0.223 & 0.197 & 0.258 & 0.251 & 0.240 & -0.022 \\
Content transformation & 0.410 & 0.390 & 0.465 & 0.374 & 0.402 & 0.112 \\
Recommendation/decision & 0.227 & 0.240 & 0.199 & 0.125 & 0.194 & 0.022 \\
State inference & -0.276 & -0.300 & -0.237 & -0.282 & -0.298 & -0.222 \\
Adaptive control & -0.259 & -0.241 & -0.279 & -0.279 & -0.277 & -0.111 \\
\bottomrule
\end{tabular}
\caption*{\scriptsize Notes: Entries are Spearman rank correlations across 164 named O*NET intermediate work activities. The primary outcome is the OpenAI public work-related U.S. message share averaged over April--June 2026. The June 2026 column uses the latest public month. The full-period column averages July 2024--June 2026. The per-task-edge column divides the 2026 Q2 work-related message share by the number of Atlas task edges in the IWA. The residual column uses residual log message share after projecting log work-related message share on log task-edge count. The all-message column uses the OpenAI public all-message IWA series rather than the work-related series. Atlas measures are built from U.S.-conditioned task labels. Channel rows are exposed task shares whose dominant channel is the named channel. AI-function rows are exposed task shares whose dominant AI function is the named function. Broad economic exposure and AI-material exposed share are aggregate measures rather than channels or AI functions.}
\end{table}

The external comparisons answer different questions. At the task and occupation levels, Atlas dimensions align most closely with established measures of the same concepts. Observed-use data provide a separate check: information-transformation measures align with work-related ChatGPT use, AI-material measures with reported AI use, and physical-execution measures with industrial-robot adoption. Eurostat, FAT and Anthropic ask whether measuring exposure separately for each country adds information beyond fixed task or industry scores. These results provide convergent evidence for the measurement design rather than a ground-truth test of feasible exposure. Realised use also depends on prices, investment, management and demand.

\FloatBarrier
\subsection{Internal validity and model-robustness checks}
\label{sec:appendix_validation}

\noindent This subsection reports the checks that operate directly on the model-produced labels and prompt outputs. Table~\ref{tab:internal_validation_summary} summarises cross-model convergence, rationale-to-label predictability, prompt-paraphrase stability and face validity.

\begin{table}[H]
\centering
\small
\setlength{\tabcolsep}{4pt}
\renewcommand{\arraystretch}{1.12}
\caption{Summary of internal validity and model-robustness checks.}
\label{tab:internal_validation_summary}
\begin{tabularx}{\textwidth}{@{}L{3.40cm}L{4.25cm}>{\raggedright\arraybackslash}X@{}}
\toprule
Check & Sample & Result \\
\midrule
\hyperref[sec:validation_convergent]{Convergent validity across model families} & $18{,}169$ context-free tasks with paired labels & Exposed/non-exposed agreement: $75.1\%$. Exact exposure-level agreement: $48.1\%$. Within one level: $95.0\%$. \\
\addlinespace[5pt]
\hyperref[sec:validation_reasoning]{Rationale-to-label predictability} & $1{,}000$ country-task observations, with $250$ from each exposure level & An independent model recovers exposed/non-exposed status from the task and rationale in $99.0\%$ of cases. Exact exposure-level agreement: $71.5\%$. \\
\addlinespace[5pt]
\hyperref[sec:validation_paraphrase]{Prompt-paraphrase stability} & Three prompt paraphrases applied to $1{,}000$ context-free tasks & Exposed/non-exposed agreement with the original labels: $88.7\%$, $87.0\%$ and $84.3\%$. Within one level: $96\%$--$98\%$. \\
\addlinespace[5pt]
\hyperref[sec:validation_face]{Face validity} & Label distributions, $2{,}320{,}863$ rationales, $18{,}797$ same-task country pairs and $11$ anchor occupations & Keyword rules flag $2{,}654$ potential rationale--label contradictions ($0.11\%$). Near-verbatim rationale reuse occurs in $0.4\%$ of country pairs. Distributions and anchor occupations provide descriptive checks. \\
\bottomrule
\end{tabularx}
\caption*{\scriptsize Notes: Linked check names lead to their supporting analyses. These checks assess agreement, stability and internal coherence, not accuracy against independently observed task-country outcomes. The cross-model and prompt-paraphrase exercises use context-free tasks. Rationale-to-label predictability conditions on the generated explanation and does not establish that it faithfully records the model's internal reasoning. Keyword flags identify potential contradictions, not confirmed errors.}
\end{table}

\subsubsection{Convergent validity across model families}
\label{sec:validation_convergent}
\begin{figure}[!htbp]
\centering
\caption{Cross-model convergent validity: main labels vs.\ \texttt{gpt-5.4-mini} over the full context-free task universe.}
\includegraphics[width=0.98\textwidth]{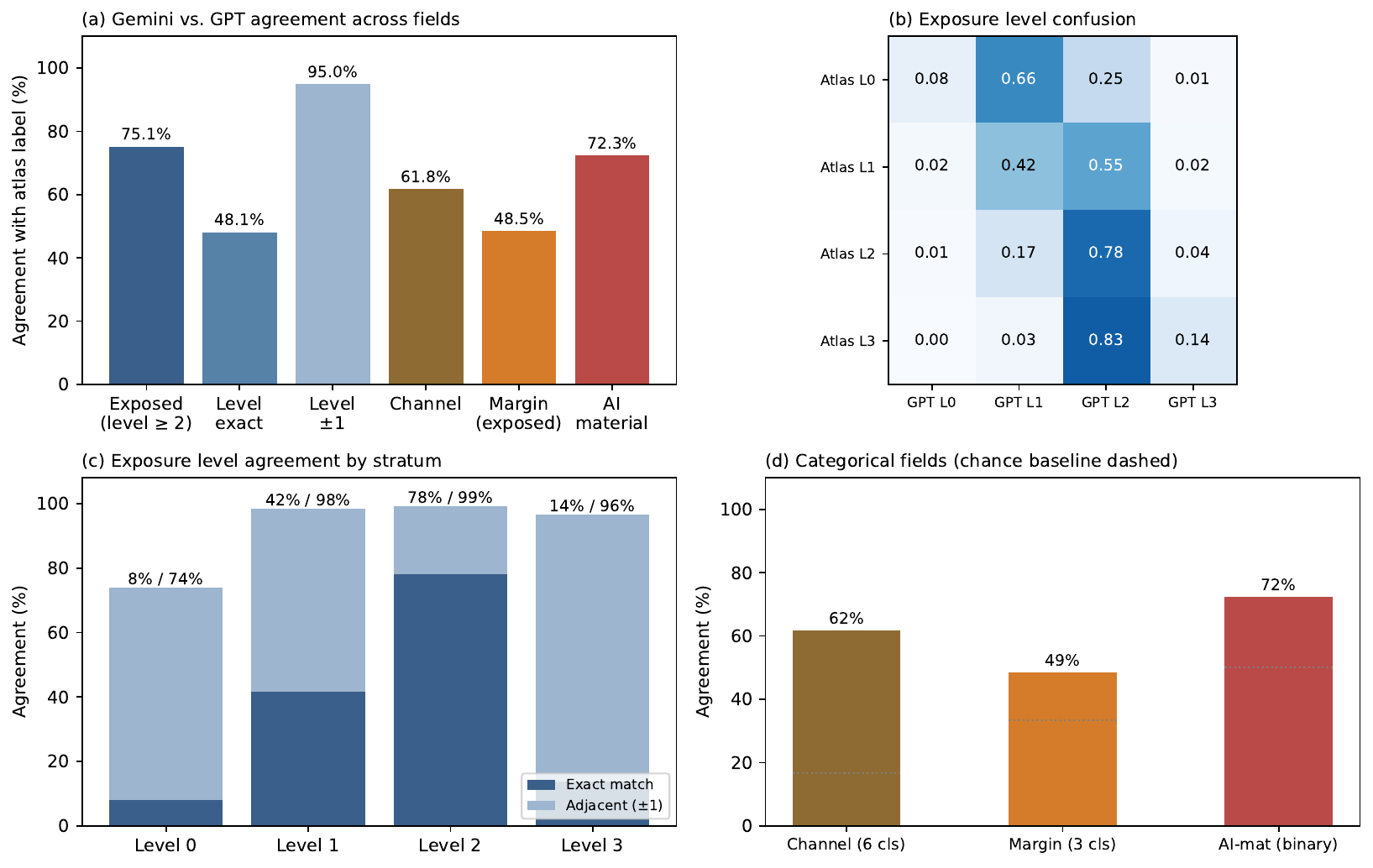}
\caption*{\scriptsize Notes: Both models receive the same context-free prompt, task text, and JSON response schema. The comparison includes 18,169 tasks with valid paired exposure labels, covering 96.7\% of the original task universe. Each statistic uses tasks with valid paired labels for the relevant field. Panel~(a) reports agreement by label field. Panel~(b) shows the exposure-level confusion matrix, panel~(c) reports agreement by original exposure level, and panel~(d) compares categorical-field agreement with chance agreement under equally likely categories. The channel comparison includes the five technology channels and \textit{none}, giving six categories and an equal-category chance benchmark of $1/6$.}
\label{fig:validation_cross_model_validity}
\end{figure}

Convergent validity asks whether the broad classification structure is reproduced across model families when the schema and prompt are held fixed. Of the $18{,}797$ tasks submitted to the main context-free run, $18{,}177$ yielded usable labels. We re-label these tasks with OpenAI's \texttt{gpt-5.4-mini} under the same context-free system prompt and JSON response schema. Figure~\ref{fig:validation_cross_model_validity} reports results for the $18{,}169$ tasks with valid exposure predictions from both models. Of these paired classifications, $95.0\%$ differ by no more than one exposure level, $75.1\%$ agree on exposed versus non-exposed status, and $48.1\%$ agree on the exact four-level label.

The disagreement pattern is systematic. GPT compresses labels toward the middle of the four-level scale, rating many original level-0 tasks as level~1 or 2 and many original level-3 tasks as level~2. This compression reduces exact and binary agreement while leaving $95.0\%$ of paired classifications within one level. Exact agreement for the other fields is $61.8\%$ for dominant channel, $72.3\%$ for AI materiality, and $48.5\%$ for labour margin among exposed tasks.

Taken together, the independent model reproduces the broad classification structure, while exact levels and the exposed-versus-non-exposed boundary carry greater model-family sensitivity. Using context-free tasks isolates model-family sensitivity from country conditioning; the exercise therefore evaluates the baseline task classification rather than the full country-conditioned label set.

\FloatBarrier

\subsubsection{Rationale-to-label predictability}
\label{sec:validation_reasoning}
We draw a stratified sample of $1{,}000$ country-task observations, with $250$ observations from each exposure level. For each observation, an independent model family receives the task description and the original rationale, with the original label withheld, and predicts the exposure level, dominant technology channel, and labour margin under the paper's schema. The independent model recovers the exposed-versus-not-exposed label on $99.0\%$ of observations; only $10$ of $1{,}000$ predictions cross the exposed threshold. Exposure level is within one step of the original label for all observations, with exact-level agreement at $71.5\%$. Dominant-channel agreement is $67.7\%$ across all $1{,}000$ observations and $72.8\%$ among the $500$ exposed observations. Labour-margin agreement among exposed observations is $74.8\%$. Disagreements concentrate at adjacent exposure levels, especially around the level-0/level-1 and level-2/level-3 boundaries. The check therefore supports the internal coherence of the headline exposure label, while showing that fine-grained fields should be read with more caution.

\begin{figure}[!htbp]
\centering
\caption{Rationale-to-label predictability under an independent model family.}
\includegraphics[width=0.98\textwidth]{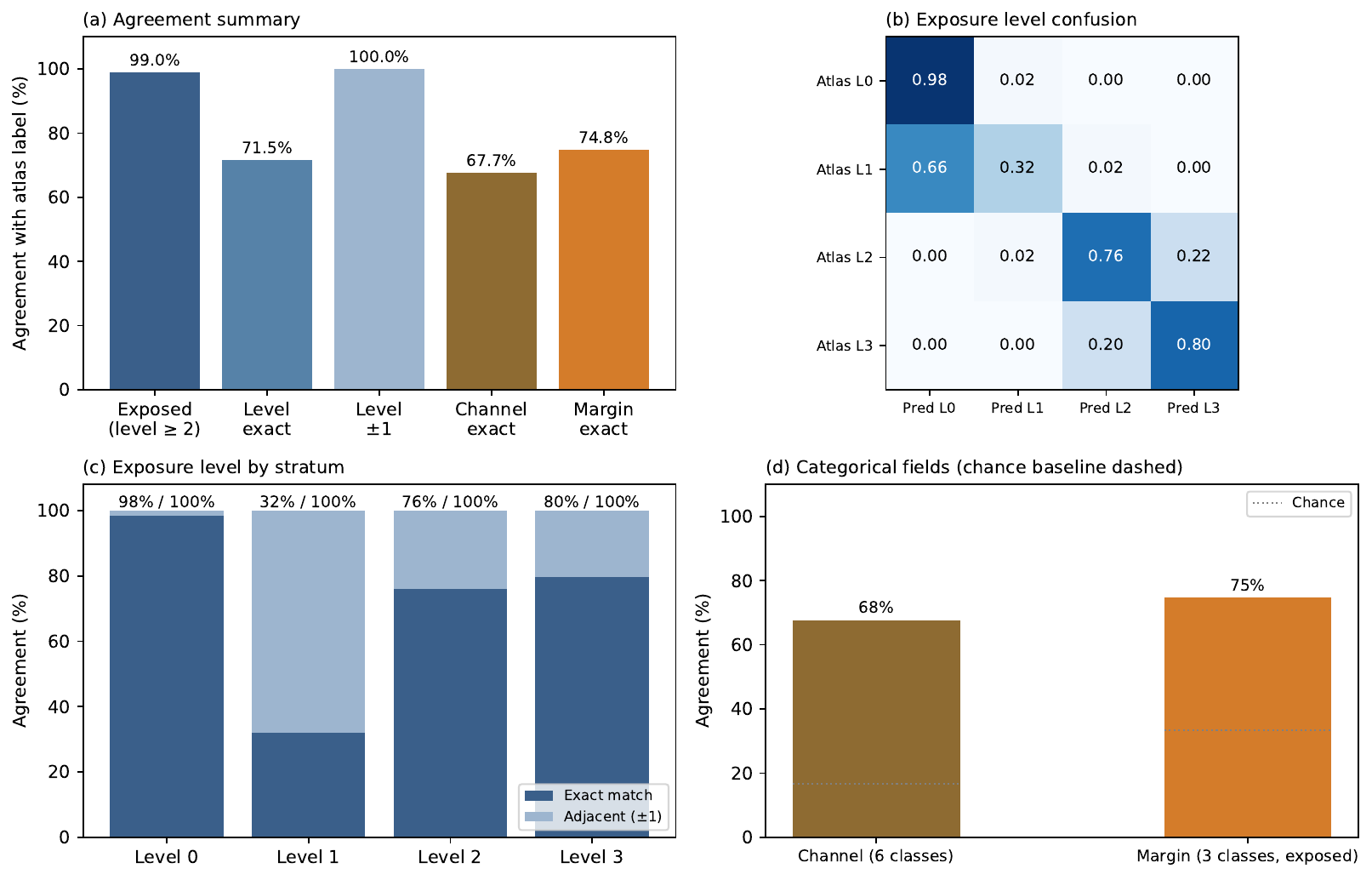}
\caption*{\scriptsize Notes: The independent model is \texttt{gpt-5.4-mini}. It receives the task description and original rationale, with the original label withheld, for $1{,}000$ country-task observations stratified equally across exposure levels. Panel~(a) reports agreement by label field; panel~(b) reports the exposure-level confusion matrix; panel~(c) reports exposure agreement by original level; panel~(d) compares channel and margin agreement with random baselines. Channel agreement uses all $1{,}000$ observations and includes \textit{none} as a sixth category. Margin agreement uses the $500$ exposed observations. The equal-category chance benchmarks are $1/6$ for channel and $1/3$ for margin.}
\label{fig:validation_rationale_predictability}
\end{figure}

\subsubsection{Prompt-paraphrase stability}
\label{sec:validation_paraphrase} We next test whether superficial changes to the prompt wording change the labels. An independent model (GPT-5.4, used only for paraphrasing) generates three reworded versions of the context-free system prompt. The paraphrases preserve all schema keys, value spaces, numeric thresholds, and consistency rules, while changing wording, sentence structure, and section order. We apply the three paraphrases to a stratified sample of $1{,}000$ tasks using the same \texttt{gemini-3.1-flash-lite} model and response schema as the original labelling run. Agreement with the original exposed-versus-not-exposed label is $88.7\%$, $87.0\%$, and $84.3\%$ across the three paraphrases; exposure level is within one step of the original label for $96\%$--$98\%$ of tasks. Across the three paraphrases, $99.8\%$ of tasks sit within one exposure level of each other. The remaining disagreements mostly occur at adjacent exposure levels. Large movements across the ordinal scale are uncommon, although exact levels and the exposed threshold are more sensitive.
\begin{figure}[!htbp]
\centering
\caption{Paraphrase stability under three rewordings of the system prompt.}
\includegraphics[width=0.98\textwidth]{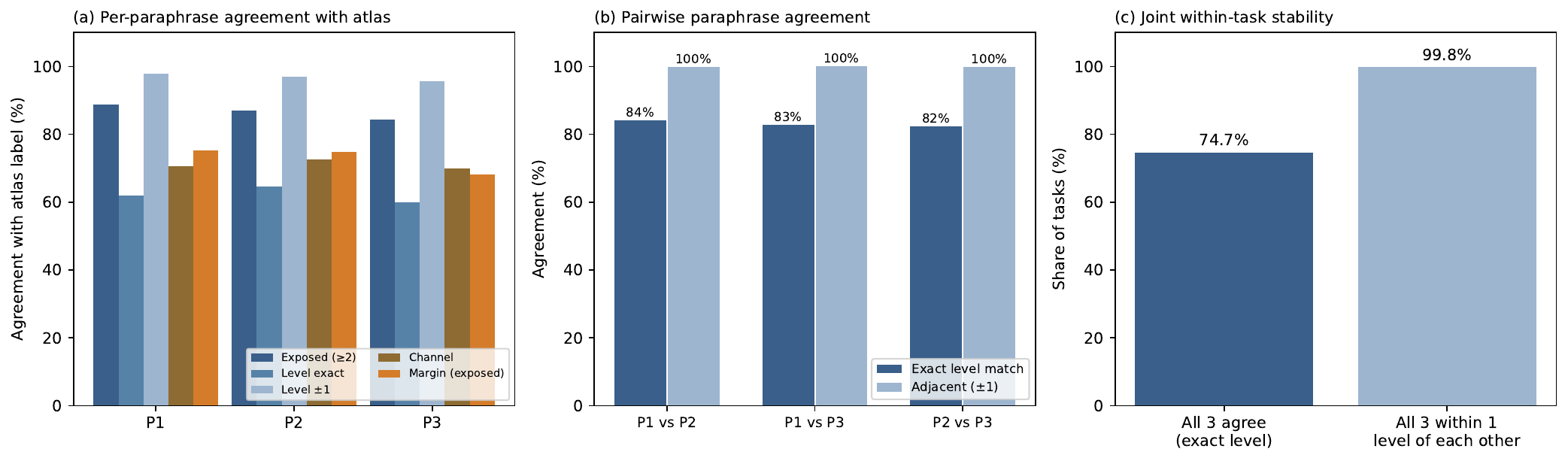}
\caption*{\scriptsize Notes: Three paraphrases of the context-free system prompt are applied to $1{,}000$ tasks stratified equally across exposure levels. The model and response schema match the original run. Panel~(a) compares each paraphrase with the original labels; panel~(b) reports pairwise exposure-level agreement across paraphrases; panel~(c) reports joint within-task stability across all three paraphrases.}
\label{fig:validation_paraphrase_stability}
\end{figure}

\subsubsection{Face validity}
\label{sec:validation_face}
\begin{figure}[!htbp]
\centering
\caption{Country-paired rationales are rarely near-verbatim.}
\includegraphics[width=0.98\textwidth]{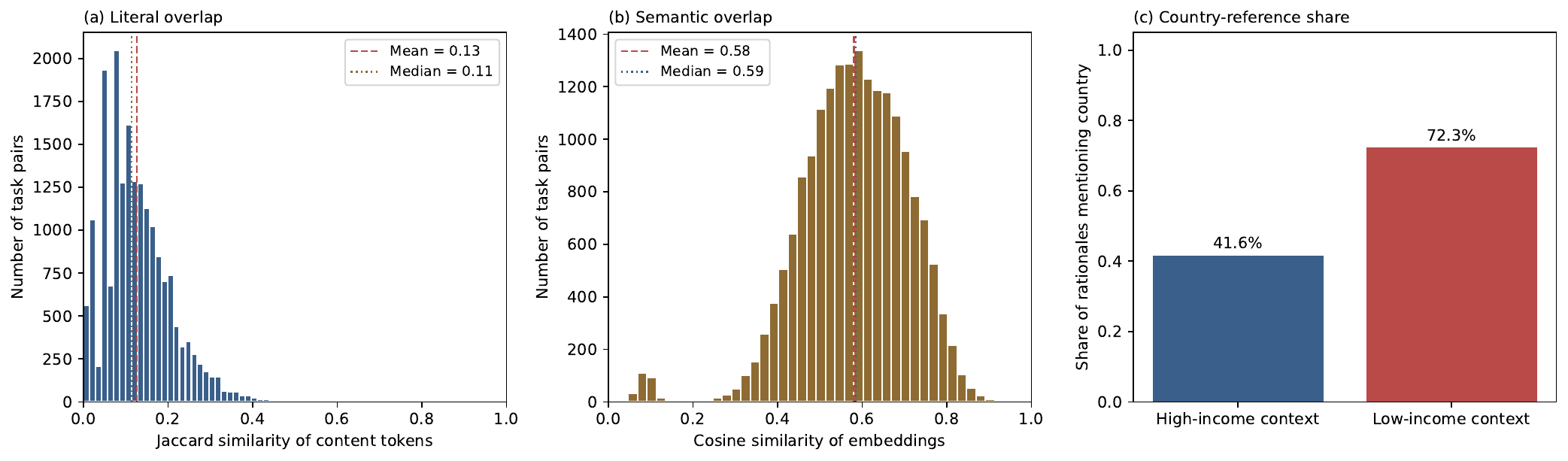}
\caption*{\scriptsize Notes: The analysis uses 18,797 same-task high-income/low-income rationale pairs. Each task contributes one randomly drawn high-income rationale and one randomly drawn low-income rationale, so measurements are paired within task. Panel~(a) reports Jaccard similarity on stopword-filtered content tokens; panel~(b) reports cosine similarity using OpenAI \texttt{text-embedding-3-large} embeddings; panel~(c) reports whether the rationale names the prompted country. Low word overlap with moderate semantic similarity indicates that rationales remain on-task while changing their country-level reasoning.}
\label{fig:validation_country_conditioning_divergence}
\end{figure}

\begin{figure}[!htbp]
\centering
\caption{Distributional check of the task-level labels.}
\includegraphics[width=0.92\textwidth]{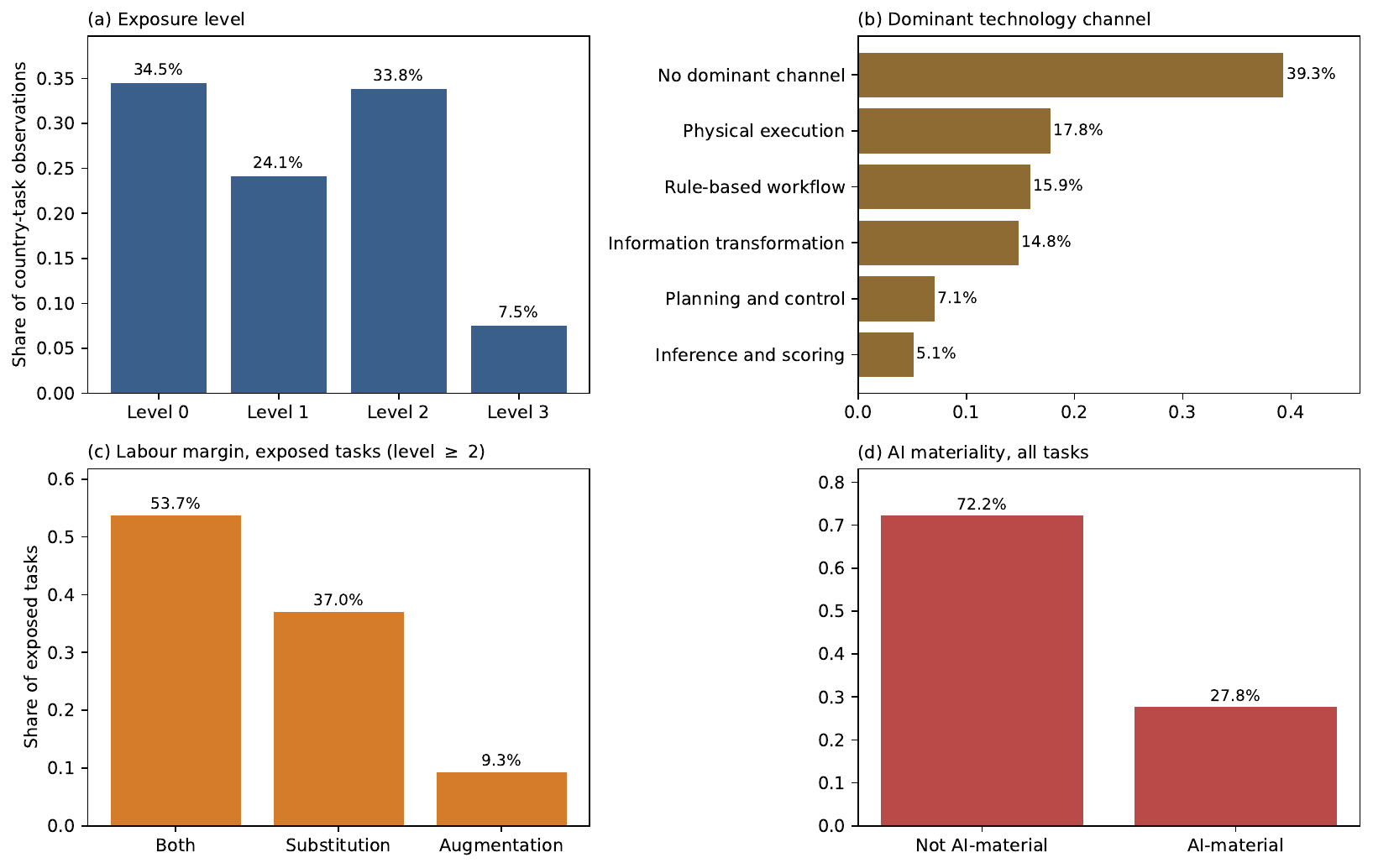}
\caption*{\scriptsize Notes: Panels report marginal distributions across the full $2{,}330{,}776$-observation task-country dataset. Panel~(a) reports exposure levels; panel~(b) reports dominant technology channels; panel~(c) reports labour margins among exposed tasks; panel~(d) reports AI materiality. Exposure means level 2 or 3.}
\label{fig:validation_distribution_overview}
\end{figure}

\begin{figure}[!htbp]
\centering
\caption{Label distributions by World Bank income group.}
\includegraphics[width=0.98\textwidth]{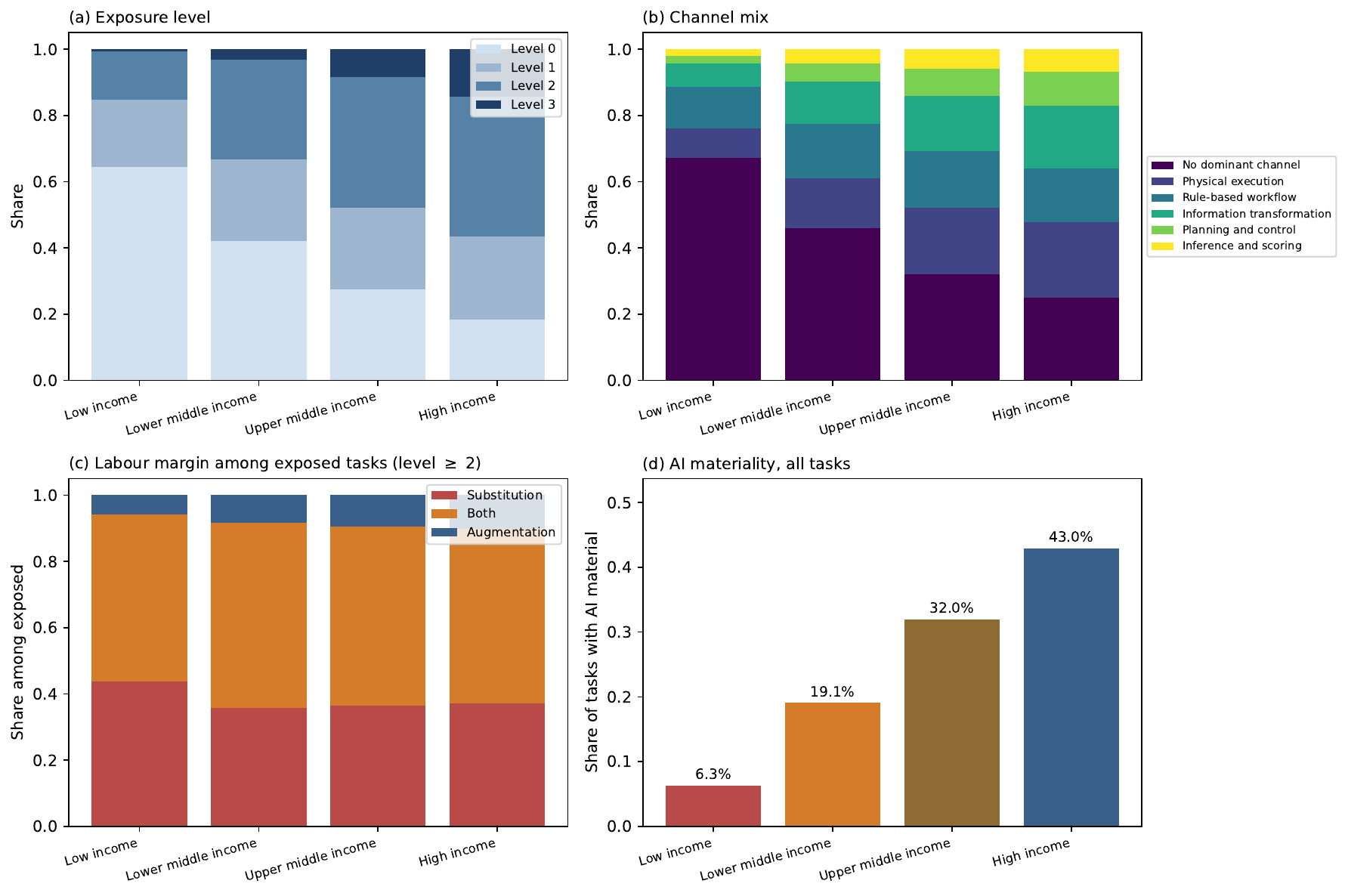}
\caption*{\scriptsize Notes: Panels report the same four label fields as Figure~\ref{fig:validation_distribution_overview}, split by World Bank income group. Panel~(a) reports exposure levels; panel~(b) reports dominant technology channels; panel~(c) reports labour margins among exposed tasks; panel~(d) reports AI materiality. Exposure means level 2 or 3.}
\label{fig:validation_distribution_by_income}
\end{figure}

Face validity asks whether the task-level labels and accompanying rationales make sense under direct inspection, and whether their distributions across the task-country dataset match the measures defined for the task-country classification. The check targets structurally implausible output that direct inspection would catch. We use four diagnostics. First, label-field distributions check for degenerate or mechanically bunched outputs (Figure~\ref{fig:validation_distribution_overview}, Figure~\ref{fig:validation_distribution_by_income}). Second, a rationale--label consistency check scans every observation in the task-country level dataset (Table~\ref{tab:validation_rationale_consistency}). Third, a rationale-divergence test asks whether country conditioning changes the substantive content of the rationale (Figure~\ref{fig:validation_country_conditioning_divergence}, Table~\ref{tab:validation_quadrant_examples}). Fourth, an anchor-occupation table compares our measure with occupations where the existing literature has strong priors (Table~\ref{tab:validation_anchor_occupations}). The four diagnostics do not identify a face-validity problem.

\begin{table}[H]
\centering
\footnotesize
\caption{Exposure for anchor occupations where the existing literature has strong priors.}
\label{tab:validation_anchor_occupations}
\begin{tabular}{@{}lcccccc@{}}
\toprule
& Prior & Mean & High & Low & Sub.\ & Aug.\ \\
Occupation & direction & exposure & income & income & share & share \\
\midrule
Childcare workers & Very low & 0.42 & 0.64 & 0.10 & 0.04 & 0.02 \\
Elementary school teachers & Very low & 0.68 & 0.93 & 0.25 & 0.04 & 0.03 \\
Surgeons & Very low & 0.72 & 1.04 & 0.24 & 0.00 & 0.11 \\
Waiters and waitresses & Medium & 0.94 & 1.40 & 0.32 & 0.19 & 0.00 \\
Farm labourers (crops) & Medium & 1.04 & 1.59 & 0.33 & 0.13 & 0.01 \\
Heavy-truck drivers & Medium & 1.36 & 1.86 & 0.62 & 0.30 & 0.01 \\
Software developers & High & 1.26 & 1.62 & 0.66 & 0.04 & 0.12 \\
Interpreters and translators & High & 1.50 & 1.70 & 1.07 & 0.13 & 0.03 \\
Cashiers & High & 1.72 & 2.12 & 0.97 & 0.51 & 0.01 \\
Accountants and auditors & High & 1.92 & 2.32 & 1.22 & 0.30 & 0.02 \\
Bookkeeping and audit clerks & High & 2.54 & 2.91 & 1.85 & 0.85 & 0.00 \\
\bottomrule
\end{tabular}
\caption*{\scriptsize Notes: Mean exposure is the cross-country unweighted occupation-level mean on the 0--3 scale. High- and low-income columns report the corresponding World Bank income-group means. Substitution and augmentation shares are task-weighted occupation means averaged across countries. The prior column summarises expectations from the existing exposure literature \citep{frey2017future,Webb2020,felten2021occupational,EloundouEtAl2024}; rows are grouped by prior direction and sorted by mean exposure within group.}
\end{table}

\paragraph{Distributional check} We first inspect the marginal distribution of each task-country label field. The exposure labels span all four levels without degeneracy, and the dominant-channel field is not concentrated in a single technology category. Among exposed tasks, the labour-margin split is mostly balanced-both or substitution-only, with a smaller augmentation-only share. AI is material in $27.8\%$ of all task-country observations and $63.9\%$ of exposed observations. It is therefore central to a majority of exposed tasks, while more than one-third of exposure operates through routes in which AI is not central.
\paragraph{Income-group check} We then repeat the distributional check by World Bank income group. The label distributions move in the expected direction: level-0 and no-channel shares fall with income, while AI-materiality rises from $6\%$ in low-income countries to $43\%$ in high-income countries. These gradients match the country and channel patterns reported in the main Results section, so the raw label distributions are coherent with the paper's main descriptive facts.

\paragraph{Rationale--label consistency} We next ask whether the rationale text ever directly contradicts the label attached to it. This is the most mechanical internal-coherence check because it is applied to the task-country dataset, covering all $2{,}320{,}863$ unique task-country rationales. We use five targeted rules that correspond to sharp schema conflicts: a level-3 rationale denying automation is possible; a level-0 rationale describing standard automation as already available; an \textit{augment} rationale describing replacement; a \textit{substitute} rationale describing assistive-only technology; and a not-AI-material rationale invoking AI, LLM, or machine-learning capabilities. A sentence-level negation filter suppresses matches in phrases such as ``not widely deployed'' or ``lacks a credible AI-based substitute.''

The screen flags $2{,}654$ observations, or $0.11\%$ of the task-country dataset (Table~\ref{tab:validation_rationale_consistency}). We read this as evidence against widespread mechanical inconsistency between labels and rationales. The number is not an adjudicated error rate: keyword rules can miss subtle contradictions and can also flag benign wording. Its value is that the most direct label-rationale conflicts are rare at full-dataset scale. The rationale-to-label predictability exercise in  Fig.~\ref{fig:validation_rationale_predictability} then tests the same issue with an independent model rather than keyword rules.

\begin{table}[H]
\centering
\footnotesize
\caption{Rationale--label consistency check over the task-country level dataset.}
\label{tab:validation_rationale_consistency}
\begin{tabular}{@{}lrrr@{}}
\toprule
 & Eligible rows & Keyword-flagged & Share \\
\midrule
Rule: level-3 rationale denies automation & 175{,}802 & 19 & 0.01\% \\
Rule: level-0 rationale describes automation & 794{,}423 & 549 & 0.07\% \\
Rule: augment rationale describes replacement & 95{,}375 & 1 & 0.00\% \\
Rule: substitute rationale describes assistive-only & 357{,}000 & 0 & 0.00\% \\
Rule: not-AI-material rationale invokes AI/LLM/ML & 1{,}673{,}970 & 2{,}099 & 0.13\% \\
\midrule
Unique observations flagged by any rule & 2{,}320{,}863 & 2{,}654 & 0.11\% \\
\bottomrule
\end{tabular}
\caption*{\scriptsize Notes: Each row scans eligible rationales for phrases that would indicate a potential contradiction with the attached label. Eligible rows are observations satisfying the relevant label condition. A sentence-level negation filter suppresses matches in negating clauses, so flagged rows should be read as potential inconsistencies rather than adjudicated errors.}
\end{table}

\paragraph{Country-paired rationale comparison} This check asks whether the classifier responds to different country prompts by repeating a common rationale and changing only the country reference. We compare rationales for the same O*NET task under one high-income and one low-income country prompt.

We use two measures because the two rationales can differ in wording while still making the same argument. First, we compute Jaccard overlap on stopword-filtered content tokens, which measures direct word reuse. Second, we embed each rationale using OpenAI \texttt{text-embedding-3-large} and compute cosine similarity, which measures semantic similarity. Low Jaccard overlap with moderate cosine similarity means the rationales discuss the same task but use different reasoning.

Near-verbatim reuse is rare. Only $0.4\%$ of pairs have both high token overlap and high semantic similarity. Most pairs have low token overlap but remain semantically related, indicating that the explanations stay connected to the same task without following fixed wording. Low-income rationales also name the prompted country more often than high-income rationales. These similarity measures do not identify which contextual conditions account for the differences. The heldout same-task concept analysis reported in the main text addresses that question by testing whether recurring production conditions differ systematically between exposed and non-exposed country labels.

\begin{table}[!htbp]
\centering
\footnotesize
\caption{Illustrative rationale pairs from the country-conditioning divergence check.}
\label{tab:validation_quadrant_examples}
\begin{tabularx}{\linewidth}{@{}L{3.0cm}>{\raggedright\arraybackslash}X@{}}
\toprule
Similarity pattern & Example task pair \\
\midrule
\textbf{Low Jaccard} \newline
\textbf{Low cosine} \newline
\textit{Substantively different} \newline
$24.7\%$ of pairs &
\textbf{Task:} Buy or sell stocks, bonds, commodity futures, or other securities on behalf of investment dealers. \newline
\textit{Australia (high income):} Trading is highly automatable through algorithmic execution and AI-based predictive systems, which have largely displaced manual human order entry and routine portfolio management in the Australian financial sector. \newline
\textit{Burundi (low income):} Burundi's current financial infrastructure lacks the digital market integration and high-volume electronic trading systems required for credible task-level automation of securities trading. \newline
Jaccard $=0.09$, cosine $=0.36$. \\
\midrule
\textbf{Low Jaccard} \newline
\textbf{High cosine} \newline
\textit{Same task, different reasoning} \newline
$74.9\%$ of pairs &
\textbf{Task:} Calculate costs for billings or commissions. \newline
\textit{Saudi Arabia (high income):} Calculating costs, billings, and commissions is a deterministic, rule-based process involving structured financial data that is routinely and reliably automated by standard enterprise resource planning software. \newline
\textit{Burundi (low income):} Calculating costs is inherently deterministic and standard business software in Burundi is already capable of automating these calculations, reducing the reliance on manual human arithmetic. \newline
Jaccard $=0.12$, cosine $=0.60$. \\
\midrule
\textbf{High Jaccard} \newline
\textbf{Low cosine} \newline
\textit{Rare edge} \newline
$0.0\%$ of pairs &
No observations fall in this region in the sampled pairs. \\
\midrule
\textbf{High Jaccard} \newline
\textbf{High cosine} \newline
\textit{Near-identical rationales} \newline
$0.4\%$ of pairs &
\textbf{Task:} Perform therapeutic wound care. \newline
\textit{Greece (high income):} Therapeutic wound care requires patient-specific manual dexterity, clinical judgment, and direct physical interaction that cannot be reliably automated or meaningfully substituted by current technology. \newline
\textit{Somalia (low income):} Therapeutic wound care requires high-level physical dexterity, clinical judgment, and patient interaction that cannot be reliably or safely automated in typical Somali clinical settings by 2026. \newline
Jaccard $=0.52$, cosine $=0.79$. \\
\bottomrule
\end{tabularx}
\caption*{\scriptsize Notes: The table shows illustrative high-income/low-income rationale pairs selected to span low and high token overlap and low and high semantic similarity. Rationales are classifier justifications under country-conditioned prompts. Token overlap is measured with Jaccard similarity on content tokens; semantic similarity is measured with cosine similarity under OpenAI \texttt{text-embedding-3-large} embeddings.}
\end{table}

\subsection{Rationale-concept hypothesis generation}
\label{sec:appendix_rationale_concept_analysis}

\paragraph{Analysis goal}

Each task-country label includes a short rationale explaining the classifier's assignment. We use these rationales as a text corpus for hypothesis generation about the production conditions that recur in country-conditioned exposure and labour-margin labels. The paper-facing analysis is restricted to three contrasts: exposed versus non-exposed labels for the same task across countries, substitution-only versus other exposed labels, and augmentation-only versus other exposed labels. The exercise is descriptive. It identifies recurring explanation concepts associated with these labels; it does not estimate the causal effect of any production condition on realised automation.

\paragraph{Target contrasts}

For each target contrast, we fixed 20 child concepts before heldout evaluation. Section~\ref{sec:appendix_rationale_concept_results} reports the heldout results.

\paragraph{Text inputs and masking}

The unit of text is the rationale attached to a task-country label. For the exposure contrast, we compare rationales for the same standardized task when it is labelled exposed in one country and non-exposed in another. For each labour-margin contrast, we form same-task pairs among exposed task-country labels, comparing substitution-only or augmentation-only assignments with the remaining exposed labels. Before concept discovery, country names and direct place identifiers are masked so that concepts reflect production-condition language rather than explicit country mentions.

\paragraph{Concept discovery and interpretation}

Within each target contrast, we embed rationale text and train a sparse-autoencoder concept model on a discovery split. Candidate features are interpreted using high-activation text examples, contrastive examples, and model-assisted summaries. We then retain a fixed set of interpretable child concepts for each contrast before evaluating them in heldout data. This procedure follows recent work using language-model representations for hypothesis generation from high-dimensional text \citep{LudwigMullainathan2024HypothesisGeneration,MovvaEtAl2025HypotheSAEs}.

\paragraph{Heldout evaluation and family display}

\begin{figure}[!htbp]
\centering
\caption{Child concepts for the exposure contrast.}
\includegraphics[width=0.95\textwidth]{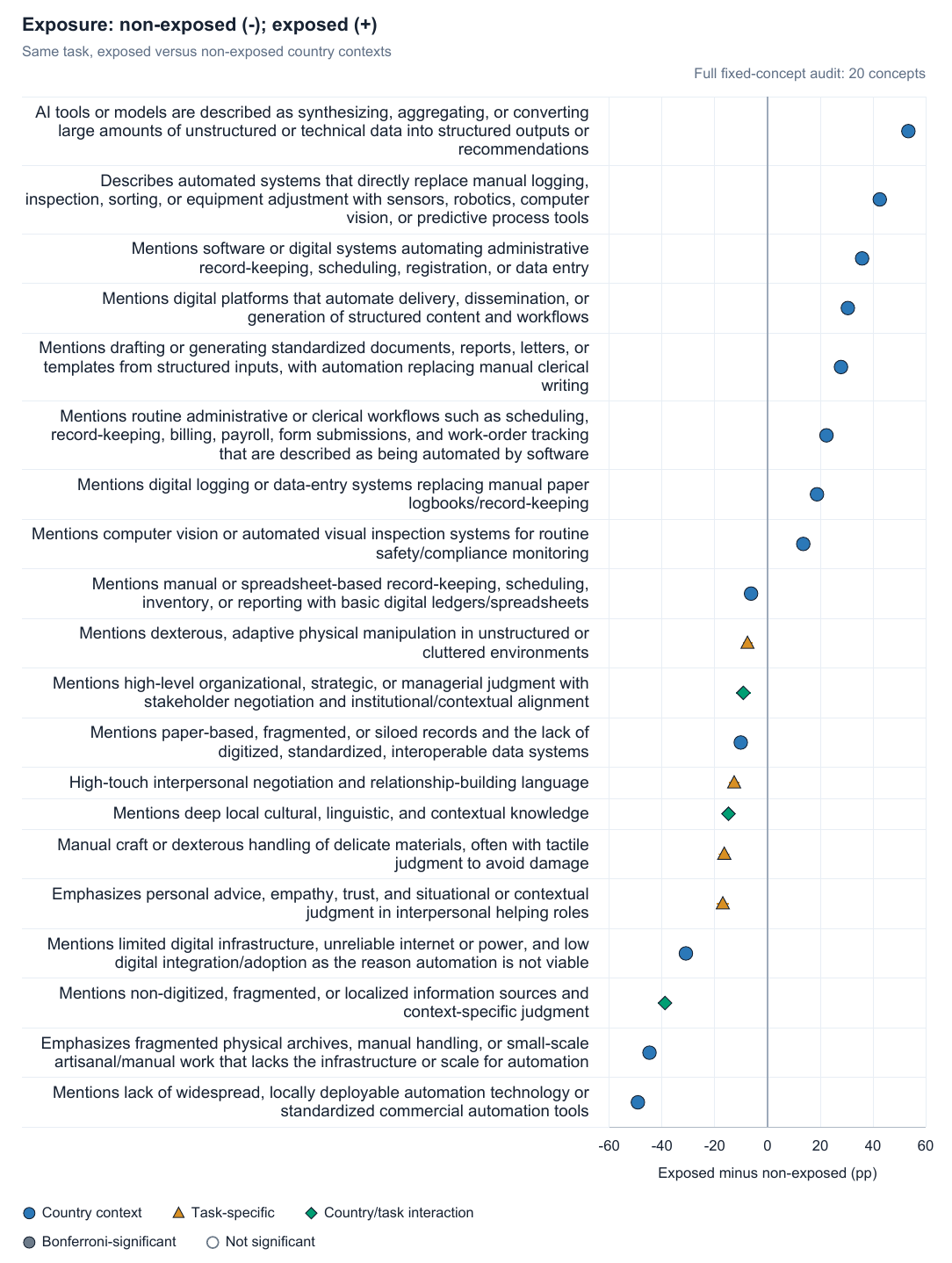}
\caption*{\footnotesize Notes: The figure reports exposed-minus-non-exposed differences in concept presence in a heldout sample of 2,000 same-task country pairs. Points are percentage-point differences and horizontal lines are 95\% confidence intervals based on paired standard errors. Significance is Bonferroni-adjusted across the 20 fixed concepts.}
\label{edfig:appendix_hypothesaes_exposure_20_concepts}
\end{figure}

\begin{figure}[!htbp]
\centering
\caption{Child concepts for the substitution-only contrast.}
\includegraphics[width=0.95\textwidth]{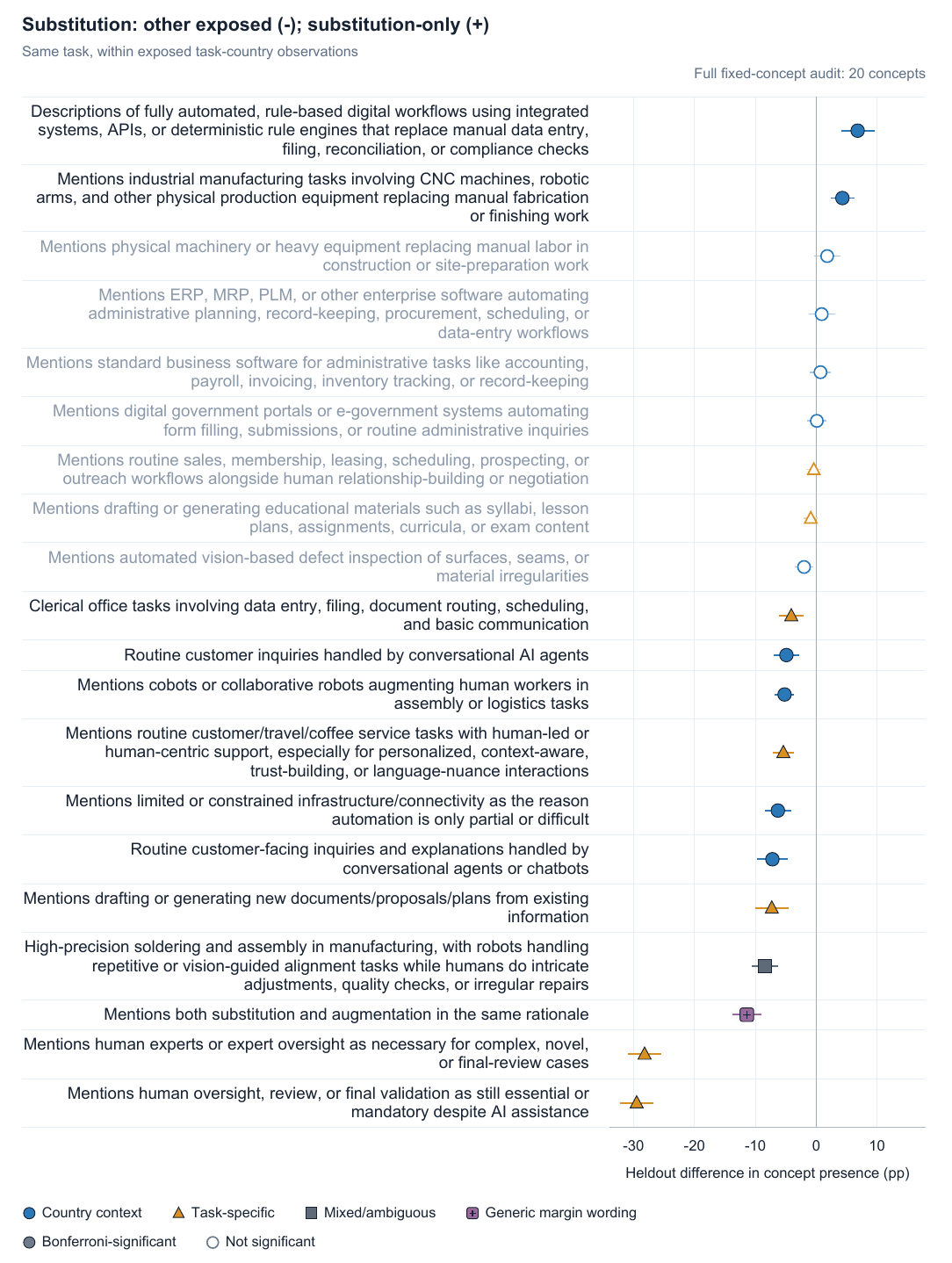}
\caption*{\footnotesize Notes: The figure reports substitution-only-minus-other-exposed differences in concept presence in a heldout sample of 2,000 same-task country pairs. Points are percentage-point differences and horizontal lines are 95\% confidence intervals based on paired standard errors. Significance is Bonferroni-adjusted across the 20 fixed concepts.}
\label{edfig:appendix_hypothesaes_substitution_20_concepts}
\end{figure}

\begin{figure}[!htbp]
\centering
\caption{Child concepts for the augmentation-only contrast.}
\includegraphics[width=0.95\textwidth]{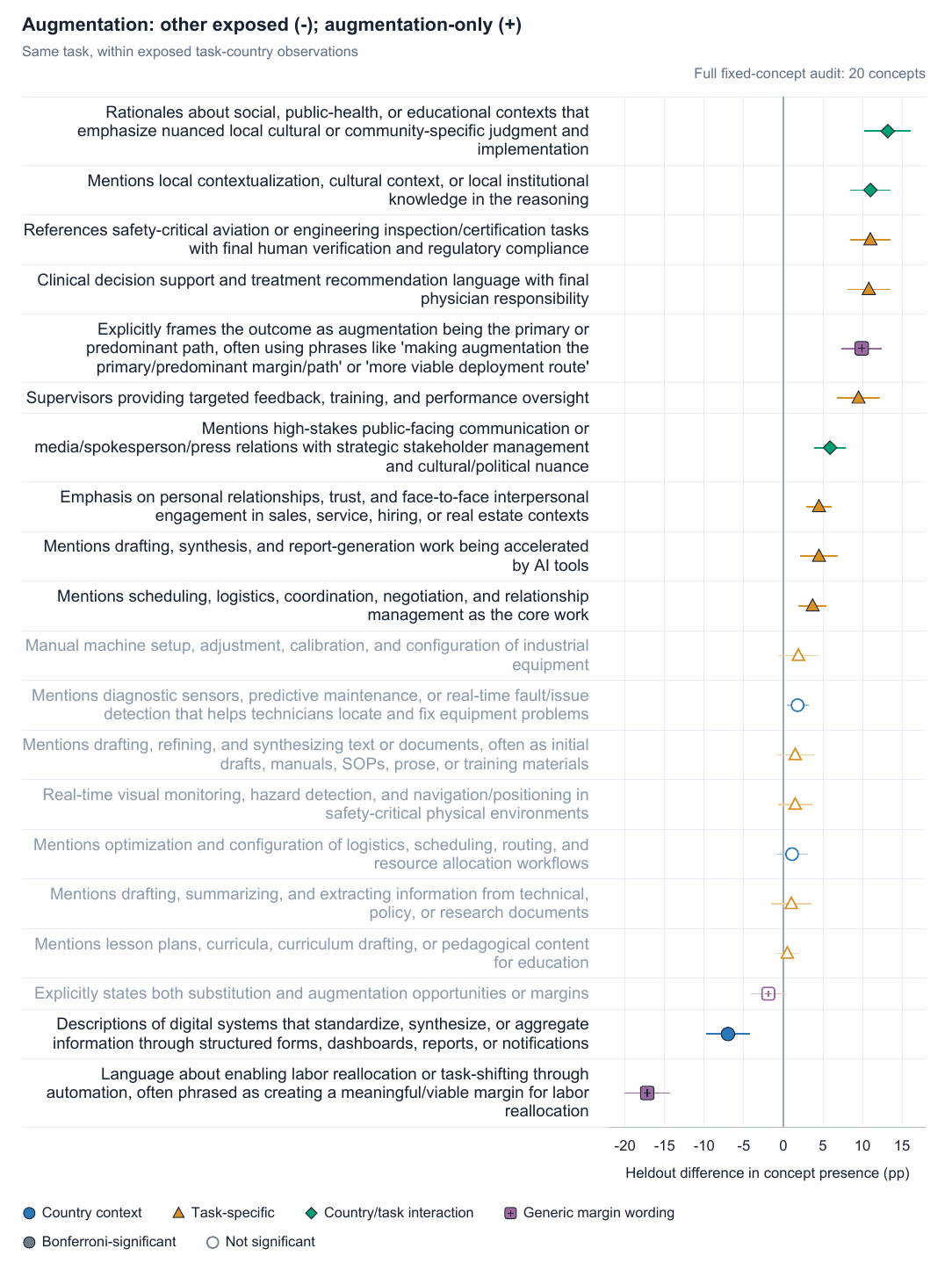}
\caption*{\footnotesize Notes: The figure reports augmentation-only-minus-other-exposed differences in concept presence in a heldout sample of 2,000 same-task country pairs. Points are percentage-point differences and horizontal lines are 95\% confidence intervals based on paired standard errors. Significance is Bonferroni-adjusted across the 20 fixed concepts.}
\label{edfig:appendix_hypothesaes_augmentation_20_concepts}
\end{figure}

No rationale used to discover or name a concept is reused for its heldout estimate. In the heldout samples, each fixed child concept is coded as present or absent in the relevant rationale text. We report exposed-minus-non-exposed, substitution-only-minus-other-exposed, and augmentation-only-minus-other-exposed differences, with paired standard errors across same-task pairs and Bonferroni adjustment within each contrast. For the compact main-text display, child concepts are grouped into parent families. Figures~\ref{edfig:appendix_hypothesaes_exposure_20_concepts}--\ref{edfig:appendix_hypothesaes_augmentation_20_concepts} report the corresponding child-concept estimates.

\FloatBarrier

\section{Supporting results and robustness checks}
\label{sec:appendix_results}

\noindent This section collects the figures and tables behind the Results, beginning with the effects of country conditioning on occupation rankings. It then reports country exposure and within-income heterogeneity, margin and pathway companions, channel and AI-function companions, occupation and industry summaries, ILOSTAT employment-composition checks, country-predictor robustness and rationale-concept findings.

\subsection{How country conditioning changes occupation rankings}
\label{sec:validation_country_contribution}
The external-validation checks in \ref{sec:validation_construct} show that country-conditioned measures align with related external evidence and add predictive information beyond fixed scores. This subsection asks a separate descriptive question: how much country conditioning changes occupation rankings relative to context-free Atlas scores, US-centric occupation scores, employment-weighted external scores and country-level scaling. The distinction matters because exposure measures are often used to identify which occupations or sectors require policy attention. Employment weighting explains part of the country ordering: an employment-weighted Felten AIOE score, built by weighting US-centric Felten occupation scores with country-specific ILOSTAT occupation shares, correlates with our country-conditioned AI-material share at Pearson $0.74$ (Spearman $0.74$) across $91$ countries.

\begin{figure}[!htbp]
\centering
\caption{Country conditioning relative to external occupation benchmarks.}
\includegraphics[width=\textwidth]{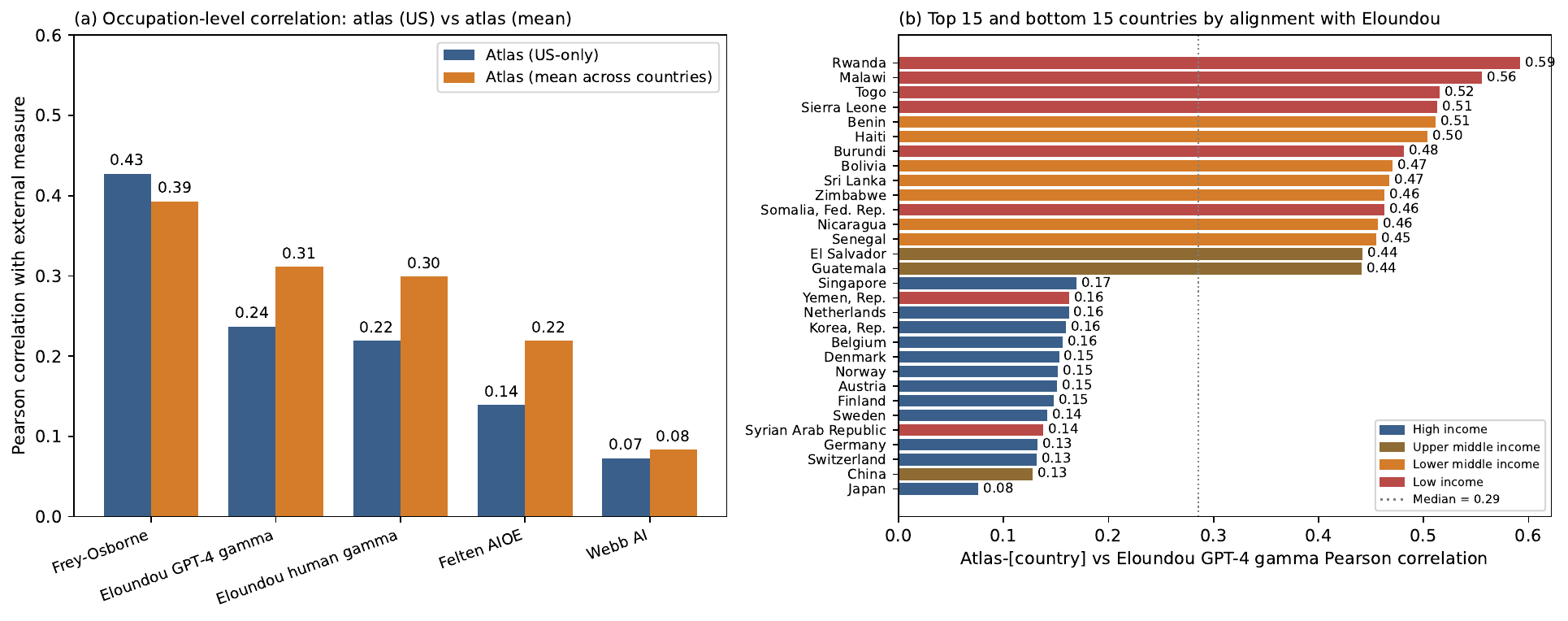}
\caption*{\scriptsize Notes: Panel~(a) compares occupation-level exposure correlations under US-conditioned labels and the unweighted cross-country mean. External measures are Frey--Osborne, Eloundou GPT-4 gamma and human gamma, Felten AIOE, and Webb AI exposure. Panel~(b) reports each country's Pearson correlation with Eloundou GPT-4 gamma for the top and bottom $15$ countries by alignment.}
\label{fig:validation_us_vs_external}
\end{figure}

\paragraph{US-conditioned and cross-country mean benchmarks} Figure~\ref{fig:validation_us_vs_external} compares our US-conditioned labels and our cross-country mean labels against external occupation scores. The US-conditioned version is closest to Frey--Osborne: the correlation is $0.427$ using US labels and $0.393$ using the cross-country mean. For Felten AIOE and both Eloundou occupation-level gamma variants, the cross-country mean is closer: Eloundou GPT-4 gamma rises from $0.237$ under US conditioning to $0.312$ under the cross-country mean, and Felten AIOE rises from $0.139$ to $0.219$. The cross-country mean summarises the $124$ country-conditioned estimates, while the individual country estimates retain institution-, industry-, and regulation-specific variation. Panel~(b) shows the size of that variation: the correlation between each country's occupation-level exposure and Eloundou GPT-4 gamma ranges from $0.08$ (Japan) to $0.59$ (Rwanda), with a cross-country median of $0.29$ and a USA value of $0.24$. Low-income countries cluster at the top of the Eloundou-agreement ranking, while several large high-income economies cluster at the bottom.

\begin{figure}[!htbp]
\centering
\caption{Country conditioning beyond additive exposure scaling.}
\includegraphics[width=\textwidth]{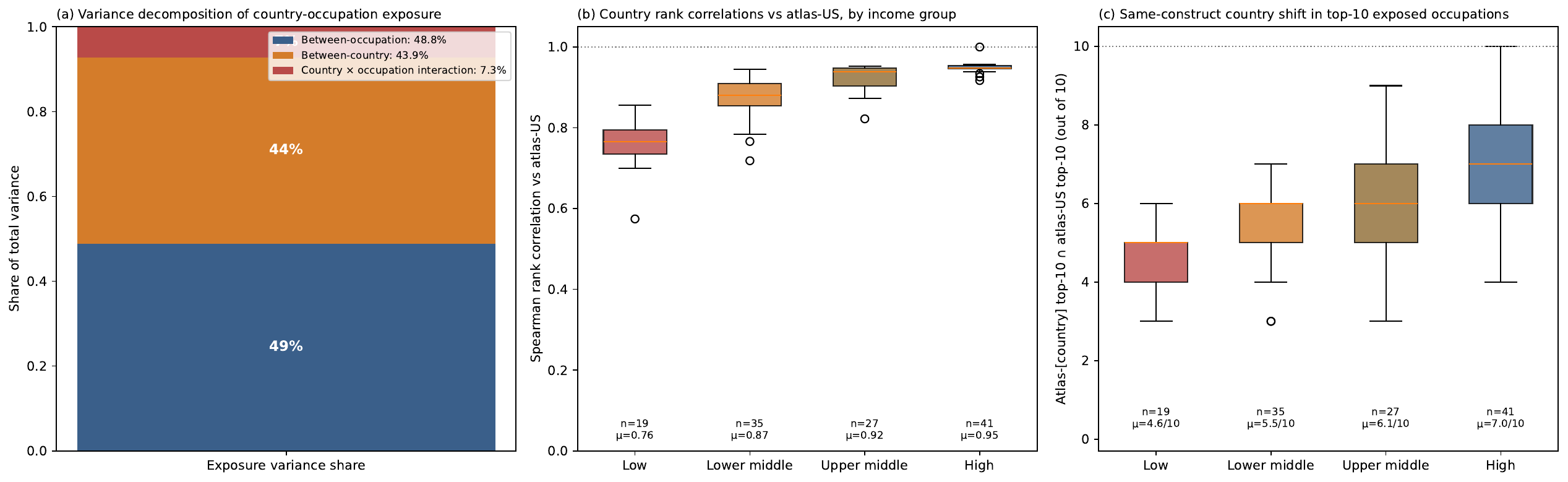}
\caption*{\scriptsize Notes: Panel~(a) decomposes variance in the $124$ country $\times$ $923$ occupation exposure matrix into occupation, country, and country--occupation components. Panel~(b) reports each country's Spearman rank correlation with the US-conditioned occupation ranking. Panel~(c) reports overlap between each country's top-$10$ exposed occupations and the US-conditioned top-$10$ list.}
\label{fig:validation_country_contribution}
\end{figure}

\paragraph{Country-conditioned occupation rankings} Figure~\ref{fig:validation_country_contribution} quantifies the country-conditioned component directly. Panel~(a) decomposes the $114{,}452$ country--occupation observations into between-occupation, between-country, and country--occupation interaction variance. Occupations explain $48.8\%$ of total variance and countries explain $43.9\%$. The remaining $7.3\%$ is the non-additive country--occupation component, which cannot be reproduced by combining fixed occupation effects with an additive country shift.  Panels~(b) and~(c) show where this component matters. The Spearman correlation between each country's occupation ranking and the US-conditioned ranking falls from $0.95$ in high-income countries to $0.76$ in low-income countries. Top-$10$ overlap falls in the same direction: high-income countries share $7.0$ of the $10$ occupations on average with the US-conditioned top-$10$ list, while low-income countries share only $4.6$. Country conditioning therefore changes which occupations rise to the top of the exposure ranking, especially far from the US reference setting.

Employment weights explain part of the cross-country variation, but they do not reproduce the country-conditioned occupation rankings. Because a country-wide scalar adjustment cannot reorder occupations, the rank and top-$10$ results show that country conditioning changes which occupations appear most exposed, especially in lower-income settings. This finding aligns with documented cross-country heterogeneity in technology adoption and skill demand \citep{CominHobijn2010,alabdulkareem2018_skill_polarization,LewandowskiMadonPark2025DevelopmentStages,AutorSalomons2018}, and with evidence that the same occupation label can carry different task and skill content across economies and over time \citep{DasHilgenstock2022,Deming2017social}.

Holding the occupation and industry mappings fixed, replacing context-free labels with country-conditioned labels produces mean absolute exposure differences of $0.553$ points across $5{,}208$ country--ISCO-2 cells and $0.497$ points across $2{,}976$ country--ISIC-2 cells on the $0$--$3$ scale. Mean within-country rank correlations are $0.917$ for occupations and $0.743$ for industries. Both comparisons cover $124$ countries, using $42$ occupation groups and the $24$ industry divisions with both sets of labels.

\Needspace{0.62\textheight}
\subsection{Country exposure and within-income heterogeneity}
\label{sec:appendix_country_heterogeneity}
\leavevmode\par\nobreak
\begin{figure}[H]
\centering
\caption{Within-income-group variability across country-level summary metrics.}
\includegraphics[width=0.98\textwidth]{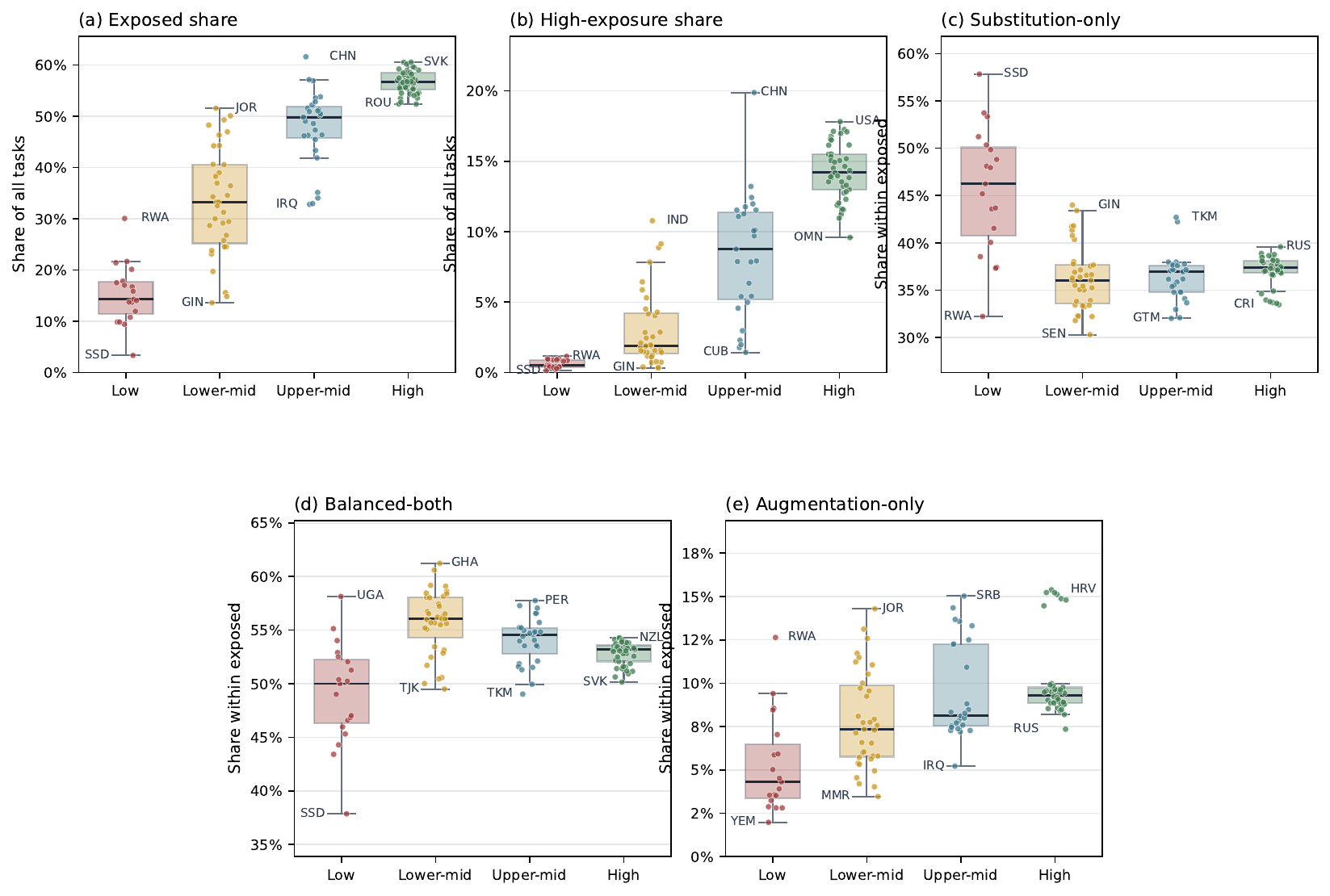}
\caption*{\scriptsize Notes: Each panel reports one country-level metric: exposed share (level $\geq 2$, share of all tasks); high-exposure share (level $= 3$, share of all tasks); substitution-only share, balanced-both share, and augmentation-only share (each within exposed tasks). Box and whiskers report the within-income-group distribution across the 122-country sample: boxes show medians and interquartile ranges, and whiskers extend to 1.5 times the interquartile range. Coloured points are countries; the highest and lowest country within each income group is labelled by ISO3 code. At fixed income tier, countries differ on all five metrics, with the widest within-tier spread in the lower-middle and upper-middle tiers.}
\label{fig:appendix_within_income_variability}
\end{figure}

\clearpage
\begin{table}[H]
\centering
\caption{Country exposure by region and income tier.}
\label{tab:country_opening_summary}
\footnotesize
\setlength{\tabcolsep}{3pt}
\begin{tabular}{@{}L{0.23\textwidth}C{0.08\textwidth}C{0.11\textwidth}C{0.11\textwidth}C{0.10\textwidth}C{0.09\textwidth}C{0.10\textwidth}@{}}
\toprule
& & & & \multicolumn{3}{c}{Within exposed} \\
\cmidrule(lr){5-7}
Income group & N & Exposed & High & Sub. & Both & Aug. \\
\midrule
\multicolumn{7}{@{}l}{\textbf{East Asia \& Pacific}} \\
High income & 6 & 0.58 & 0.16 & 0.37 & 0.53 & 0.09 \\
Upper middle income & 4 & 0.53 & 0.13 & 0.38 & 0.54 & 0.08 \\
Lower middle income & 6 & 0.32 & 0.04 & 0.38 & 0.54 & 0.08 \\
\addlinespace[0.35em]
\multicolumn{7}{@{}l}{\textbf{Europe \& Central Asia}} \\
High income & 24 & 0.56 & 0.14 & 0.37 & 0.53 & 0.10 \\
Upper middle income & 7 & 0.51 & 0.09 & 0.37 & 0.52 & 0.11 \\
Lower middle income & 3 & 0.37 & 0.03 & 0.38 & 0.52 & 0.10 \\
\addlinespace[0.35em]
\multicolumn{7}{@{}l}{\textbf{Latin America \& Caribbean}} \\
High income & 3 & 0.57 & 0.12 & 0.34 & 0.51 & 0.14 \\
Upper middle income & 11 & 0.47 & 0.07 & 0.35 & 0.54 & 0.11 \\
Lower middle income & 4 & 0.32 & 0.02 & 0.34 & 0.55 & 0.11 \\
\addlinespace[0.35em]
\multicolumn{7}{@{}l}{\textbf{Middle East \& North Africa}} \\
High income & 6 & 0.56 & 0.13 & 0.36 & 0.52 & 0.11 \\
Upper middle income & 4 & 0.39 & 0.04 & 0.37 & 0.56 & 0.07 \\
Lower middle income & 4 & 0.48 & 0.06 & 0.33 & 0.55 & 0.11 \\
Low income & 2 & 0.16 & 0.01 & 0.52 & 0.45 & 0.02 \\
\addlinespace[0.35em]
\multicolumn{7}{@{}l}{\textbf{North America}} \\
High income & 2 & 0.57 & 0.17 & 0.38 & 0.53 & 0.09 \\
\addlinespace[0.35em]
\multicolumn{7}{@{}l}{\textbf{South Asia}} \\
Lower middle income & 5 & 0.38 & 0.05 & 0.35 & 0.57 & 0.08 \\
Low income & 1 & 0.14 & 0.01 & 0.50 & 0.47 & 0.04 \\
\addlinespace[0.35em]
\multicolumn{7}{@{}l}{\textbf{Sub-Saharan Africa}} \\
Upper middle income & 1 & 0.51 & 0.12 & 0.38 & 0.55 & 0.07 \\
Lower middle income & 13 & 0.27 & 0.02 & 0.37 & 0.57 & 0.07 \\
Low income & 16 & 0.15 & 0.01 & 0.44 & 0.49 & 0.06 \\
\bottomrule
\end{tabular}
\caption*{\footnotesize Notes: Exposed and high-exposure shares use all tasks as the denominator; substitution-only, balanced-both, and augmentation-only shares are computed within exposed tasks. Values are first computed by country and then averaged within region $\times$ income cells. Two countries without World Bank income classifications are omitted. Regional labels follow the paper's seven-region convention.}
\end{table}

Table~\ref{tab:country_opening_summary} reports the country-level exposure and margin summaries by region and income group. Read with Figure~\ref{fig:appendix_within_income_variability}, it shows that exposure rises strongly with income, while meaningful within-tier dispersion remains.

\FloatBarrier
\Needspace{0.37\textheight}
\subsection{Margin and pathway companions}
\label{sec:appendix_pathway_findings}
\leavevmode\par\nobreak

\begin{figure}[H]
\centering
\caption{Adjacent income-group transition matrices for all-task pathway states.}
\includegraphics[width=0.94\textwidth]{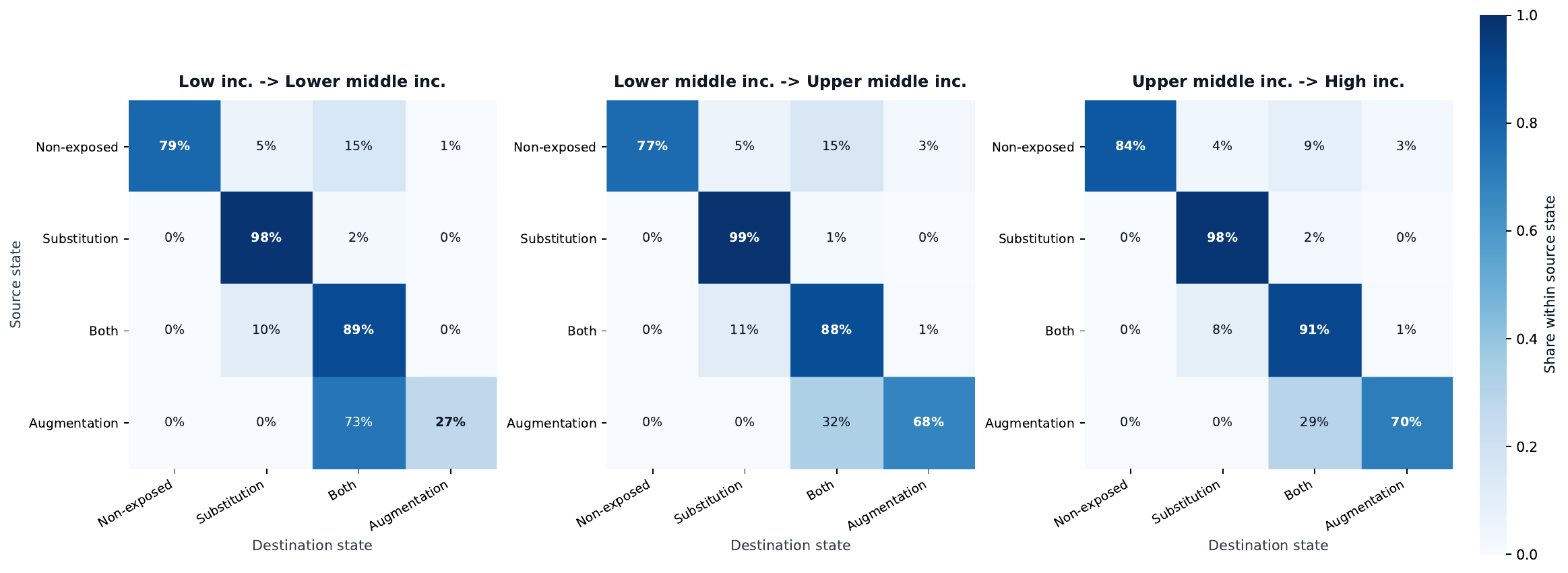}
\caption*{\scriptsize Notes: Each panel reports one adjacent income-group step, and rows sum to $100\%$ within the source state. The sample contains $18{,}797$ tasks, each assigned its modal pathway within each income group. Off-diagonal cells are tasks whose modal pathway state changes between adjacent income groups.}
\label{fig:appendix_income_group_pathway_heatmaps}
\end{figure}

\FloatBarrier
\subsubsection{Polarisation index as an alternative within-exposed summary}
\label{sec:appendix_polarisation_p}
Within exposed work, each country has three mutually exclusive margin shares: substitution-only, balanced-both, and augmentation-only. We summarise the mass outside the balanced-both category with a polarisation index, $P_c = \mathrm{sub}_c + \mathrm{aug}_c = 1 - \mathrm{bal}_c$. This adapts the idea of mass at the extremes from the labour-economics polarisation literature \citep{goos2007_lousy,AutorDorn2013}. Here, the extremes are substitution and augmentation. The complementary tilt index $T_c = \mathrm{sub}_c / (\mathrm{sub}_c + \mathrm{aug}_c)$ records which extreme dominates within the polarised mass. Because substitution dominates that mass in most countries, $P$ is the more useful one-number summary.

\begin{figure}[!htbp]
\centering
\caption{Polarisation $P$ by development level and income tier.}
\begin{subfigure}[t]{0.76\textwidth}
    \centering
    \caption{Development gradient.}
    \includegraphics[width=\textwidth]{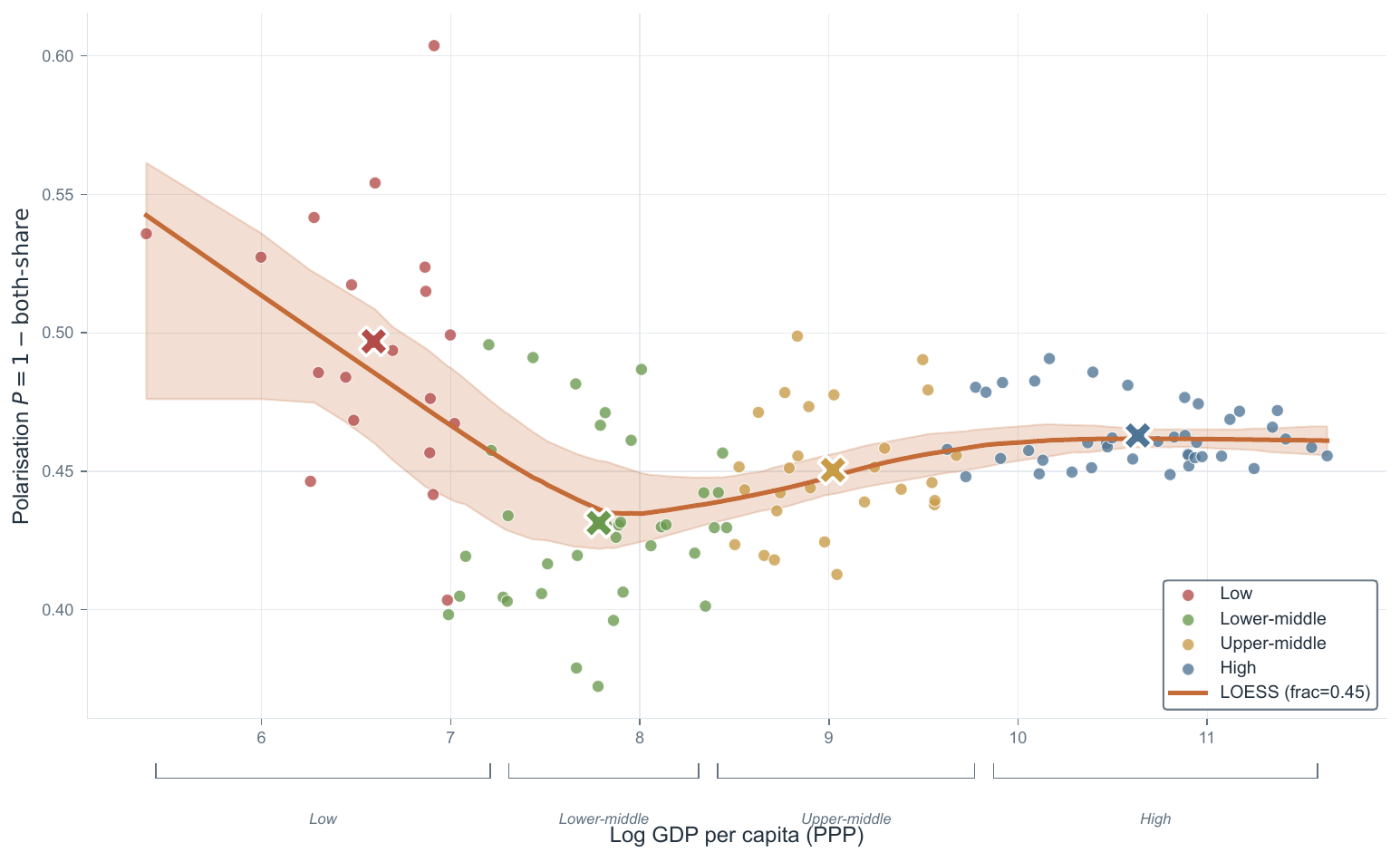}
\end{subfigure}

\vspace{0.35em}

\begin{subfigure}[t]{0.76\textwidth}
    \centering
    \caption{Within-tier dispersion.}
    \includegraphics[width=\textwidth]{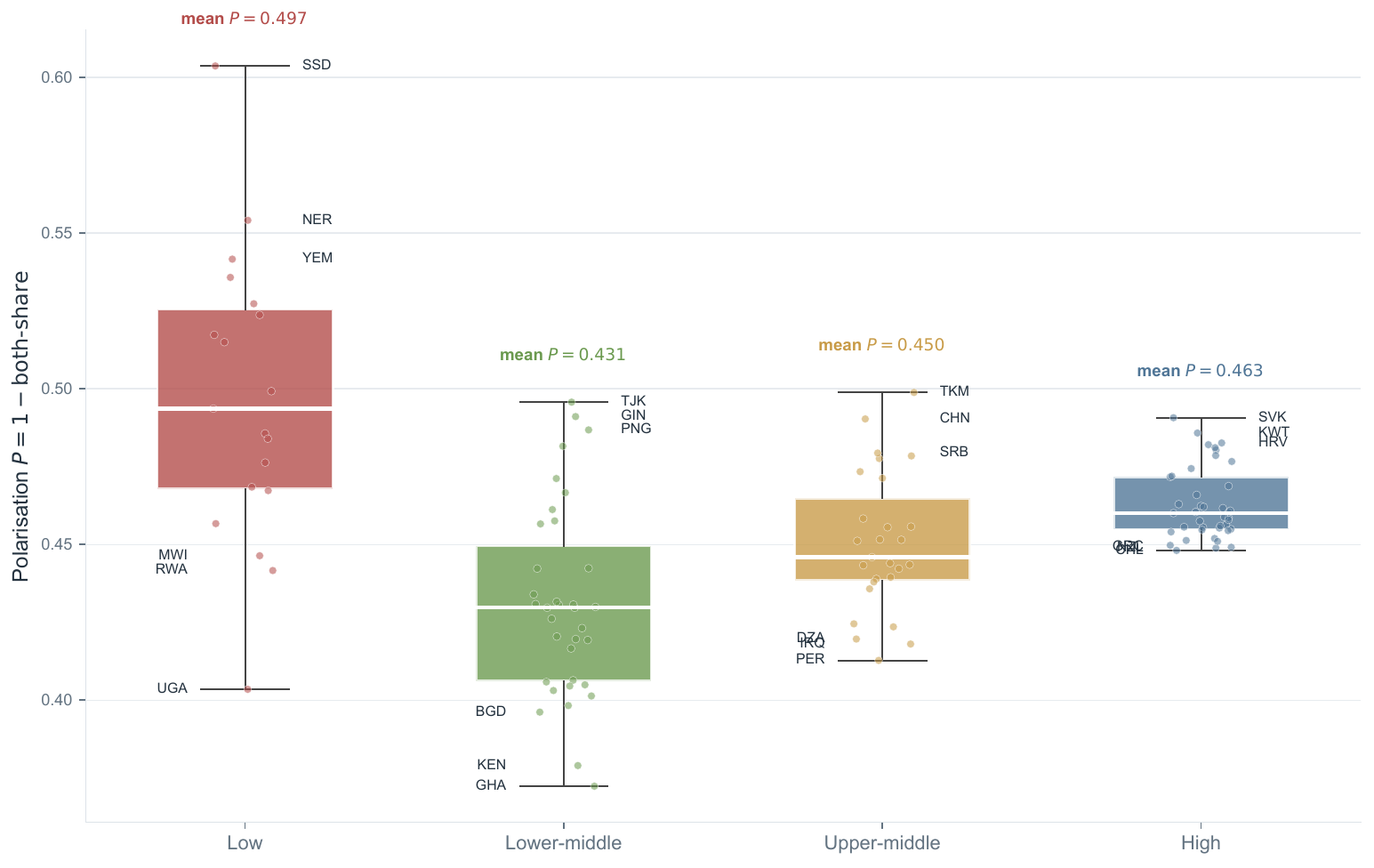}
\end{subfigure}
\caption*{\scriptsize Notes: Polarisation is $P_c = \mathrm{sub}_c + \mathrm{aug}_c = 1 - \mathrm{bal}_c$, where $\mathrm{bal}_c$ is the balanced-both share within exposed tasks. Panel~(a) plots $P$ against log GDP per capita with a LOESS smooth and $95\%$ bootstrap interval from $200$ country-level resamples. Panel~(b) reports boxplots by World Bank income group, with country dots and labelled within-tier extremes.}
\label{edfig:appendix_polarisation_p}
\end{figure}

Figure~\ref{edfig:appendix_polarisation_p} reports both the development gradient in polarisation $P$ and its within-tier dispersion. The relationship is non-monotone: $P$ is highest at the low-income end of the gradient ($\approx 0.50$), falls through lower-middle ($\approx 0.43$, the trough), and then rises gradually back to $\approx 0.46$ across upper-middle and high-income countries. Low-income countries span roughly $0.40$--$0.60$ on $P$ (the widest spread); high-income countries span only $0.45$--$0.49$ (the narrowest). Read together, the panels show that within-tier dispersion narrows as income rises, so margin polarisation is more stable among richer economies, where the balanced-both category absorbs a steadier share of exposed work.

\begin{figure}[!htbp]
\centering
\caption{Mutually exclusive pathway decomposition across country, occupation, and industry summaries.}
\includegraphics[width=0.96\textwidth]{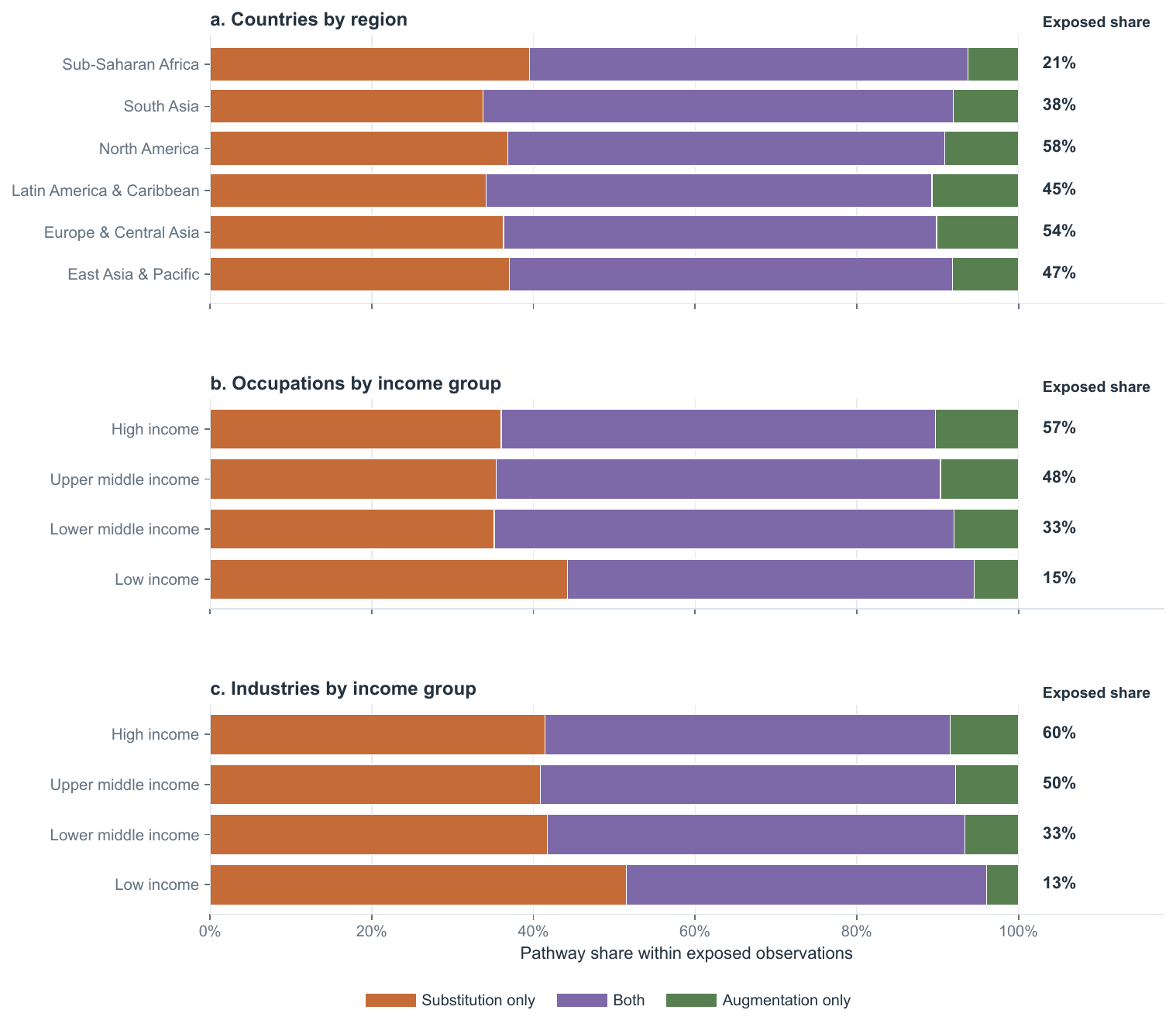}
\caption*{\scriptsize Notes: Panel a reports regional country summaries, panel b reports the weighted occupation summaries by income group, and panel c reports the bottom-up industry summaries by income group. Bars partition exposed observations into substitution-only, balanced-both, and augmentation-only shares under the same mutually exclusive pathway definition used in the main text. Balanced-both exposure remains a large part of exposed work at all three reporting levels.}
\label{fig:joint_pathway_decomposition}
\end{figure}

\FloatBarrier
\subsection{Channel and AI-function companions}
\label{sec:appendix_benchmark_heterogeneity_support}

\begin{figure}[!t]
\centering
\caption{AI materiality, AI function mix, and labour-margin composition.}
\includegraphics[width=0.98\textwidth]{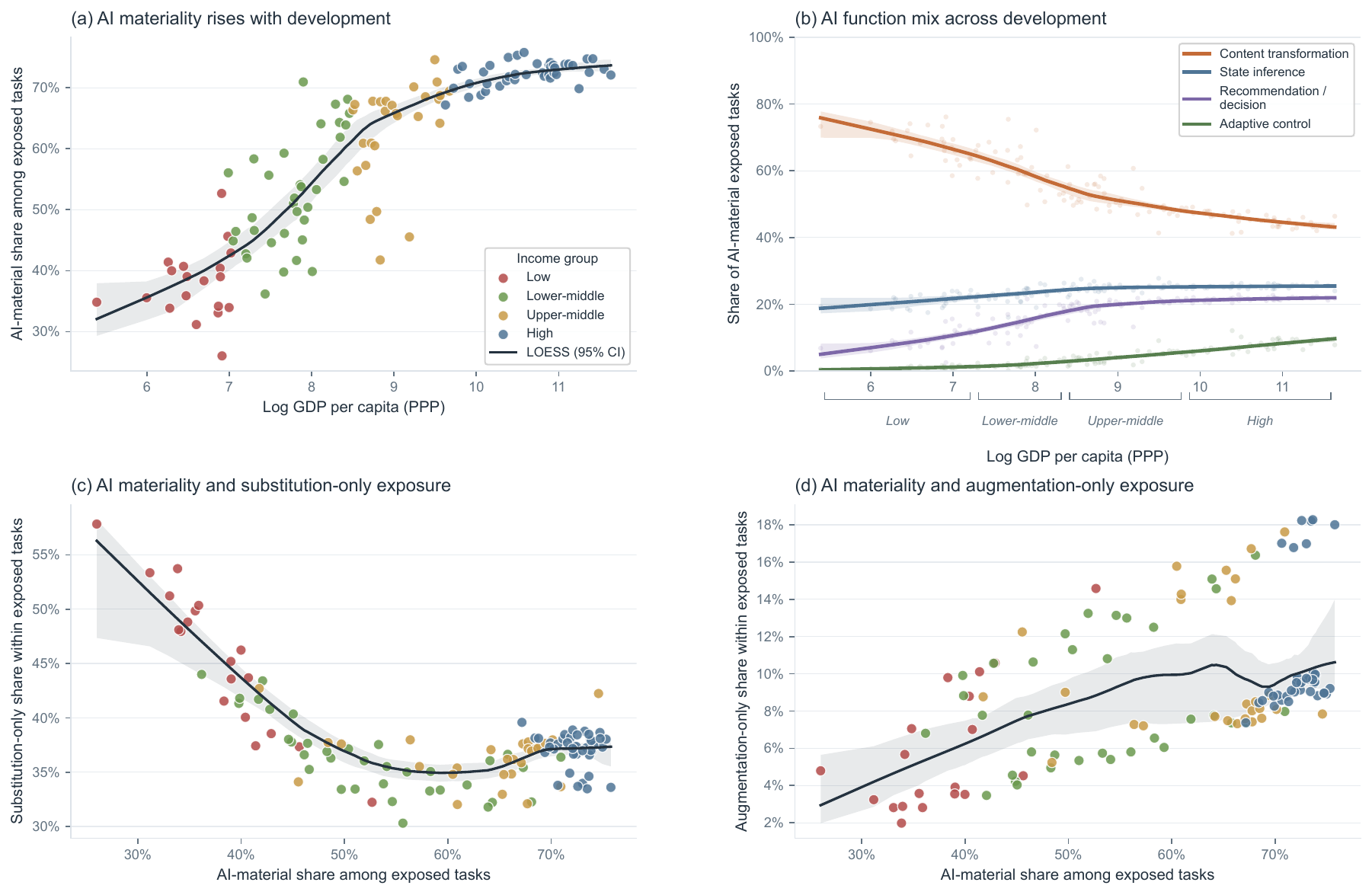}
\vspace{-0.35em}
\caption*{\scriptsize Notes: Panel~(a) plots AI-material share among exposed tasks against log GDP per capita. Panel~(b) decomposes AI-material exposure by dominant AI function. Panels~(c) and~(d) relate AI materiality to substitution-only and augmentation-only shares within exposed tasks. Points are countries, colours denote income group and curves are LOESS fits with $95\%$ bootstrap intervals.}
\label{fig:ai_materiality_margins}
\end{figure}

\begin{figure}[!htbp]
\centering
\caption{Channel exposure as a share of all tasks across country development.}
\includegraphics[width=0.90\textwidth]{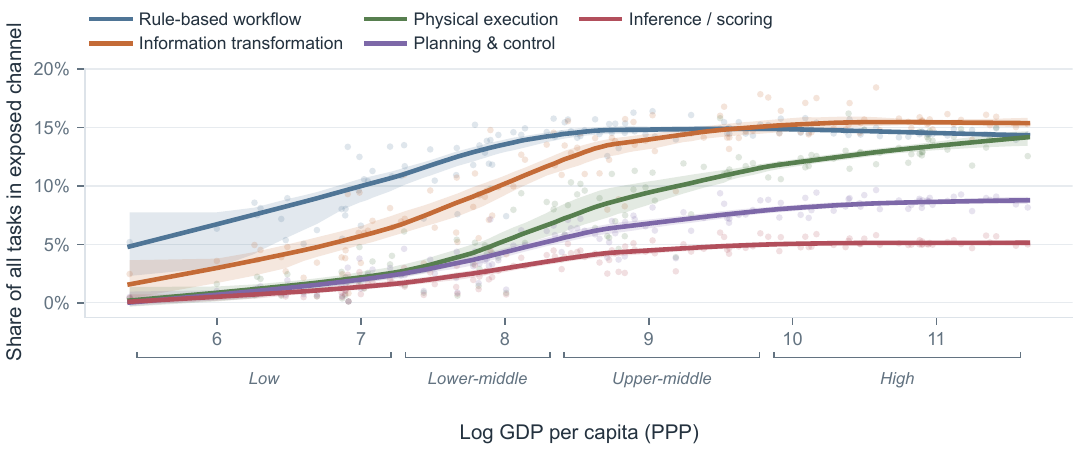}
\caption*{\scriptsize Notes: Lines report the country-level share of all tasks exposed through each dominant channel. Each line is a LOESS smooth against log GDP per capita over the $122$-country sample with complete channel and GDP data. Shaded bands are $95\%$ percentile bootstrap intervals from $200$ country-level resamples; faded points are country observations.}
\label{fig:appendix_channel_rotation_all_tasks}
\end{figure}

\begin{figure}[!htbp]
\centering
\caption{Channel composition of exposed work, broken out by World Bank income group.}
\includegraphics[width=0.97\textwidth]{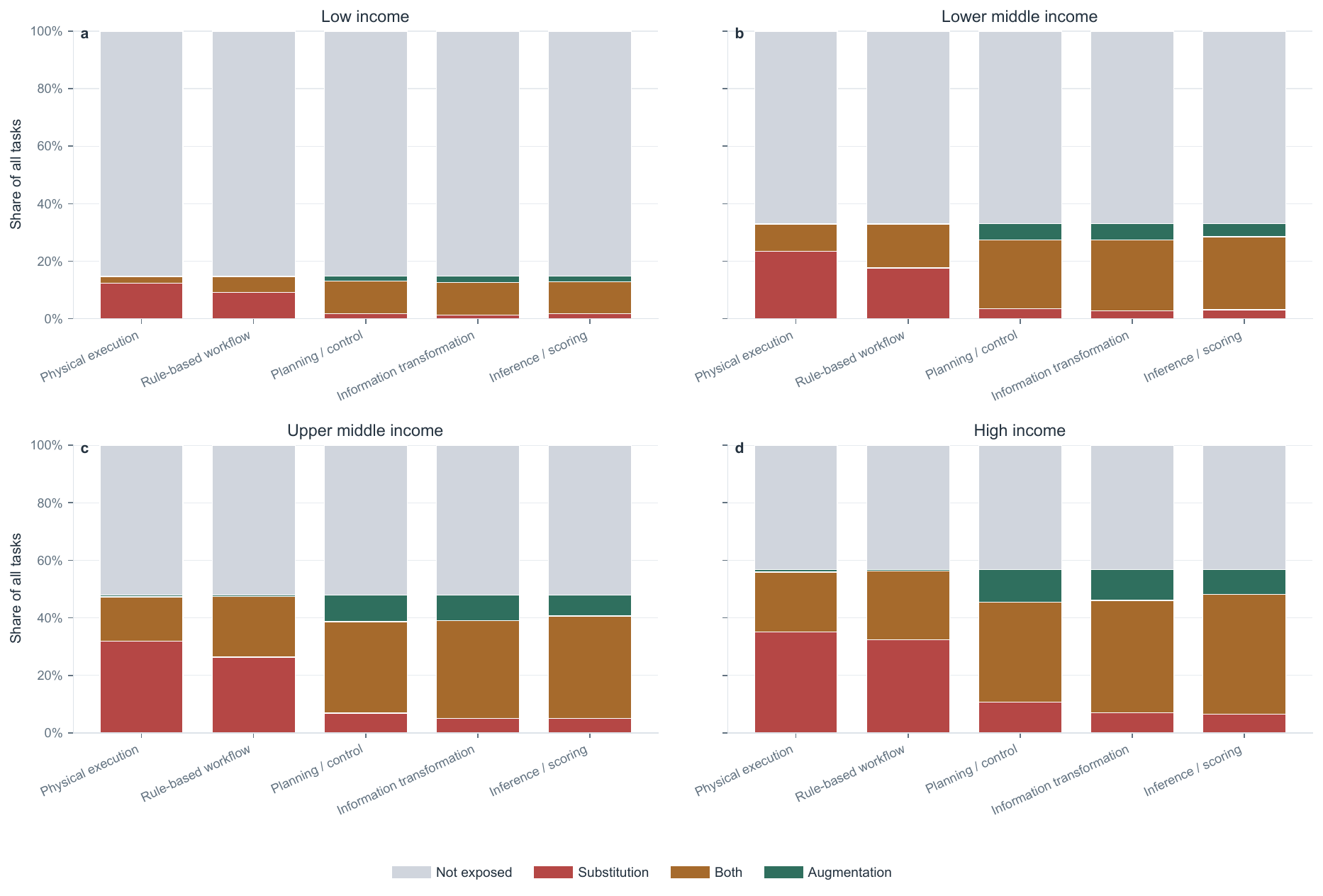}
\caption*{\scriptsize Notes: Each panel reports one World Bank income group. Within each dominant channel, coloured segments show substitution-only, balanced-both, and augmentation-only shares; the grey segment is non-exposed task mass. Bars sum to $100\%$ of tasks within each income group.}
\label{fig:appendix_channel_composition_income}
\end{figure}

\begin{figure}[!htbp]
\centering
\caption{AI-function mix within AI-material exposed tasks by income group.}
\includegraphics[width=0.88\textwidth]{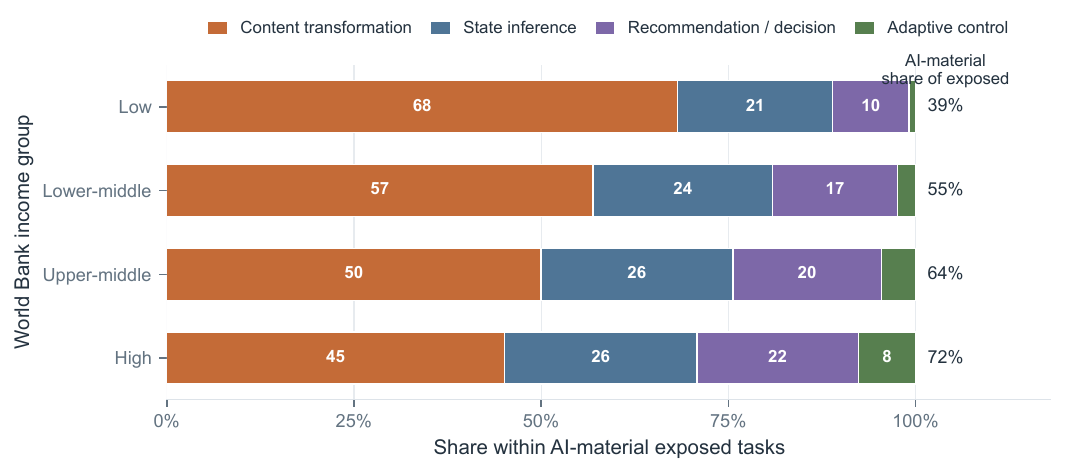}
\caption*{\scriptsize Notes: Rows are World Bank income groups. Stacked bars decompose AI-material exposed task-country observations into four dominant AI functions. Labels at right report the AI-material share among exposed observations in the same income group. The sample covers the $122$ countries with classified income groups.}
\label{fig:appendix_ai_function_income_mix}
\end{figure}

Figs.~\ref{fig:ai_materiality_margins}--\ref{fig:appendix_ai_function_income_mix} report AI materiality and labour margins, all-task channel trends, income-group channel composition and the AI-function mix.

\FloatBarrier
\subsection{Occupation and industry companions}
\label{sec:appendix_occupation_industry_companions}

This subsection aggregates the task-country labels into the ISCO and ISIC units reported in the main text. The figures report the income-group occupation and industry patterns discussed in the main text.

\begin{figure}[!htbp]
\centering
\caption{Weighted two-digit ISCO exposed shares by income group.}
\includegraphics[width=0.72\textwidth]{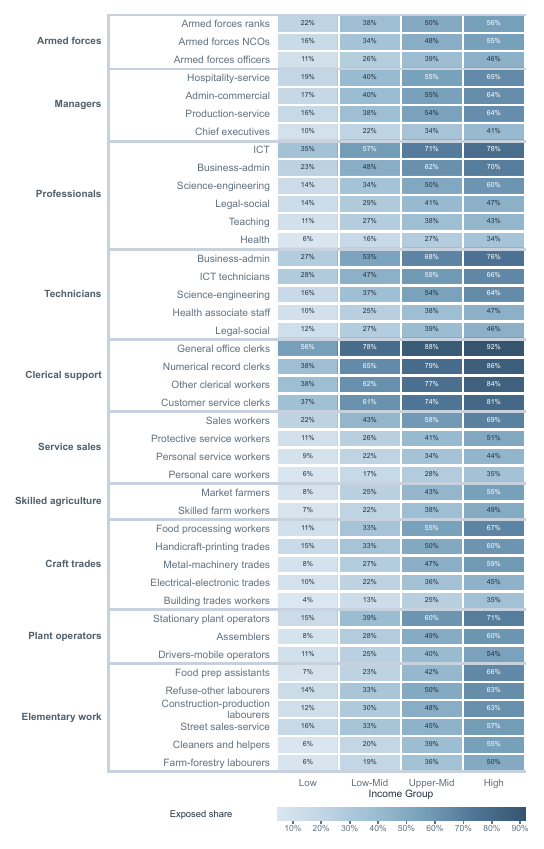}
\vspace{-0.6em}
\caption*{\scriptsize Notes: Columns are World Bank income groups. Rows are weighted two-digit ISCO occupation groups after SOC$\to$ISCO linkage. Left tags mark one-digit ISCO major groups; outlined blocks show their constituent two-digit rows. Cells report exposed shares (the share of linked tasks at exposure levels 2 or 3 after linkage).}
\label{fig:isco_income_rotation}
\end{figure}

\begin{figure}[!htbp]
\centering
\caption{Detailed two-digit ISIC task exposure by income group.}
\includegraphics[width=0.97\textwidth,height=0.84\textheight,keepaspectratio]{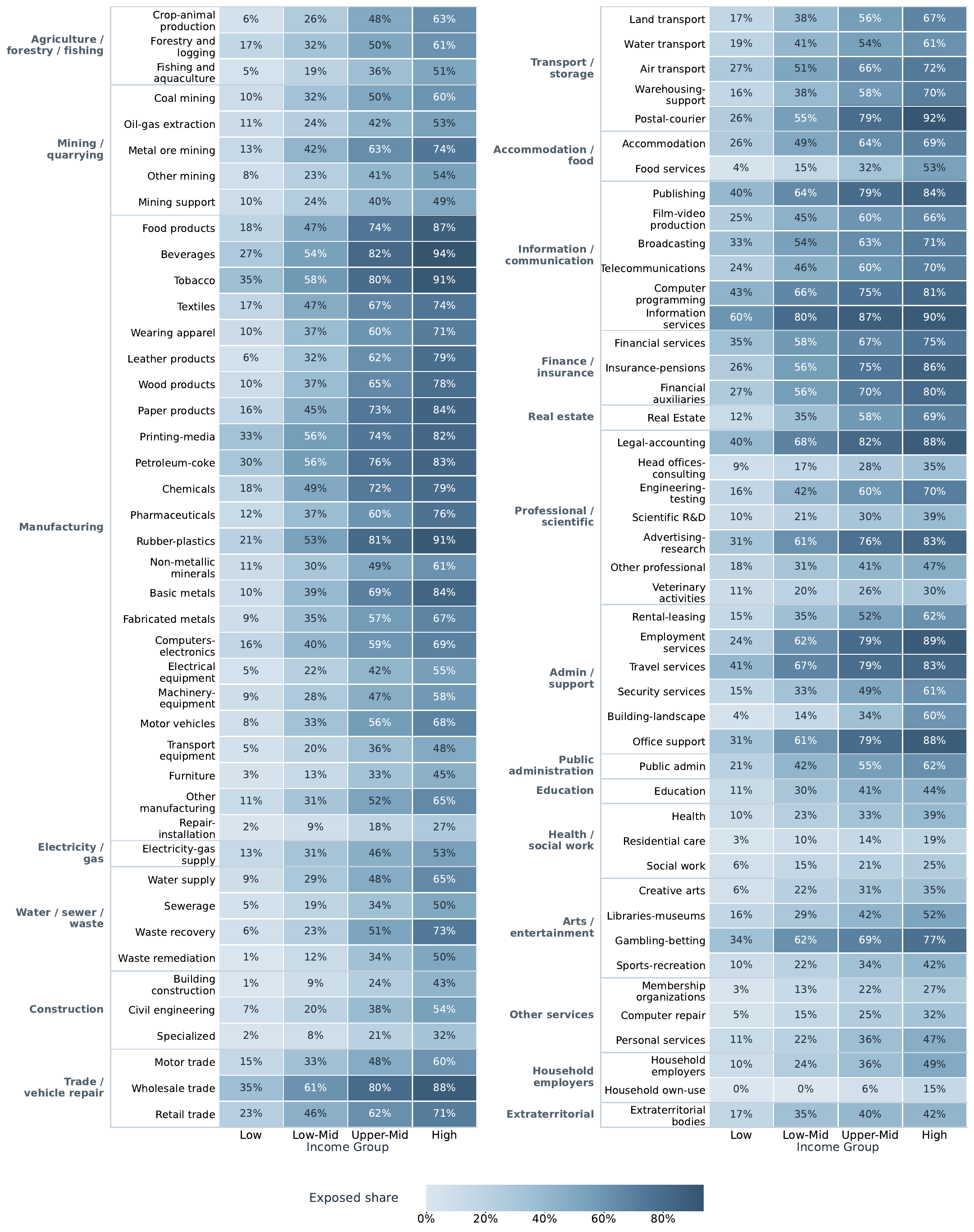}
\caption*{\scriptsize Notes: Rows are the $88$ supported two-digit ISIC divisions, grouped by official section. Within each country, exposure is computed at the retained four-digit class level, averaged equally to two-digit divisions, and then averaged equally across countries within each income group. Cells report exposed shares at exposure levels 2 or 3.}
\label{fig:industry_isic_income_main}
\end{figure}

\FloatBarrier
\subsection{Top substitution and augmentation occupations and industries by income group}
\label{sec:appendix_aggregation_route}
This subsection lists the occupation and industry rankings behind the main-text discussion of substitution-only and augmentation-only exposure. For each income group, the tables report the top entries selected by exposed share multiplied by the relevant labour-margin share among exposed tasks. The occupation table uses the weighted ISCO route; the industry table uses the bottom-up ISIC route used for the sector results. The purpose is descriptive: to show which occupations and sectors drive the substitution and augmentation patterns discussed in the main text.

\begin{table}[htbp]
\centering
\scriptsize
\setlength{\tabcolsep}{2pt}
\caption{Top-three substitution-only and augmentation-only weighted ISCO occupations within each income group.}
\label{tab:isco_income_pockets_joint}
\resizebox{0.98\textwidth}{!}{%
\begin{tabular}{@{}L{0.10\textwidth}L{0.125\textwidth}C{0.07\textwidth}C{0.06\textwidth}C{0.05\textwidth}L{0.12\textwidth}@{\hspace{0.1em}}L{0.125\textwidth}C{0.07\textwidth}C{0.06\textwidth}C{0.05\textwidth}L{0.12\textwidth}@{}}
\toprule
Income group & \multicolumn{5}{c}{Sub-only pocket} & \multicolumn{5}{c}{Aug-only pocket} \\
\cmidrule(lr){2-6}\cmidrule(lr){7-11}
 & Occupation & Exp. & Share & \% AI & Channel & Occupation & Exp. & Share & \% AI & Channel \\
\midrule
Low & Billing and Posting Clerks & 0.69 & 0.78 & 10 & \shortstack{Rule\\workflow} & Database Architects & 0.45 & 0.13 & 70 & \shortstack{Rule\\workflow} \\
 & Tellers & 0.66 & 0.81 & 9 & \shortstack{Rule\\workflow} & Atmospheric and Space Scientists & 0.31 & 0.13 & 81 & \shortstack{Info\\trans.} \\
 & Payroll and Timekeeping Clerks & 0.54 & 0.82 & 6 & \shortstack{Rule\\workflow} & Airline Pilots, Copilots, and Flight Engineers & 0.23 & 0.18 & 31 & \shortstack{Plan.\\control} \\
Low/Mid & Tellers & 0.90 & 0.77 & 24 & \shortstack{Rule\\workflow} & Air Traffic Controllers & 0.37 & 0.28 & 62 & \shortstack{Plan.\\control} \\
 & Billing and Posting Clerks & 0.89 & 0.74 & 31 & \shortstack{Rule\\workflow} & Database Architects & 0.67 & 0.15 & 80 & \shortstack{Info\\trans.} \\
 & Brokerage Clerks & 0.84 & 0.68 & 33 & \shortstack{Rule\\workflow} & Transportation Engineers & 0.37 & 0.24 & 77 & \shortstack{Info\\trans.} \\
Upper/Mid & Tellers & 0.97 & 0.80 & 32 & \shortstack{Rule\\workflow} & Air Traffic Controllers & 0.51 & 0.32 & 75 & \shortstack{Plan.\\control} \\
 & Billing and Posting Clerks & 0.95 & 0.80 & 45 & \shortstack{Rule\\workflow} & Transportation Engineers & 0.55 & 0.29 & 86 & \shortstack{Info\\trans.} \\
 & Word Processors and Typists & 0.87 & 0.80 & 51 & \shortstack{Rule\\workflow} & Electrical Engineers & 0.56 & 0.26 & 87 & \shortstack{Info\\trans.} \\
High & Tellers & 0.99 & 0.86 & 39 & \shortstack{Rule\\workflow} & Air Traffic Controllers & 0.61 & 0.33 & 84 & \shortstack{Plan.\\control} \\
 & Billing and Posting Clerks & 0.95 & 0.86 & 65 & \shortstack{Rule\\workflow} & Transportation Engineers & 0.65 & 0.30 & 93 & \shortstack{Plan.\\control} \\
 & Word Processors and Typists & 0.91 & 0.87 & 60 & \shortstack{Rule\\workflow} & Urban and Regional Planners & 0.65 & 0.27 & 96 & \shortstack{Info\\trans.} \\
\bottomrule
\end{tabular}
}
\caption*{\footnotesize Note: Rows report the top three occupations on each side within each income group. Entries are selected separately using exposed share multiplied by the relevant pathway share among exposed tasks. Exp. is the linked occupation-level exposed share; Share is the sub-only or aug-only share within exposed tasks; \% AI is the AI-material share among exposed tasks.}
\end{table}

\begin{table}[htbp]
\centering
\scriptsize
\caption{Top-three substitution-only and augmentation-only bottom-up ISIC sectors within each income group.}
\label{tab:isic_income_pockets_joint}
\resizebox{\textwidth}{!}{%
\begin{tabular}{L{0.10\textwidth}L{0.15\textwidth}cccc@{\hspace{0.6em}}L{0.15\textwidth}cccc}
\toprule
Income group & Sub. sector & Exp. & Share & \% AI & Channel & Aug. sector & Exp. & Share & \% AI & Channel \\
\midrule
Low & Printing and media & 0.25 & 0.52 & 35 & \shortstack{Rule\\workflow} & Apparel & 0.21 & 0.05 & 29 & \shortstack{Rule\\workflow} \\
 & Beverages & 0.23 & 0.54 & 19 & \shortstack{Rule\\workflow} & Beverages & 0.23 & 0.04 & 19 & \shortstack{Rule\\workflow} \\
 & Apparel & 0.21 & 0.52 & 29 & \shortstack{Rule\\workflow} & Other manufacturing & 0.16 & 0.05 & 35 & \shortstack{Rule\\workflow} \\
Low/Mid & Printing and media & 0.45 & 0.47 & 52 & \shortstack{Rule\\workflow} & Apparel & 0.47 & 0.09 & 45 & \shortstack{Rule\\workflow} \\
 & Chemicals & 0.37 & 0.54 & 32 & \shortstack{Rule\\workflow} & Other manufacturing & 0.35 & 0.09 & 44 & \shortstack{Rule\\workflow} \\
 & Beverages & 0.47 & 0.42 & 36 & \shortstack{Rule\\workflow} & Beverages & 0.47 & 0.06 & 36 & \shortstack{Rule\\workflow} \\
Upper/Mid & Paper products & 0.52 & 0.61 & 38 & \shortstack{Physical\\exec.} & Apparel & 0.65 & 0.10 & 55 & \shortstack{Rule\\workflow} \\
 & Chemicals & 0.55 & 0.52 & 44 & \shortstack{Physical\\exec.} & Other manufacturing & 0.51 & 0.10 & 55 & \shortstack{Rule\\workflow} \\
 & Petroleum and coke & 0.74 & 0.38 & 64 & \shortstack{Physical\\exec.} & Beverages & 0.62 & 0.08 & 52 & \shortstack{Rule\\workflow} \\
High & Paper products & 0.63 & 0.66 & 46 & \shortstack{Physical\\exec.} & Apparel & 0.75 & 0.11 & 66 & \shortstack{Info\\trans.} \\
 & Petroleum and coke & 0.83 & 0.42 & 76 & \shortstack{Physical\\exec.} & Pharmaceuticals & 0.73 & 0.10 & 66 & \shortstack{Physical\\exec.} \\
 & Chemicals & 0.66 & 0.52 & 56 & \shortstack{Physical\\exec.} & Beverages & 0.71 & 0.10 & 68 & \shortstack{Rule\\workflow} \\
\bottomrule
\end{tabular}
}
\caption*{\footnotesize Note: Rows report the top three sectors on each side within each income group. Entries are selected separately using exposed share multiplied by the relevant pathway share among exposed tasks. Exp. is the industry-level exposed share; Share is the substitution-only or augmentation-only share within exposed tasks; \% AI is the AI-material share among exposed tasks. The retained bottom-up ISIC graph is narrower than the occupation layer.}
\end{table}

\subsection{Observed employment composition from ILOSTAT}
\label{sec:appendix_ilostat_employment}
This subsection reports two employment-composition exercises. The first reweights occupation exposure using total employment across ISCO-08 major groups. For this exercise, we collapse the weighted SOC$\rightarrow$ISCO bridge to the same level. The second uses sex-specific employment data at the two-digit ISCO-08 occupation and ISIC Rev.~4 industry levels to compare female and male exposure.

Table~\ref{tab:appendix_ilostat_coverage} reports the data level, country-year selection rule, and final sample for each exercise. The broad occupation reweighting and the sex-specific analyses use separate ILOSTAT extracts and coverage rules.

\begin{table}[H]
\centering
\scriptsize
\setlength{\tabcolsep}{2pt}
\renewcommand{\arraystretch}{1.08}
\caption{ILOSTAT samples used in the employment-composition analyses.}
\begin{tabularx}{\textwidth}{@{}L{0.16\textwidth}L{0.23\textwidth}X C{0.08\textwidth} C{0.08\textwidth}@{}}
\toprule
Analysis & ILOSTAT breakdown & Coverage rule & Usable countries & Paper sample \\
\midrule
Broad occupation reweighting
& Total employment by ISCO-08 major group
& Latest 2015--2025 year with at least eight major groups and positive employment; shares normalised within country-year.
& 152 & 91 \\

Gender: occupations
& Female and male employment by two-digit ISCO-08 group
& Latest common female--male year in 2018--2024; shares normalised within sex and country.
& 135 & 88 \\

Gender: industries
& Female and male employment by two-digit ISIC Rev.~4 division
& Latest common female--male year in 2018--2024; shares normalised within sex and country.
& 105 & 72 \\
\bottomrule
\end{tabularx}
\caption*{\scriptsize Note: `Usable countries' reports the countries satisfying each source-specific coverage rule before matching to the atlas and applying the paper sample restrictions. `Paper sample' reports the countries used in the corresponding analysis. The fixed-effect gender analysis further normalises female and male shares over cells observed for both sexes within each country.}
\label{tab:appendix_ilostat_coverage}
\end{table}

\begin{figure}[!htbp]
\centering
\caption{Employment-weighted versus linkage-weighted occupation exposure by country.}
\includegraphics[width=0.83\textwidth]{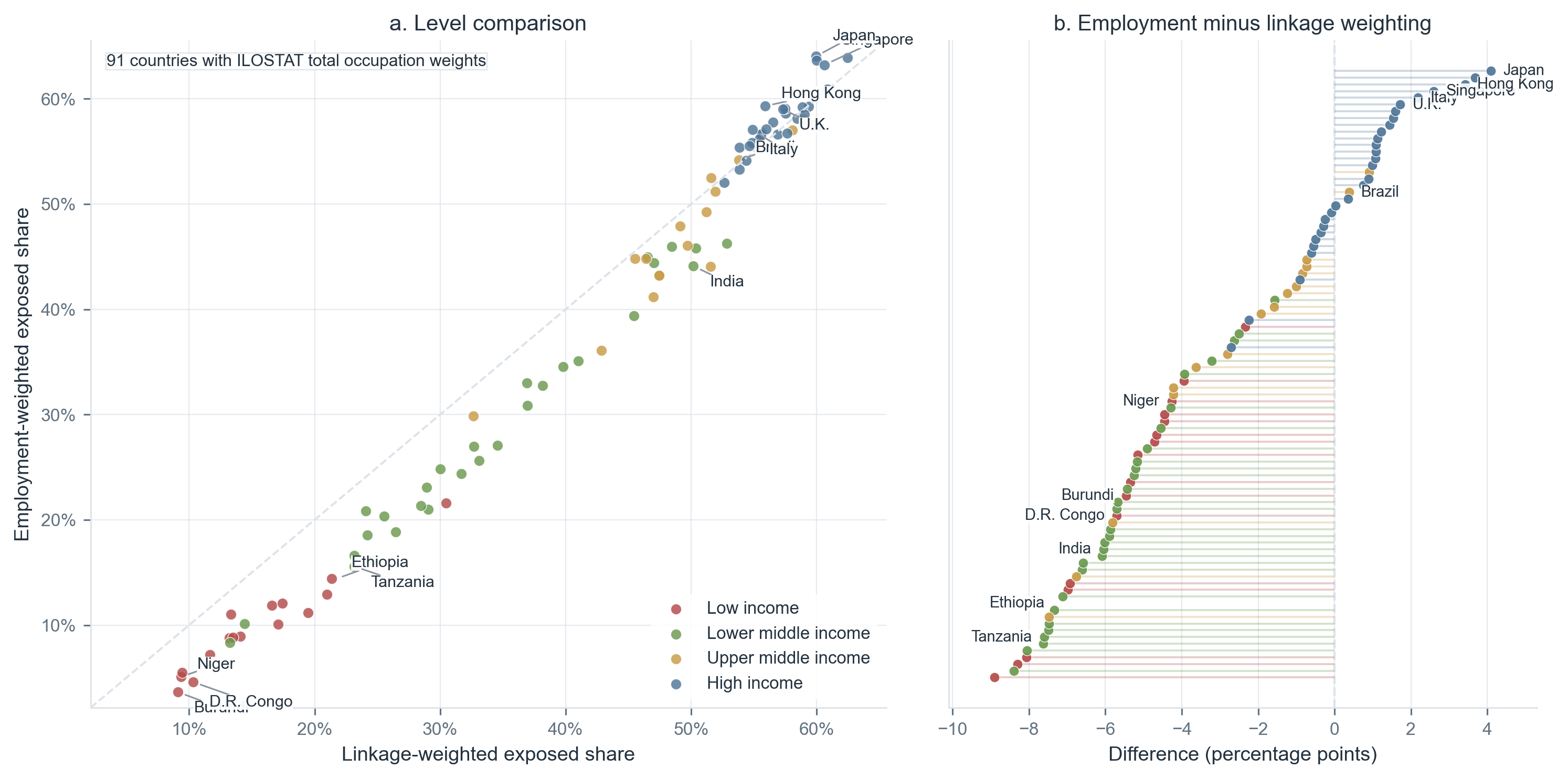}
\caption*{\scriptsize Notes: Each point is one of the 91 countries in the atlas--ILOSTAT overlap sample with usable total-employment occupation weights. The x-axis reports linkage-weighted exposed share from the task$\to$SOC$\to$ISCO bridge; the y-axis reports the same summary reweighted by observed ISCO-08 major-group employment shares. The 45-degree line marks equality.}
\label{fig:appendix_ilostat_country_weighting}
\end{figure}

Employment weighting preserves the broad occupation-layer gradient. High-income countries remain higher-exposure and low-income countries remain lower-exposure on average. But employment composition still moves some country summaries meaningfully because exposed major groups do not account for the same share of workers everywhere.

The gender analysis uses separate, more detailed ILOSTAT extracts than the major-group exercise above. The main-text gender analysis uses two-digit female and male employment shares for 88 countries in the occupation analysis and 72 countries in the industry analysis. Female and male shares are normalised separately over the cells observed for each sex. The comparison therefore asks whether employed women and men are distributed across occupation or industry cells with different automation-margin profiles.

For country $c$, sex $s$, cell $j$, and margin $m$, the employment-weighted contribution is
\[
E_{csm}=\sum_j s_{cjs}x_{cjm},
\]
where $s_{cjs}$ is the employment share of sex $s$ in occupation or industry cell $j$, and $x_{cjm}$ is the corresponding country-conditioned share of linked task mass assigned to margin $m$. The plotted gap is \(D_{cm}=E_{cFm}-E_{cMm}\), so positive values indicate higher female employment-weighted exposure and negative values indicate higher male employment-weighted exposure.

To separate employment sorting from country-specific exposure differences, define \(g_{cj}=s_{cjF}-s_{cjM}\), and let \(\bar{x}_{jm}\) denote the simple mean exposure of cell \(j\) across countries. The country-level gap can then be written as
\[
D_{cm}
=
\underbrace{\sum_j g_{cj}\bar{x}_{jm}}_{\text{employment sorting}}
+
\underbrace{\sum_j g_{cj}(x_{cjm}-\bar{x}_{jm})}_{\text{country-specific exposure}}.
\]
The employment-sorting component applies each country's female-minus-male employment profile to a common occupation or industry exposure profile. The second captures whether country-specific exposure differences reinforce or offset that employment profile. The two components sum exactly to the country-level gap.

Table~\ref{tab:gender_gap_decomposition_by_income} reports these components by income group for occupations and industries.

\begin{table}[!htbp]
\centering
\caption{Accounting decomposition of female-minus-male employment-weighted exposure gaps.}
\label{tab:gender_gap_decomposition_by_income}
\footnotesize
\setlength{\tabcolsep}{3.8pt}
\begin{tabular}{@{}lllrrrr@{}}
\toprule
 &  &  &  & \multicolumn{3}{c}{Mean gap (percentage points)} \\
\cmidrule(lr){5-7}
Domain & Margin & Income & Countries & Total & Employment sorting & Country-specific exposure \\
\midrule
Occupation & Substitution-only & Low & 14 & 1.07 & 1.76 & -0.69 \\
 &  & Lower middle & 26 & 1.70 & 1.99 & -0.29 \\
 &  & Upper middle & 16 & 1.85 & 1.82 & 0.03 \\
 &  & High & 32 & 2.08 & 2.12 & -0.04 \\
\addlinespace[2pt]
Occupation & Augmentation-only & Low & 14 & -0.09 & -0.13 & 0.05 \\
 &  & Lower middle & 26 & 0.07 & 0.02 & 0.05 \\
 &  & Upper middle & 16 & -0.01 & 0.12 & -0.13 \\
 &  & High & 32 & -0.15 & -0.19 & 0.03 \\
\addlinespace[2pt]
Industry & Substitution-only & Low & 4 & 1.05 & 2.77 & -1.72 \\
 &  & Lower middle & 21 & 1.59 & 1.72 & -0.14 \\
 &  & Upper middle & 16 & 0.99 & 0.89 & 0.10 \\
 &  & High & 31 & -2.59 & -1.77 & -0.82 \\
\addlinespace[2pt]
Industry & Augmentation-only & Low & 4 & -0.34 & -0.82 & 0.47 \\
 &  & Lower middle & 21 & -0.44 & -0.75 & 0.31 \\
 &  & Upper middle & 16 & -1.08 & -0.82 & -0.25 \\
 &  & High & 31 & -0.89 & -0.75 & -0.14 \\
\bottomrule
\end{tabular}
\caption*{\scriptsize Notes: Positive values indicate higher female employment-weighted exposure and negative values indicate higher male employment-weighted exposure. The total gap equals the employment-sorting and country-specific exposure components for every country. The employment-sorting component applies each country's female-minus-male employment weights to the cross-country mean exposure of each occupation or industry cell. The country-specific exposure component uses deviations from that cell mean. Income-group means preserve the exact additive decomposition.}
\end{table}

For occupations, the mean substitution gap is \(1.77\) percentage points. The employment sorting component contributes \(1.97\) points, while country-specific exposure differences contribute \(-0.20\) points. The occupation result therefore mainly reflects gender sorting across occupations with different average substitution exposure.

The industry result is more income-dependent. Substitution exposure is female-weighted in low- and middle-income economies but male-weighted in high-income economies. In high-income economies, the mean industry gap is \(-2.59\) percentage points, with a \(-1.77\)-point employment sorting component and a \(-0.82\)-point country-specific component. Augmentation gaps are small across occupations and male-weighted across industries. The low-income industry estimates cover four countries and should be interpreted cautiously.

The decomposition in Table~\ref{tab:gender_gap_decomposition_by_income} underlies panels~(a) and~(b) of the main-text gender panels and extends the comparison to augmentation-only exposure. It shows that the gender gap is mainly a composition result: women and men are employed in different occupation and industry cells, and those cells carry different substitution and augmentation profiles. The result should not be read as evidence that automation technologies are intrinsically female- or male-biased. It shows how existing gender segregation interacts with the labour margins of exposed work.

\subsection{Country-predictor robustness}
\label{sec:appendix_residual_country_explanations}
These checks ask whether the country-predictor ranking changes with sample size, denominator, attribution rule, or correlated covariates. We report a residual-fit check in text and two figure companions: the earlier random-forest permutation-importance version and a linear Shapley $R^2$ companion for the 68-country predictor sample. Additional checks show that the wider-coverage $90$-country specification preserves the same broad capability cluster, that adding the balanced-both margin to the within-exposed composition gives the same qualitative reading, and that winsorising the country covariates at the 1st/99th and 5th/95th percentiles leaves the leading capability cluster unchanged. For the augmentation-only outcome, the leading capital-intensity result is also unchanged in leave-one-country-out refits. These analyses describe country-level patterns and do not change the task labels. Table~\ref{tab:appendix_country_covariate_dictionary} defines the country covariates used in the main and appendix screens.

\begin{table}[!htbp]
\centering
\footnotesize
\caption{Country covariates used in the country-covariate screens.}
\label{tab:appendix_country_covariate_dictionary}
\begin{tabularx}{\textwidth}{>{\raggedright\arraybackslash}p{0.19\textwidth}>{\raggedright\arraybackslash}X>{\raggedright\arraybackslash}p{0.26\textwidth}>{\raggedright\arraybackslash}p{0.12\textwidth}}
\toprule
Figure label & Construction & Source & Year rule \\
\midrule
Log GDP per capita & Natural log of GDP divided by population, using the country-list GDP and population fields. & World Development Indicators \citep{worldbank_wdi} & 2024 \\
Human capital & PWT human-capital index, \texttt{hc}, based on schooling and returns to education. & Penn World Table 10.01 \citep{FeenstraInklaarTimmer2015PWT} & 2019 \\
Years of schooling & Average years of schooling among adults aged 15--64. & Barro--Lee educational-attainment data \citep{BarroLee2013Education} & 2015 \\
Capital intensity & Natural log of real capital stock per worker, \(\log(\texttt{rkna}/\texttt{emp})\), where \texttt{emp} is persons engaged. & Penn World Table 10.01 \citep{FeenstraInklaarTimmer2015PWT} & 2019 \\
Investment (\% GDP) & Gross fixed capital formation as a percentage of GDP. & World Development Indicators \citep{worldbank_wdi} & Latest non-missing, 2018--2024 \\
Government effectiveness & Government effectiveness percentile rank. & Worldwide Governance Indicators \citep{worldbank_wgi} & 2024 \\
Regulatory quality & Regulatory quality percentile rank. & Worldwide Governance Indicators \citep{worldbank_wgi} & 2024 \\
Internet users (\%) & Individuals using the Internet as a percentage of population. & World Development Indicators \citep{worldbank_wdi} & Latest non-missing, 2018--2024 \\
Goods trade (\% GDP) & Merchandise-trade value divided by GDP. & CEPII BACI HS22 2023 trade data \citep{GaulierZignago2010BACI} and World Bank GDP \citep{worldbank_wdi} & 2023 trade; latest GDP \\
\bottomrule
\end{tabularx}
\caption*{\scriptsize Notes: The labels match the country-covariate figures. The main 68-country random-forest specification uses all rows in the table. The wider-coverage 90-country specification drops human capital, capital intensity, and investment to reduce complete-case sample loss. PWT variables are constructed national-accounts and education measures rather than survey-question variables. The PWT rows use 2019 values, the latest year in PWT 10.01 and a pre-COVID baseline. All logarithms are natural logs.}
\end{table}

Before imposing the full complete-case restriction, government effectiveness and regulatory quality are each available for 118 countries. In the 68-country main sample they correlate at \(0.96\), with variance-inflation factors of \(22.1\) and \(19.2\), respectively. We retain both because they measure different institutional concepts, but interpret them as part of a common governance and development cluster rather than as separate marginal effects.

As a residual-fit check, we regress raw country exposure and substitution-only share, and their GDP- or income-group residuals, on the same broad capability covariates. The structural covariate set explains the raw cross-country gradient well, with adjusted $R^2$ of about $0.93$ for mean exposure and $0.92$ for substitution-only share. Fit falls after conditioning on development stage: after removing income-group means, the same covariates explain about $0.11$ of exposure residuals and $0.19$ of substitution-only residuals. This supports the interpretation that the capability cluster captures much of the rich-poor gradient, while leaving substantial within-income country variation.

The wider-coverage $90$-country specification addresses sample composition. The richer main-text specification retains more structural content, but its complete-case sample is smaller. Dropping the lowest-coverage structural variables expands country coverage and leaves GDP, digital connectivity, governance, and schooling as the dominant available cluster. A within-exposed version that adds the balanced-both margin gives the same qualitative reading: substitution-only and balanced-both shares track the development-capability cluster, while augmentation-only exposure is more closely associated with capital intensity.

\begin{figure}[!htbp]
\centering
\caption{Permutation-importance companion for country predictors.}
\includegraphics[width=0.96\textwidth]{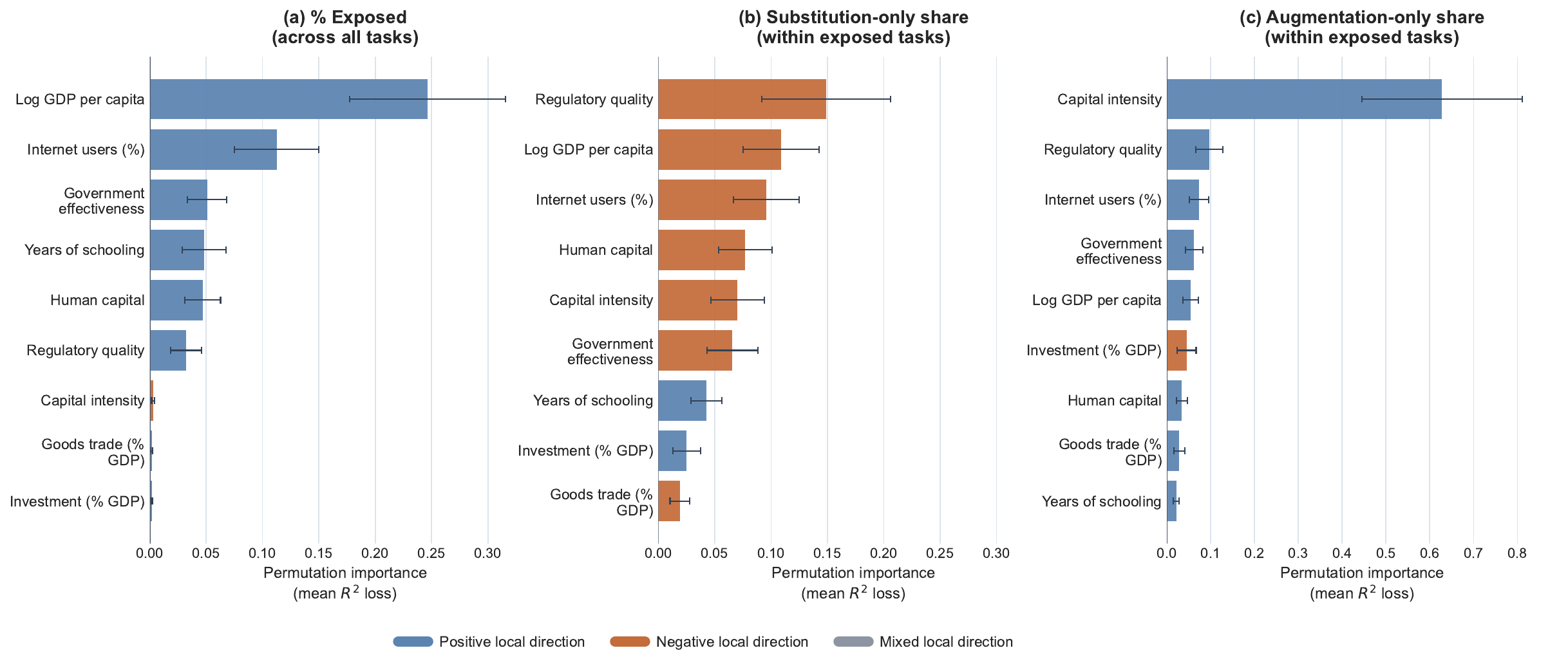}
\caption*{\scriptsize Notes: Bars report the earlier random-forest permutation-importance ranking in the same 68-country complete-case sample as Figure~\ref{fig:country_covariate_feature_importance}. Outcomes are exposed share among all tasks and substitution-only or augmentation-only shares within exposed tasks. Colours use the same five-forest accumulated-local-effect direction rule as Figure~\ref{fig:country_covariate_feature_importance}: positive or negative requires sign agreement in at least four of five forests; otherwise the direction is mixed. This companion is reported because unconditional permutation importance can be sensitive when predictors are correlated.}
\label{fig:appendix_country_covariate_permutation_importance}
\end{figure}

\begin{figure}[!htbp]
\centering
\caption{Linear Shapley companion for the country-predictor specification.}
\includegraphics[width=0.96\textwidth]{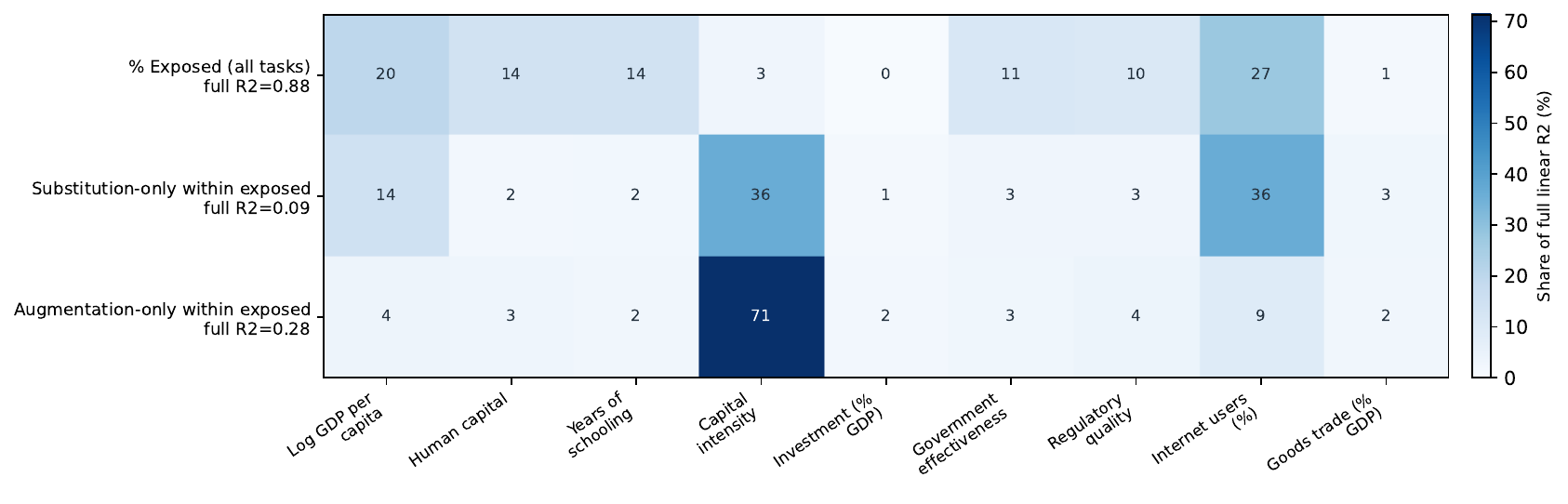}
\caption*{\scriptsize Notes: Cells report exact dominance-analysis contributions from linear models on the same 68-country complete-case sample as the main country-predictor figure. Contributions average each covariate's incremental $R^2$ over all possible orderings. The outcomes are exposed share, substitution-only share within exposed tasks, and augmentation-only share within exposed tasks. This is a linear variance-decomposition companion, distinct from the TreeSHAP attribution used for the random-forest figures.}
\label{fig:appendix_country_predictor_linear_shapley}
\end{figure}

\FloatBarrier
\subsection{Rationale-concept results}
\label{sec:appendix_rationale_concept_results}

Forty-five of the 60 fixed concepts pass the Bonferroni-adjusted significance screen in the heldout samples.

\begin{table}[!htbp]
\centering
\caption{Rationale-concept contrasts.}
\label{tab:appendix_rationale_concept_target_families}
\begin{tabular}{L{6.0cm} C{2.0cm} C{2.2cm} C{2.6cm} C{1.5cm}}
\toprule
Target contrast & Fixed child concepts & Bonferroni significant & Heldout comparison size & Main panel \\
\midrule
Exposure: exposed versus non-exposed labels for the same task across countries & 20 & 20 & 2,000 same-task pairs & d \\
Substitution-only: substitution-only versus other exposed labels & 20 & 13 & 2,000 same-task pairs & e \\
Augmentation-only: augmentation-only versus other exposed labels & 20 & 12 & 2,000 same-task pairs & f \\
\midrule
Total displayed contrasts & 60 & 45 & -- & -- \\
\bottomrule
\end{tabular}
\caption*{\footnotesize Notes: The table reports the concept inventory used for the paper-facing rationale-concept analysis. ``Fixed child concepts'' are concepts selected before heldout coding. ``Bonferroni significant'' counts concepts whose heldout differences remain significant after adjustment within contrast.}
\end{table}

\clearpage
\paragraph{Illustrative paired rationales}
Table~\ref{tab:appendix_country_context_rationale_examples} reports illustrative same-task rationale pairs from the exposure contrast and shows the rationale texts used in the concept-coding analysis.
{\footnotesize
\setlength{\tabcolsep}{2.5pt}
\renewcommand{\arraystretch}{1.16}

\begin{longtable}{L{0.15\textwidth}L{0.18\textwidth}L{0.31\textwidth}L{0.31\textwidth}}
\caption{Illustrative same-task rationale pairs for the current exposure concepts.}
\label{tab:appendix_country_context_rationale_examples} \\
\toprule
Concept family & Task & Exposed-country example & Non-exposed-country example \\
\midrule
\endfirsthead

\toprule
Concept family & Task & Exposed-country example & Non-exposed-country example \\
\midrule
\endhead

\midrule
\multicolumn{4}{r}{\footnotesize Continued on next page} \\
\endfoot

\bottomrule
\multicolumn{4}{p{0.98\textwidth}}{\footnotesize Notes: Each row shows one same-task country pair from the current heldout exposure sample. The pairs were selected for readability among observations with the largest family-score difference in the displayed direction. Selection took place after the child concepts were fixed and does not affect the concept estimates. The rationale text is shortened for compactness.} \\
\endlastfoot

Capital-intensive physical automation
& Heat-treat parts, using acetylene torches.
& Belgium: automated induction heating and robotic torch manipulation can replace manual heat-treating for many industrial parts, offering higher consistency and throughput than manual acetylene torch work.
& Congo, Dem. Rep.: this manual craft task requires high-dexterity physical manipulation and on-site expert judgement, making it economically non-automatable in the current infrastructure. \\

Structured document-production workflows
& Prepare reports on students and activities as required by administration.
& China: automated systems can aggregate student data and generate coherent performance narratives, reducing teachers' manual effort while allowing them to refine or verify AI-generated insights.
& Chad: fragmented and largely manual record-keeping prevents the systematic data ingestion required for automation, limiting technology to simple digital assistance. \\

Formal administrative workflow systems
& Maintain records of engineering department activities, including expenses and equipment maintenance and repairs.
& Algeria: digital record-keeping systems automate data entry and maintenance scheduling, reducing the labour required for manual filing and tracking.
& Guinea: digital record-keeping tools assist with organization but do not materially reduce human input into record creation, verification, and entry. \\

Sensorized monitoring capacity
& Protect hotel and guest property and report damage, theft, and found articles.
& Serbia: modern video analytics allow automated detection of suspicious behaviour or missing items, reducing constant manual monitoring while helping guards respond to validated alerts.
& Rwanda: the task relies on human situational judgement, physical presence, and security discretion that are not currently automatable at scale in typical hotel operations. \\

Relational and trust-based markets
& Interview people and keep track of their responses.
& Tunisia: digital survey tools and transcription models automate data capture and structuring, reducing manual recording and processing, although human-led interviewing remains useful for rapport.
& Guinea: the task relies on human rapport, local-language nuance, and trust-building, which are not currently automatable at scale. \\

Missing digital and data complements
& Record production and test data for each food-product batch.
& Dominican Republic: digitized ERP systems allow automated data capture from production machinery, reducing reliance on manual record-keeping.
& South Sudan: data recording remains manual or paper-based, without the digital infrastructure needed for widespread automation or material labour reallocation. \\

Localized knowledge and fragmented information
& Read news flashes to inform audiences of important events.
& Cote d'Ivoire: AI-driven text-to-speech and automated news summarization can replace human readers in routine bulletins and help journalists tailor news segments.
& South Sudan: the task relies on human presence, language nuance, and local credibility, which cannot be credibly substituted or materially augmented in typical settings. \\

\end{longtable}
}

\FloatBarrier

\FloatBarrier
{\small
\bibliographystyle{unsrtnat}
\bibliography{references,additional_references}
}

\end{document}